%% file: efosiv3.tex
\documentstyle[12pt,epsf,epsfig]{article}
\def\ga{\mathrel{\raise.3ex\hbox{$>$\kern-.75em\lower1ex\hbox{$\sim$}}}}
\def\la{\mathrel{\raise.3ex\hbox{$<$\kern-.75em\lower1ex\hbox{$\sim$}}}}
\def\gyr{{\rm \, G\kern-0.125em yr}}
\def\gev{{\rm \, Ge\kern-0.125em V}}
\def\tev{{\rm \, Te\kern-0.125em V}}
\def\beq{\begin{equation}}
\def\eeq{\end{equation}}
\def\ss{\scriptscriptstyle}
\def\scs{\scriptstyle}
\def\mb{m_{\widetilde B}}
\def\mst{m_{\tilde\tau_R}}
\def\stau{\tilde \tau}
\def\mchi{m_{\tilde \chi}}
\def\msf{m_{\tilde f}}
\def\m12{m_{1\!/2}}

\def\mf{m_{\ss{f}}}

\def\ohsq{\Omega_{\widetilde\chi}\, h^2}
\def\ch{{\widetilde \chi}}
\def\st{{\widetilde \tau}_{\scriptscriptstyle\rm R}}
\def\sm{{\widetilde \mu}_{\scriptscriptstyle\rm R}}
\def\sel{{\widetilde e}_{\scriptscriptstyle\rm R}}
\def\sl{{\widetilde \ell}_{\scriptscriptstyle\rm R}}

\def\tsq{|{\cal T}|^2}
\def\tcm{\theta_{\rm\scriptscriptstyle CM}}
\def\half{{\textstyle{1\over2}}}
\def\neq{n_{\rm eq}}
\def\qeq{q_{\rm eq}}
\def\slash#1{\rlap{\hbox{$\mskip 1 mu /$}}#1}%
\def\mw{m_W}
\def\mz{m_Z}
\def\mhb{m_{H}}
\def\mhl{m_{h}}
\newcommand\f[1]{f_#1}
\def\nl{\hfill\nonumber\\&&}

\begin{document}
\begin{titlepage}
\pagestyle{empty}
\baselineskip=18pt
\rightline{hep-ph/9905481}
\rightline{CERN-TH/99-146}
\rightline{MADPH-99-1117}
\rightline{TPI--MINN--99/28}
\rightline{UMN--TH--1801/99}
\rightline{May 1999}
\vskip.25in
\begin{center}

{\large{\bf Calculations of Neutralino-Stau Coannihilation
Channels and
 the Cosmologically Relevant Region of MSSM Parameter Space}}
\end{center}
\begin{center}
\vskip 0.1in
{John Ellis}

{\it Theory Division, CERN, CH-1211 Geneva 23, Switzerland}

{Toby Falk}

{\it Department of Physics, University of Wisconsin, Madison, WI~53706,
USA}

Keith A.~Olive

{\it Theoretical Physics Institute,
{School of Physics and Astronomy,
University of Minnesota, Minneapolis, MN 55455, USA}\\}

and

Mark Srednicki

{\it Department of Physics,
University of California, Santa Barbara, CA 93106, USA}
\vskip 0.1in
{\bf Abstract}
\end{center}
\baselineskip=18pt \noindent

Assuming that the lightest supersymmetric particle (LSP)
is the lightest neutralino $\ch$,
we present a detailed exploration of
neutralino-stau ($\ch - {\tilde \tau}$) 
coannihilation channels, including analytical expressions
and numerical results. We also include ${\ch }$ coannihilations
with the $\tilde e$ and $\tilde \mu$.
We evaluate the implications of coannihilations for the
cosmological relic density of the LSP, which is assumed to be stable,
in the constrained  minimal supersymmetric extension of the Standard Model
(CMSSM), in which the soft supersymmetry-breaking parameters are
universal at the supergravity GUT scale. We evaluate the
changes due to coannihilations in the region of the MSSM parameter
space that is consistent with the cosmological upper limit
on the relic LSP density. In particular, we find that
the upper limit on $m_{\ch}$ is increased from
about $200$~GeV to about $600$~GeV in the CMSSM, 
and estimate a qualitatively similar increase for gauginos in
the general MSSM. 
\end{titlepage}
\baselineskip=18pt
\section{Introduction}

One of the most appealing candidates for the cold dark matter
in the Universe is the lightest supersymmetric particle (LSP).
This is stable if the quantum number $R \equiv (-1)^{3B + L + 2S}$
is conserved \cite{pf}, as in the minimal supersymmetric extension of
the Standard Model (MSSM) \cite{MSSM}, and hence a candidate relic from
the Big Bang. Stringent upper limits on the relative abundances of
anomalous heavy isotopes suggest that the relic LSP should be electrically
neutral with no strong interactions, so as to ensure that it does not
bind to nuclei. Weakly-interacting candidates for the LSP, within the
MSSM, include the sneutrinos $\tilde \nu_i$ and the lightest neutralino
$\ch$. LEP limits on $Z^0 \rightarrow {\rm invisible~neutral~particles}$
suggest that $m_{\tilde \nu_i} \ga M_{Z^0}/2$, in which case direct
searches for dark matter particles along with cosmological constraints,
remove any sneutrino from consideration\footnote{However, sneutrinos may
be
acceptable as dark matter if the MSSM is extended to include additional
lepton number violating superpotential terms~\cite{mur}.} as the dark
matter in the MSSM
\cite{fkosi}. Thus supersymmetric dark matter is commonly thought to
consist of $\ch$ neutralino particles \cite{hg,ehnos}.

It is a remarkable feature of $\ch$ dark matter that its
cosmological relic density naturally~\cite{natural} falls in the range
allowed by
cosmology and preferred by astrophysics in generic domains
of MSSM parameter space~\cite{ehnos}. This is in agreement with general
arguments that the mass of a cold dark matter particle whose
relic density is fixed at freeze-out from thermal equilibrium
should not be more than $\sim \sqrt{T_0 \times M_P} \sim 1$~TeV, where
$T_0 \sim 2.73$~K is the present cosmic microwave background
temperature and $M_P \sim 1.2 \times 10^{19}$~GeV. This
general argument suggests that such a dark matter particle
should be detectable in experiments at the LHC~\cite{Dimopoulos}, which is
also the conclusion reached by studies of the physics reach 
of the LHC in the parameter space of the MSSM~\cite{CMS}.

It has been appreciated for some time that the relic density at
freeze-out may be sensitive to coannihilation processes involving the
LSP and heavier sparticles, ${\slash \ch}$~\cite{gs,co2}. The relative
importance of such
coannihilation effects is controlled essentially by the ratio of
coannihilation and annihilation cross sections: $\sigma_{\ch \slash 
  \ch}/ \sigma_{\ch \ch}, \sigma_{\slash \ch \slash \ch} / \sigma_{\ch
  \ch}$ and the ratio of number densities, which is determined by a
Boltzmann factor: exp$(m_{\slash \ch} - m_{\ch}) / T_f$, where $T_f$
is the freeze-out temperature. Since, typically, $T_f = {\cal O}
(m_{\ch} / 20)$, this latter factor might suggest that coannihilation
effects would normally be important only for $m_{\slash \ch} - m_{\ch}
\sim T_f \sim m_{\ch} / 20$. However, $\sigma_{\ch \ch}$ is often
suppressed by mass and/or phase-space factors in the non-relativistic
limit, in which case coannihilation processes with larger $\sigma_{\ch
\slash  \ch}, \sigma_{\slash \ch \slash  \ch}$ may assume greater
relative importance. This was indeed found to be the case for
coannihilations in the region of MSSM parameter space where the LSP is
mainly a neutral higgsino that is only slightly lighter than the
lighter chargino and the second-lightest neutral
higgsino~\cite{gs,co2}. 

This is also what we found in a large domain of the
MSSM parameter region where the LSP is approximately a
bino $\tilde B$ and the next-to-lightest supersymmetric
particle (NLSP) is the $\tilde \tau_R$~\cite{efo}. Moreover,
coannihilations with slightly heavier sparticles such as the
${\tilde e}_R$ and ${\tilde \mu}_R$ are also important.
The essential reason is that the non-relativistic
threshold $S$-wave contributions
to many of the $\ch {\tilde \ell}_R$
and ${\tilde \ell}_R {\tilde \ell}^*_R$ coannihilation
channels are not suppressed by fermion mass factors,
so that $\sigma_{\ch \slash \ch}, \sigma_{\slash \ch \slash \ch}
\gg \sigma_{\ch \ch}$. This is in contrast to higgsino coannihilation
where $\sigma_{\ch \slash \ch}, \sigma_{\slash \ch    
\slash  \ch} \sim \sigma_{\ch \ch}$
above the
$W^{\pm}$ threshold.

In a previous paper~\cite{efo}, we listed many of the important
coannihilation channels, reported on calculations of their
cross sections, and emphasized their significance
for the cosmologically-allowed region of MSSM parameter
space. In particular, we highlighted the fact that the
cosmological upper limit on $m_\ch$ is increased from
$\sim 200$~GeV, as previously estimated~\cite{up}, to $\sim 600$~GeV.
This was demonstrated explicitly in the constrained MSSM
(CMSSM), in which all the soft supersymmetry-breaking mass
parameters $m_0, m_{1/2}$ and $A$are universal at some
input supergravity GUT scale. 

In this paper, we amplify and extend this previous discussion by
detailing the method we have used to
calculate coannihilation cross sections and providing 
simplified analytic expressions. These and the numerical
results we present serve to explain which coannihilation
channels are the most important. We also go beyond our
previous discussion by discussing the dependence of
coannihilation effects on such MSSM parameters 
as $A$ and $\tan \beta$. We use these results to
analyze in more detail not only the impact of
coannihilation effects on the cosmological upper limit
on $m_\ch$, but also the concomitant bounds on other
MSSM parameters~\cite{efos,efos2,efgos}, such as that on $\tan \beta$,
which follows in particular from the lower limit on the Higgs mass
provided by direct searches at LEP.

The layout of this paper is as follows. In Section 2,
we provide a general discussion of relic-density
calculations, which serves as a framework for our
analysis of coannihilation effects. 
Then, in Section 3, we review
the standard relic density analysis neglecting neutralino-stau
coannihilation in the MSSM and CMSSM.
In Section 4, we discuss in some detail
coannihilation effects in the CMSSM.   
Section 5 explores the implications of our coannihilation
results for the cosmological upper limit on the mass
of the LSP and other MSSM parameters, and Section 6
draws some conclusions. Useful analytic formulae
resulting from our calculations are listed in an
Appendix.

\section{Basic Aspects of Relic Density Calculations}

In many cases of interest, the density of relics
left over from the early
Universe may be determined relatively simply, once the
relevant annihilation cross sections have been calculated
and used to obtain an
annihilation rate. As the Universe expands,
a rate or Boltzmann equation is solved to determine a
freeze-out density, and the relic density subsequently scales with
the inverse of the comoving volume, and hence with the entropy density.
In the case of Dirac neutrinos, it is sufficient to
calculate an
$S$-wave cross section to obtain a good approximation to the 
exact result~\cite{lwh}. In the MSSM framework discussed here, however,
the LSP is a neutralino. Since neutralinos are Majorana fermions, the $S$-wave
annihilation cross sections into fermion-antifermion pairs
are suppressed by the masses of the final-state
fermions, and it is therefore necessary to compute the $P$-wave
contribution to the cross section~\cite{hg,ehnos}. 

The rate equation for a stable particle with density $n$ is 
\beq
{dn \over dt} = -3 {\dot R \over R} n - \langle \sigma v_{\rm rel} 
\rangle (n^2 - \neq^2) \;,
\label{rate}
\eeq
where $\neq$ is the equilibrium number density and $\langle \sigma v_{\rm rel}
\rangle$ is the thermally averaged product of the annihilation cross
section $\sigma$ and the relative velocity $v_{\rm rel}$.
In the early Universe, we can write $\dot R/R = (8\pi G_N \rho/3)^{1/2}$,
where $\rho = \pi^2 g(T) T^4/30$ is the energy density in radiation and 
$g(T)$ is the number of relativistic degrees of freedom.
Conservation of the entropy density $s = 2 \pi^2 h(T) T^4/45$ implies that
$\dot R/R = - \dot T/T - h'\dot{T}/3h$ where 
$h' \equiv dh/dT$.  Generally, we have $h(T)
\approx g(T)$. Defining $x \equiv T/m$ and $q \equiv n/T^3h$, we can
write
\beq
{dq \over dx} = m \left({\textstyle{4\pi^3\over 45}} G_N g\right)^{-1/2} 
                \left(h + {\textstyle{1 \over 3}}mxh'\right)
                \langle\sigma v_{\rm rel} \rangle
                (q^2 - \qeq^2) \; .
\label{rate2}
\eeq
The effect of the $h'$ term was discussed in detail in~\cite{swo}, and is
most important when the mass $m$ is between 2 and 10 GeV.  Since we
only consider neutralinos that are significantly more massive, we neglect
it below.

For $x \gg 1$, neutralinos are in thermal and chemical equilibrium and 
$q = \qeq \sim$ constant, since $\neq \sim T^3$. When $x = {\cal O}(1)$, 
$q \simeq \qeq \sim e^{-1/x}/x^{3/2}$ until freeze-out,
after which $q$ is again approximately
constant. For annihilations governed by weak-strength interactions,
freeze-out occurs when $x \sim 1/20$.
The final relic density is determined by integrating (\ref{rate}) down to
$x = 0$, and is given by
\beq 
\rho_\chi = m q(0)h(0)T_0^3
\eeq
More generally, when coannihilations are important, there are several
particle species $i$, with different masses, and
each with its own number density $n_i$ and 
equilibrium number density $n_{{\rm eq},i}$.
In this case \cite{gs}, the rate equation (\ref{rate}) still applies,
provided $n$ is
interpreted as the total number density, 
\beq
n \equiv \sum_i n_i \;,
\label{n}
\eeq
$\neq$ as the total equilibrium number density, 
\beq
\neq \equiv  \sum_i n_{{\rm eq},i} \;,
\label{neq}
\eeq
and the effective annihilation cross section as
\beq
\langle\sigma_{\rm eff} v_{\rm rel}\rangle \equiv
\sum_{ij}{ n_{{\rm eq},i} n_{{\rm eq},j} \over \neq^2}
\langle\sigma_{ij} v_{\rm rel}\rangle \;.
\label{sv2}
\eeq
In eq.~(\ref{rate2}),  $m$ is now understood as the mass of the lightest
particle under consideration.

For $T \la m_i$, 
the equilibrium number density of each species is given by \cite{swo,gg}
\begin{eqnarray}
n_{{\rm eq},i} &=& g_i\int {d^3p\over(2\pi)^3} \; e^{-E/T} 
\nonumber \\
               &=& {g_i m_i^2 T \over 2\pi^2} K_2(m_i/T) \;,
\nonumber \\
               &=& g_i \left({m_i T \over 2\pi}\right)^{3/2} \exp(-m_i/T)
                       \left(1 + {15 T\over 8m_i}+ \ldots \right) \;,
\label{neqi}
\end{eqnarray}
where $g_i$ is a spin degeneracy factor and 
$K_2(x)$ is a modified Bessel function.
We have made the approximation of
Boltzmann statistics for the annihilating particles, which is
excellent in practice.

We now wish to compute 
$\langle\sigma_{12} v_{\rm rel}\rangle$ 
for the process $1+2\to 3+4$ in an efficient manner.
Suppose we have determined the squared transition matrix element $\tsq$
(summed over final spins and averaged over initial spins) and expressed
it as a function of the Mandelstam variables $s$, $t$, $u$.
We now wish to express $\tsq$ in terms of $s$ and the 
scattering angle $\tcm$ in the center-of-mass frame.  We have
\begin{equation}
t-u=-{(m_1^2-m_2^2)(m_3^2-m_4^2)\over s} + 4 p_1(s) p_3(s) \cos\tcm \;,
\label{t-u}
\end{equation}
where $p_i(s)$ is the magnitude of the 3-momentum of particle~$i$
in the CM frame, given by
\begin{eqnarray}
p_1(s) = p_2(s) &=& \left[{s\over4}-{{m_1^2+m_2^2}\over2}
                         +{(m_1^2-m_2^2)^2\over 4s}\right]^{1/2} \;,
\label{p1} \\
p_3(s) = p_4(s) &=& \left[{s\over4}-{{m_3^2+m_4^2}\over2}
                         +{(m_3^2-m_4^2)^2\over 4s}\right]^{1/2} \;.
\label{p3}
\end{eqnarray}
We can then use $s+t+u = m_1^2 + m_2^2 + m_3^2 + m_4^2$ to write
\begin{eqnarray}
t &=& \half[m_1^2+m_2^2+m_3^2+m_4^2-s+(t-u)]\;,
\label{t} \\
u &=& \half[m_1^2+m_2^2+m_3^2+m_4^2-s-(t-u)]\;,
\label{u}
\end{eqnarray}
Using (\ref{t-u})--(\ref{u}), we can write $\tsq$ as a function
of $s$ and $\cos\tcm$.

We now define\footnote{An extra factor of ${1\over2}$ must be included
  in the case of identical final state particles.}
\begin{eqnarray}
w(s) &\equiv& {1\over4}
              \int {d^3 p_3\over(2\pi)^3 E_3}\,{d^3 p_4\over(2\pi)^3 E_4}
              \,(2\pi)^4\delta^4(p_1+p_2-p_3-p_4)\; \tsq
\nonumber \\
&=& {1\over32\pi}\,{p_3(s)\over s^{1/2}} \int_{-1}^{+1}d\cos\tcm\,\tsq \;.
\label{w}
\end{eqnarray}
In terms of $w(s)$, the total annihilation cross section
$\sigma_{12}(s)$ is given by
$\sigma_{12}(s) = w(s)/s^{1/2}p_1(s)$.  Our $w(s)$
is also the same as $w(s)$ in~\cite{swo,fkosi,efo}, 
which is written as $W/4$ in~\cite{eg}.

So far all this is exact.  To reproduce the usual partial wave expansion,
we expand $\tsq$ in powers of $p_1(s)/m_1$.  
The odd powers vanish upon
integration over $\tcm$, while the zeroth and second order terms 
correspond to the usual $S$ and $P$ waves, respectively.
We see from (\ref{t-u}) that each factor of $p_1(s)$ 
is accompanied by a factor of $\cos\tcm$,
so we have
\begin{equation}
\int_{-1}^{+1}d\cos\tcm\,\tsq =
 \left(\tsq_{\cos\tcm\,\to\, +1/\sqrt3} \; + \;
       \tsq_{\cos\tcm\,\to\, -1/\sqrt3}\right) + {\cal O}(p_1^4)\;.
\label{itsq}
\end{equation}
We can therefore evaluate the $S$ and $P$ wave contributions to
$w(s)$ simply by evaluating $\tsq$
at two different values of $\cos\tcm$;
no integrations are required.

The proper procedure for thermal averaging has been discussed 
in~\cite{swo,gg} for the case of $m_1=m_2$, and 
in~\cite{fkosi,eg} for the case of $m_1\ne m_2$,
with the result
\begin{equation}
\langle\sigma_{12} v_{\rm rel}\rangle 
        = {1\over 2 m_1^2 m_2^2 T K_2(m_1/T) K_2(m_2/T)}
          \int_{(m_1+m_2)^2}^\infty ds\,K_1(\sqrt{s}/T)p_1(s)w(s) \;.
\label{sv}
\end{equation}
Using the asymptotic expansion 
$K_n(x)=(2x/\pi)^{-1/2}e^{-x}[1+(4n^2-1)/(8x)+\ldots]$
of the Bessel functions, changing the integration variable from $s$
to $y = (s^{1/2}-m_1-m_2)/T$, and then expanding in powers of $T$,
we find
\begin{eqnarray}
\langle\sigma_{12} v_{\rm rel}\rangle
 &=& {w(s_0)\over m_1 m_2} 
                   - {3(m_1+m_2)\over 2 m_1 m_2}
                     \left[{w(s_0)\over m_1 m_2} - 2 w'(s_0)\right]T 
                   + {\cal O}(T^2) 
\nonumber \\
 &=& {1\over m_1 m_2}\left[1-{3(m_1+m_2)T\over2m_1m_2}\right]
                  w(s)\bigr|_{s\to (m_1+m_2)^2 + 3(m_1+m_2)T} + {\cal O}(T^2) \;,
\nonumber \\
 &\equiv& a_{12} + b_{12} \, x + {\cal O}(x^2) \;,
\label{sv3}
\end{eqnarray}
where $x = T/m_1$ (assuming $m_1<m_2$), and, in the first line, 
$s_0 = (m_1+m_2)^2$.  
We extract $a_{12}$ and $b_{12}$ 
from the transition amplitudes
listed in the Appendix by performing the substitutions 
(\ref{t-u})-(\ref{sv3}) for each final state.   We set
$x=0$ to get $a_{12}$, and then compute $b_{12}$ by setting $x$
to a numerical value small enough to render the ${\cal O}(x^2)$ terms negligible.
We compute
$a_{\rm eff}$ and $b_{\rm eff}$ by performing the sum over initial
states as in eq.~(\ref{sv2}).  We then
integrate the rate equation (\ref{rate2}) numerically to obtain the relic
LSP abundance. To a fair approximation, the relic density can simply be
written as~\cite{ehnos,gs}
\begin{equation}
  \label{eq:ohsq}
  \Omega h^2 \approx 
{10^9 \gev^{-1} \over g_{\ss f}^{1/2} M_{\rm pl}(a_{\rm eff}+
b_{\rm eff} x_{\ss f}/2)x_{\ss f}},
\end{equation}
where the freeze-out temperature $T_{\!f}\sim m_\ch/20$, and $g_{\ss f}$
is the number of 
relativistic degrees of freedom at $T_{\!f}$.   Note that this implies that the
ratio of relic densities computed with and without coannihilations is, 
roughly,
\begin{equation}
  \label{eq:R}
 R\equiv{\Omega^0\over\Omega} 
 \approx \left({\hat\sigma_{\rm eff}\over\hat\sigma_0}\right)
\left({x_{\!\ss f}\over x_{\!\ss f}^{0}}\right),
\end{equation}
where $\hat\sigma\equiv a + b x/2$ and sub- and superscripts 0 denote
quantities computed ignoring coannihilations.  The ratio ${x_{\!\ss
    f}^0 / x_{\!\ss f}}\approx 1+x_{\!\ss f}^0 \ln (g_{\rm
  eff}\sigma_{\rm eff}/g_1\sigma_0)$, where 
$g_{\rm eff}\equiv\sum_i g_i (m_i/m_1)^{3/2}e^{-(m_i-m_1)/T}$.
For the case of three degenerate slepton NLSPs, $g_{\rm eff}=\sum_i g_i =8$ and 
${x_{\!\ss f}^0 / x_{\!\ss f}}\approx 1.2$.

The non-relativistic expansion (\ref{itsq}) is known to be inaccurate
near $s$-channel poles and final-state thresholds \cite{gs,gg}, where
the cross section can depend strongly on the initial LSP momenta, and
where LSPs in the Boltzmann tail can access resonances and final
states forbidden to zero-momentum LSPs. In the CMSSM,
well-studied examples are neutralino annihilation on the $Z$ and light
Higgs poles, where a detailed treatment of the thermal averaging
\cite{gs,gg} is necessary to compute accurately the neutralino relic
abundance.  Examples of final-state thresholds include ${\bar t} t$
and $hZ$. 

However, none of these effects are significant for the analysis of
this paper.  The regions with $\mchi\la\mz/2$ and $\mchi\la m_h/2$
have now been excluded by LEP chargino searches.  Annihilation on the
$H$ and $A$ poles can be important for $\tan\beta\ga 40$~\cite{bk},
where the heavy Higgs masses can be close to $2\mchi$ for very large
ranges of $\m12$ (since the heavy Higgs and bino masses scale
similarly with $\m12$).  However, we do not consider in this paper
such large values of $\tan\beta$, where stau mixing may be important
as well.  In the CMSSM, the only non-negligible threshhold occurs at
$\mchi \sim m_t$, which is visible as a kink near $\m12=400$ GeV in
Figs.\ref{fig:ss}-\ref{fig:ss10.3}.  However, as can be seen there,
the effect of the top threshhold (and hence also that of
sub-threshhold neutralino annihiliation into tops) on the total
annihilation rate is tiny, because the contribution from the ${\bar t}
t$ final state is suppressed by the large stop masses.  Likewise, the
$H$ and $A$ masses are much too large in the CMSSM for heavy Higgs
final states to be of relevance, except perhaps again for the very
large values of $\tan\beta>40$ (not considered here), where the heavy
Higgs masses are smaller. Finally, other thresholds such as $h Z$ are
suppressed by the smallness of the Higgsino component of the LSP,
which is a very pure bino in the CMSSM.

As for such effects in coannihilations
(via sleptons in our case), we find that coannihilations are important
in the CMSSM 
for fairly large values of $\m12\ge 300$ GeV.  The relevant CMSSM
regions are thus above the light thresholds (e.g.,
$\sl^{\,i}\,\sl^{\,i^{\scs *}}\rightarrow W^+W^-$) and still well
below the heavy threshholds (e.g., $\sl^{\,i}\,\sl^{\,i^{\scs
    *}}\rightarrow \gamma H$) for slepton annihilation and
slepton-neutralino coannihilation, since only staus with masses $\mst\sim\mchi$
affect the neutralino relic density.  The process $\sl^{\,i}\,\sl^{\,i^{\scs
    *}}\rightarrow t\bar t$ is
numerically irrelevant (see Figs.\ref{fig:ss}-\ref{fig:ss10.3}).  
Therefore the partial-wave expansion (\ref{itsq}) is a valid
approximation in our analysis.

\section{The MSSM Without Coannihilations}

In the MSSM, the identity of the LSP is determined by the
following tree-level parameters:
the supersymmetry-breaking gaugino mass $m_{1/2}$ (assuming 
gaugino mass universality
at the GUT scale), the supersymmetric Higgs mixing mass $\mu$, and the
ratio of Higgs vacuum expectation values (vev's), $\tan \beta$. The
annihilation cross section and
hence the relic density depend on the identity of the LSP.
For $m_{1/2} \gg \mu$, the LSP is mostly a higgsino
state, whilst in the
opposite limit, $\mu \gg m_{1/2}$, the LSP is mostly 
a bino ${\tilde B}$. The negative results of previous SUSY particle searches 
at LEP have been able to impose strong constraints
on the MSSM parameter space~\cite{efos,efos2,efgos}.  Roughly speaking, 
one may conclude that $m_{1/2}$ and $|\mu| > 80$~GeV.

In the region of the MSSM parameter space where the LSP 
is mainly a ${\tilde B}$, annihilations
to fermion-antifermion pairs proceed mainly through sfermion exchange.
As we discuss in more detail below, this process is $P$-wave suppressed, 
which implies a great reduction in the annihilation rate at the low
temperatures at which the $\tilde B$'s freeze out.
Over much of this region, it is possible to obtain a cosmologically
significant relic density, as we review below. By contrast, in the region where 
the LSP is mainly a higgsino,
annihilations to $W^\pm$ pairs are dominant above threshold.   This
process is not $P$-wave suppressed, and,
as a result, the relic
density is suppressed in this higgsino region, unless one considers
either very heavy higgsinos ($m_{\tilde H} \ga 500$ GeV), or higgsinos with
masses below the $W$ threshold. Further, in much of the higgsino region of parameter
space, either the second-lightest neutralino or the
higgsino-like chargino is very close in mass to the LSP.  In this case,
it was shown~\cite{gs,co2} that coannihilations
were important
in determining the relic density of LSPs. Indeed, when combined with
the current LEP limits, there remains very little (if any) parameter space
remaining where a light higgsino LSP could contribute a sufficient relic
density ($\Omega h^2 > 0.1$) to be of interest as the dominant
component of astrophysical dark matter~\cite{efgos}.   At very low
$\tan\beta$ (1.2-1.6), some solutions with $\Omega h^2 > 0.1$ may 
exist just below the W threshold \cite{eg}.  However, the LEP Higgs 
bounds (though dependent on quantities such as $m_{\tilde t}$) make it
extremely difficult to achieve
$\tan\beta$ this low, as discussed, for example,
in~\cite{efos,efos2,efgos}.

It is also of interest to consider a constrained version of the MSSM
(CMSSM), in which
all the soft supersymmetry-breaking scalar masses, $m_0$, are unified at
the GUT scale. In this case, the conditions which determine electroweak symmetry
breaking also fix $|\mu|$ and the pseudoscalar MSSM Higgs mass at the weak
scale. For all choices of $m_{1/2}$ and $m_0$ consistent with
LEP mass bounds, the lightest neutralino is predicted to be a ${\tilde B}$, modulo
thin fringe strips of parameter space close to where the electroweak
symmetry is not dynamically broken\footnote{In fact, unification of the
gaugino 
  masses at the GUT scale by itself implies a  ${\tilde B}$-like LSP
  over the bulk of the parameter space~\cite{f1}.}.
Although ${\tilde B}$'s  typically have an interesting relic 
density, this is no longer true if the ${\tilde B}$ mass happens to lie near
$m_Z/2$ or $m_h/2$, in which case there are large contributions to the
annihilation through direct $S$-channel resonance exchange. However,
since LEP limits on the chargino
mass can be translated into bounds on $m_{1/2}$, 
these resonant cases are now all but excluded for small values of $\tan
\beta$, as we discuss further in Section 4.

Since we are mainly interested the $\tilde{B}$ LSP candidate, we 
focus our
discussion on this case, in order to be more specific.
 The thermally-averaged cross section for $\tilde{B} \tilde{B} \to f
\bar{f}$ takes the generic form
\begin{equation}
\langle \sigma v \rangle = 
{g_1^4 \over 128 \pi} 
\left(1 - {\mf^2 \over \mb^2}\right)^{1/2} 
\left[ (Y_L^2 + Y_R^2)^2 \left({\mf^2 \over \Delta_f^2}\right) + 
(Y_L^4 + Y_R^4)  \left({4 \mb^2 \over \Delta_f^2}\right) 
\Bigl(1 + {\cal O}(\mf^2/\mb^2)\Bigr) x \right],
\label{eqn:sigv}
\end{equation}
where $Y_{L(R)}$ is the hypercharge of $f_{L(R)}$, $\Delta_f \equiv
\msf^2 + \mb^2 - \mf^2$, and we have shown only the leading $P$-wave
contribution proportional to $x \equiv T/\mb$.  As advertised, the
$S$-wave piece is proportional to the fermion mass-squared and hence is 
negligible, except  perhaps for the top quark, and this has the net
effect of reducing the neutralino annihilation cross-section by ${\cal
  O}(x_{\! f})$.  The general form of (\ref{eqn:sigv}) leads to
an upper bound on the possible mass of the LSP within the MSSM,
due to the cosmological relic density~\cite{up}. 
Specifically, in the case of a bino LSP, the upper limit on
$m_{\tilde B}$ comes about as
follows.  The assumption that the ${\tilde B}$ is the LSP requires, in
particular, that $\mb < \msf$.  In order to minimize the relic
density, we must maximize the cross section, which is done by setting
$\msf = \mb$.  The cross section is then approximately inversely
proportional to $\mb^2$. The 
cosmological upper limit on $\Omega_{\tilde B}
h^2$ translates into a lower limit on $\langle \sigma v \rangle$ which 
then, in turn, yields an upper limit to $\mb$.  In the MSSM, this limit is
$\mb \la 300$ GeV, when all sfermion masses are taken to be equal at the
weak scale, though the limit can be weakened when sfermion
mixing~\cite{fkmos}  or CP-violating
phases are included~\cite{fkosi2}.

In the CMSSM, the argument is somewhat similar, although $\mb$
and the sfermion masses are now no longer entirely
independent, because it is assumed in the CMSSM
that there is a common scalar
mass $m_0$ at the GUT scale. For a given value of the common
gaugino
mass $\m12$ at the GUT scale, the relic ${\tilde B}$ density falls with
decreasing $m_0$, since 
$\msf^2=m_0^{2} + C_{f} \m12^2 + {\cal O}(m_Z^2)$, where $C_f$ is a 
positive numerical
coefficient that is calculable via the renormalization-group evolution of
the sfermion masses.
Therefore, the cosmological upper limit on $\Omega_{\tilde B} h^2$
translates at fixed $\m12$ into an upper limit on $m_0$.
At low to moderate $\tan\beta$, this upper limit is typically not
larger than  $m_0 \sim 170$ GeV, unless
one is sitting on a direct-channel pole, i.e., when $\mb \sim m_Z/2$ or
$m_h/2$, in which case
$s$-channel annihilation is dominant and there is no upper limit
on $m_0$. However, we are interested in an upper bound on $\mb$, and hence
in masses far from the light Higgs and $Z$ poles. We recall that $\mb$
scales with
$\m12$, and it transpires for $\m12 \ga 400$ GeV that
$\mb$  exceeds the mass of the lightest sfermion, which is
typically the ${\tilde \tau}_R$, for $m_0$ small enough to satisfy the
cosmological bound as traditionally computed (i.e., neglecting
coannihilations) \cite{upper}. Thus, the LSP is no longer a neutralino
for such large values of $\m12$, and
hence an upper bound $\mb \la 200$~GeV~\cite{upper} can be
established.~\footnote{This upper
bound can be strengthened by requiring that the global minimum
of the effective potential of the MSSM conserve electric charge and
color~\cite{bbc,af}.}

In \cite{efos2}, it was shown that the LEP constraints on the mass of the
supersymmetric Higgs boson, when combined with the above cosmological
upper limit on the LSP mass (or $m_{1/2}$), leads to an interesting bound
on $\tan \beta$.  The argument is as follows. At fixed $\tan \beta$, due to
the radiative corrections to the Higgs mass in the
MSSM~\cite{MSSMHiggs}, it is
always possible to satisfy a given experimental constraint on the 
lightest MSSM Higgs mass by going to
large values of either $m_{1/2}$ or $m_0$. At large $m_0$, however, the
cosmological bound forces one to low $|\mu|$, so that the LSP is
higgsino-like
and annihilations are not suppressed by large sfermion masses. In the CMSSM,
this is not possible, since $\mu$ is fixed by the condition of electroweak
symmetry breaking. Even when universality among the soft Higgs masses is not
assumed, if $m_0$ is large, obtaining a reasonable relic density requires
$|\mu|$ to be low enough to ensure that the LSP is a higgsino.  However,
unless $m_{1/2}$ is very large, lowering
$|\mu|$ this far results in a lower Higgs mass. At a given value of $\tan \beta$,  the
Higgs mass bound can be translated into a lower bound on $m_{1/2}$. If this
lower bound is greater than the cosmological upper bound on $m_{1/2}$
discussed above, the corresponding value of $\tan \beta$ considered is
excluded.  Since the lower bound on $m_{1/2}$ due to the Higgs mass bound is
very dependent on $\tan \beta$, we can derive a lower bound on $\tan \beta$ by
combining the LEP and cosmological bounds~\cite{efos,efos2,efgos}. 

\section{Coannihilations in the CMSSM}

As discussed earlier,
if the masses of the next-to-lightest sparticles (NLSPs) are close to
the LSP mass: $\Delta M = {\cal O}(x_f) M$, the number densities of
the NLSPs have only slight Boltzmann suppressions with respect to the
LSP number density when the LSP freezes out of chemical equilibrium
with the thermal bath\footnote{We recall that
 2-2 scatterings with particles in the
  thermal bath keep the NLSPs in chemical equilibrium with each other
  and with the LSP, down to temperatures well below the temperature at
  which the comoving LSP number density freezes out.}. 
Moreover, it is well
known~\cite{gs} that, in such circumstances, coannihilations of NLSPs
with the LSP, along with NLSP-NLSP annihilations, may play an
important r\^ole in keeping the LSPs in chemical equilibrium with the
bath~\cite{gs}, and the number density of neutralinos can be
significantly reduced by such coannihilations.  These processes can be
particularly important when the LSP annihilation rate itself is
suppressed, as is the case for neutralinos, as discussed above.  The
case of heavy higgsinos is a well studied example~\cite{dnry}.
Analogously to that case, the ${\tilde B}$ relic density can be
reduced through coannihilation with slightly heavier ${\st}$'s or
other sleptons, as we now discuss in detail.

In the CMSSM, when $\mb$ attains the upper bound
discussed in the previous section, the ${\tilde B}$ is
degenerate in mass with the ${\tilde \tau}_R$, and quite close in mass
to the ${\tilde e}_R$ and ${\tilde \mu}_R$.  In \cite{efo}, we showed
that the effects of coannihilations between the neutralino LSP and
the ${\tilde \tau_R}$ (also including the ${\tilde \mu_R}$ and ${\tilde
  e_R}$) can have have a dramatic effect on the derived upper bound on
$m_{1/2}$ and the mass of the LSP. Such coannihilation effects
thereby also affect the derived bound on $\tan \beta$.

To compute the effective annihilation cross sections for light
sparticles in the CMSSM, we allow the index $i$ in (\ref{n}) to run
over $\st, \st^*, \sel, \sel^*, \sm$ and $\sm^*$, as well as $\ch$.
Many of the resulting 49 $\sigma_{ij}$ in (\ref{sv2}) are related, so we
can write
\begin{eqnarray}
\sigma_{\rm eff} &=& 
\sigma_{\chi\chi} r_{\chi}r_{\chi} + 4\, \sigma_{\chi\tau} r_{\chi} r_{\tau} +
8\,  \sigma_{\chi e} r_{\chi} r_{e} +2\, (\sigma_{\tau\tau}+
\sigma_{\tau\tau^*})r_{\tau}r_{\tau}+8\, (\sigma_{\tau e}+\sigma_{\tau e^*})r_{\tau}r_{e} 
+ \hfill\nonumber \\&&4\,  (\sigma_{e e}+\sigma_{e e^*})r_{e}r_{e}+ 4\,  
(\sigma_{e \mu}+\sigma_{e \mu^*})r_{e}r_{e},
\end{eqnarray}
where $r_i\equiv n_{{\rm eq},i}/n_{\rm eq}$,
we have taken the $\sel$ and $\sm$ to be degenerate in mass,
and we have neglected the electron and muon masses. We list in
Table~\ref{table:states} the sets of initial and final
states for which we compute  the annihilation cross sections.  We list
in the Appendix the transition amplitudes for the scattering processes, which 
are sufficient to compute the $a$ and $b$ coefficients, 
following the discussion in
Section 2. We have verified that the $\tau$ mass is numerically irrelevant
in our analysis, and the formulae we present are simplified 
to the $m_\tau\rightarrow 0$ 
limit, although our numerical results do include a non-zero $\tau$ mass,
including extra diagrams which are present only when $m_\tau\ne 0$, but which
are not listed in the Appendix.  We ignore the effect of $\tilde\tau$ mixing on the cross-sections, 
which may be important at large $\tan\beta$. 
The final states involving heavy Higgses are kinematically unavailable
in the regions of CMSSM parameter space relevant to our analysis.
However, for completeness, their transition amplitudes are also tabulated
in the Appendix.  
As  $\ch \stau$ coannihilation is important at large $\m12$, where the
$\widetilde B$ purity
is very high, we compute the three $\ch \stau$ amplitudes  in the
$\widetilde B$ limit, 
where $t$-channel neutralino exchange  is suppressed. 
We also have not included three body final state processes, such as
s-channel Higgs exchange to a Higgs + gauge boson pair, via a two
Higgs/two gauge boson vertex.  In addition to being phase-space
supressed, the magnitude of these contrubtions to the total cross
section is down by a factor of $\alpha$ relative to the two body final
states we have included.

\begin{table}[htb]\caption{Initial and Final States for Coannihilation:
$\{i,j=\tau,e,\mu\}$}
\begin{center}
\begin{tabular}{c|l}\hline
Initial State & Final States\\ \hline\\
$\sl^{\,i}\,\sl^{\,i^{\scs *}}$    & $\gamma\gamma,\, \, ZZ\, ,\,\gamma Z,\, W^+W^-, \,Zh\, ,
\gamma h\, ,\,h \, h, \,f\bar f,$\\[0.5ex]
                & ${\rm ZH, \gamma H, ZA, W^+H^-,hH,hA, HH,HA, AA,H^+H^-}$\\[0.5ex]
$\sl^{\,i}\,\sl^{\,j}$  & $\ell^{\,i}\ell^{\,j}$\\[0.5ex]
$\sl^{\,i}\,\sl^{\,{j}^{\scs *}},\,i\ne j$   & $\ell^{\,i}\bar \ell^{\,j}$\\[0.5ex]
$\sl^{\,i}\,\ch$   & $\ell^{\,i}\gamma, \ell^{\,i}Z, \ell^{\,i}h$\\[0.5ex]
\label{table:states}
\end{tabular}
\end{center}
\end{table}
\vspace{-.5cm}

\begin{figure}
\vspace*{-0.5in}
\begin{minipage}[b]{8in}
\epsfig{file=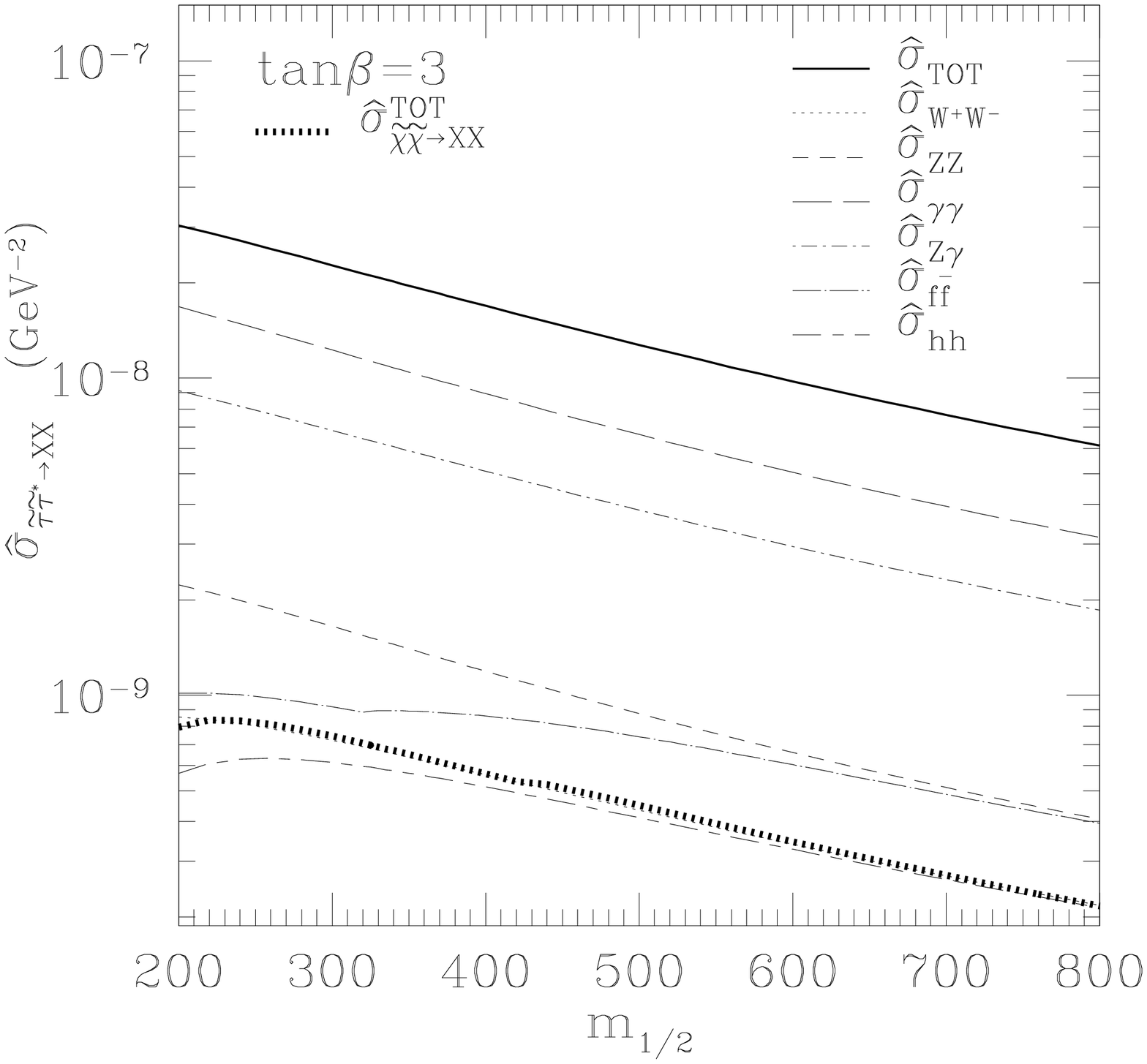,height=3.5in}
\epsfig{file=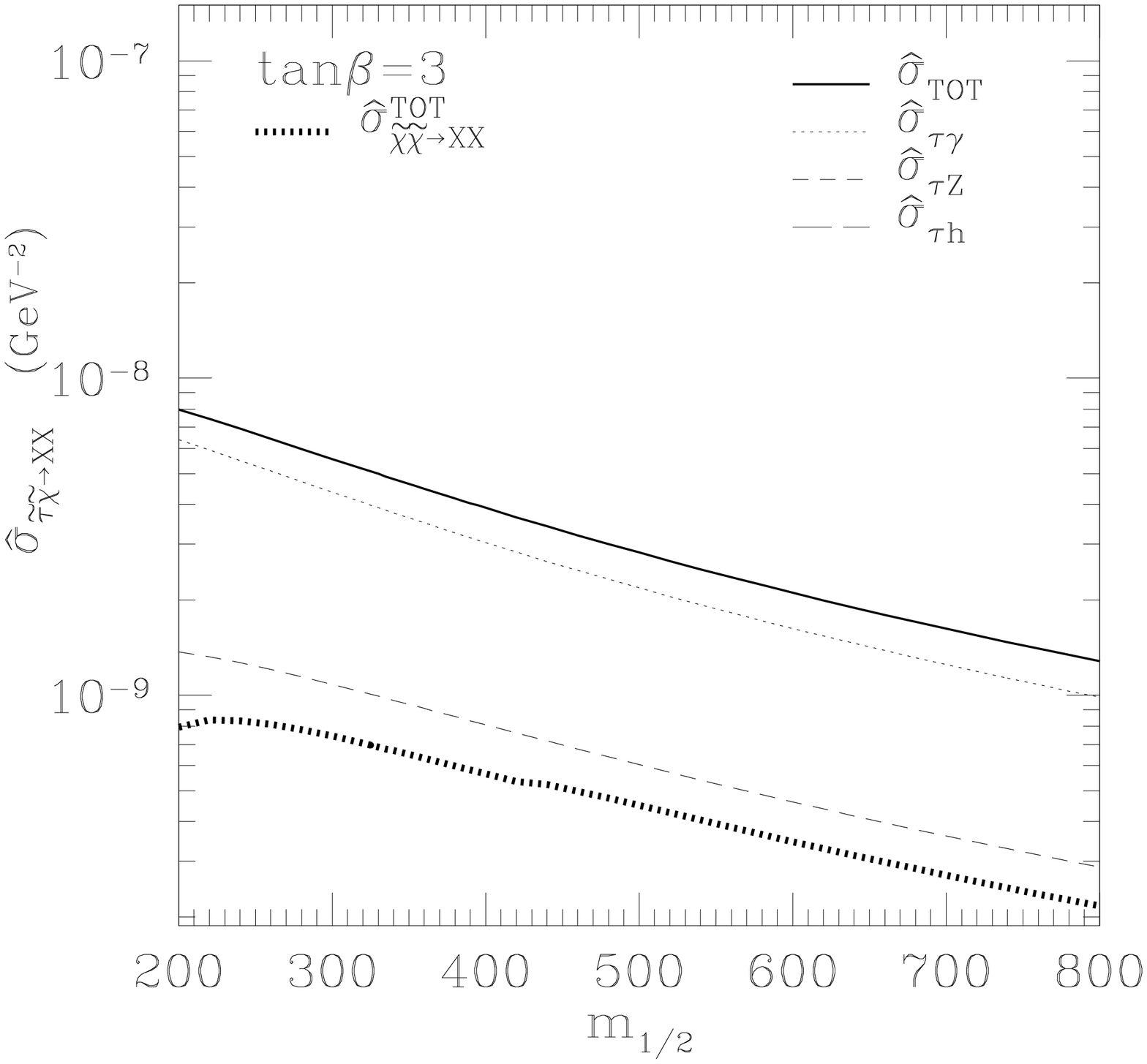,height=3.5in} \hfill
\end{minipage}
\hspace*{1.6in}
\epsfig{file=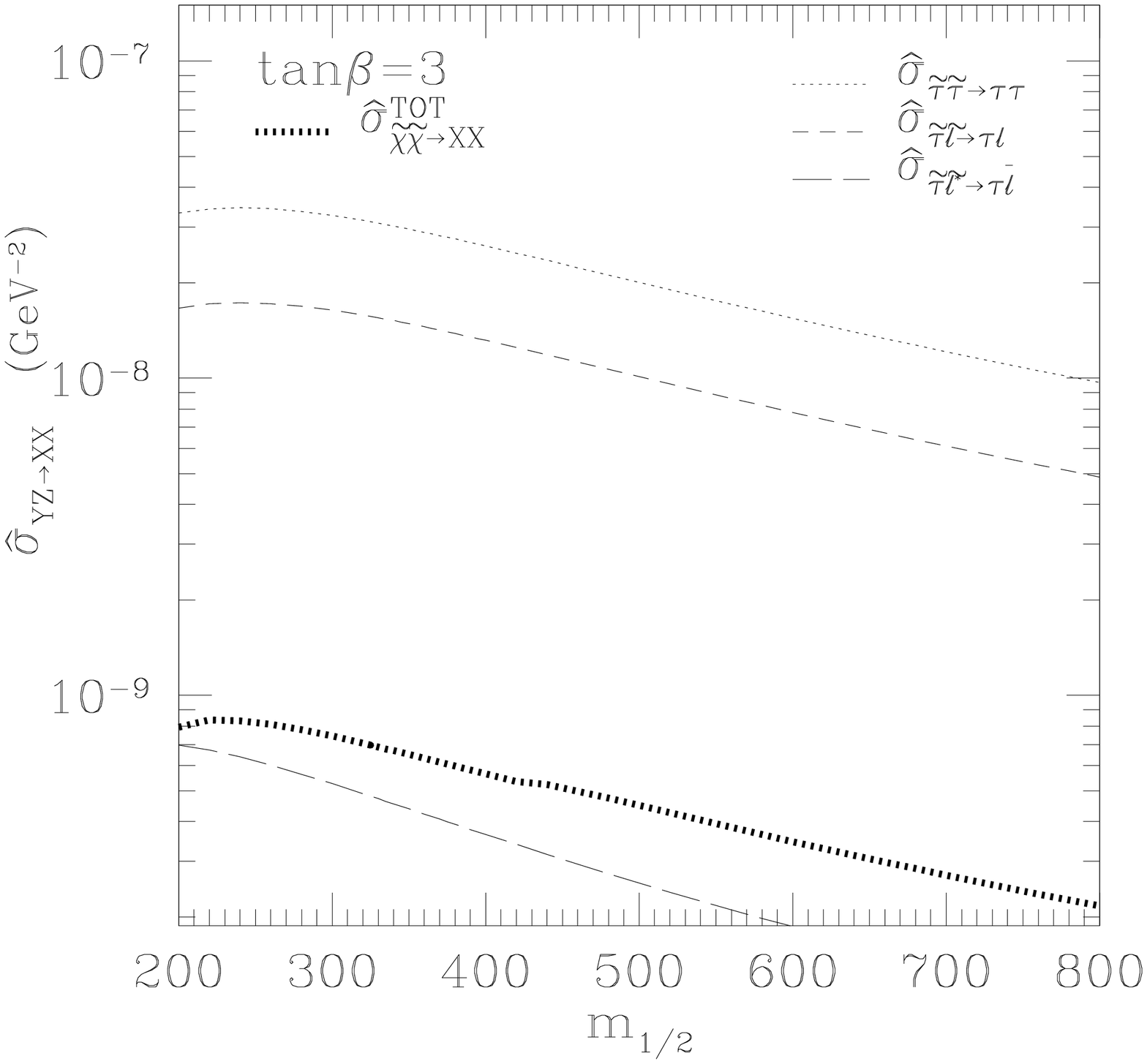,height=3.5in} 
\caption{{\it The separate contributions to the cross sections 
  $\hat\sigma\equiv a+{1\over2}b x$ for $x=T/m_\ch=1/23$ and 
  $m_0=120\gev$, as functions of $\m12$: a)$\stau\stau^*$,
  b)$\stau\ch$, and c) other interactions.  For comparison, the thick
  dotted line is the $\ch \ch $ cross section.}
\label{fig:ss}}
\end{figure}

\begin{figure}
\vspace*{-0.5in}
\begin{minipage}[b]{8in}
\epsfig{file=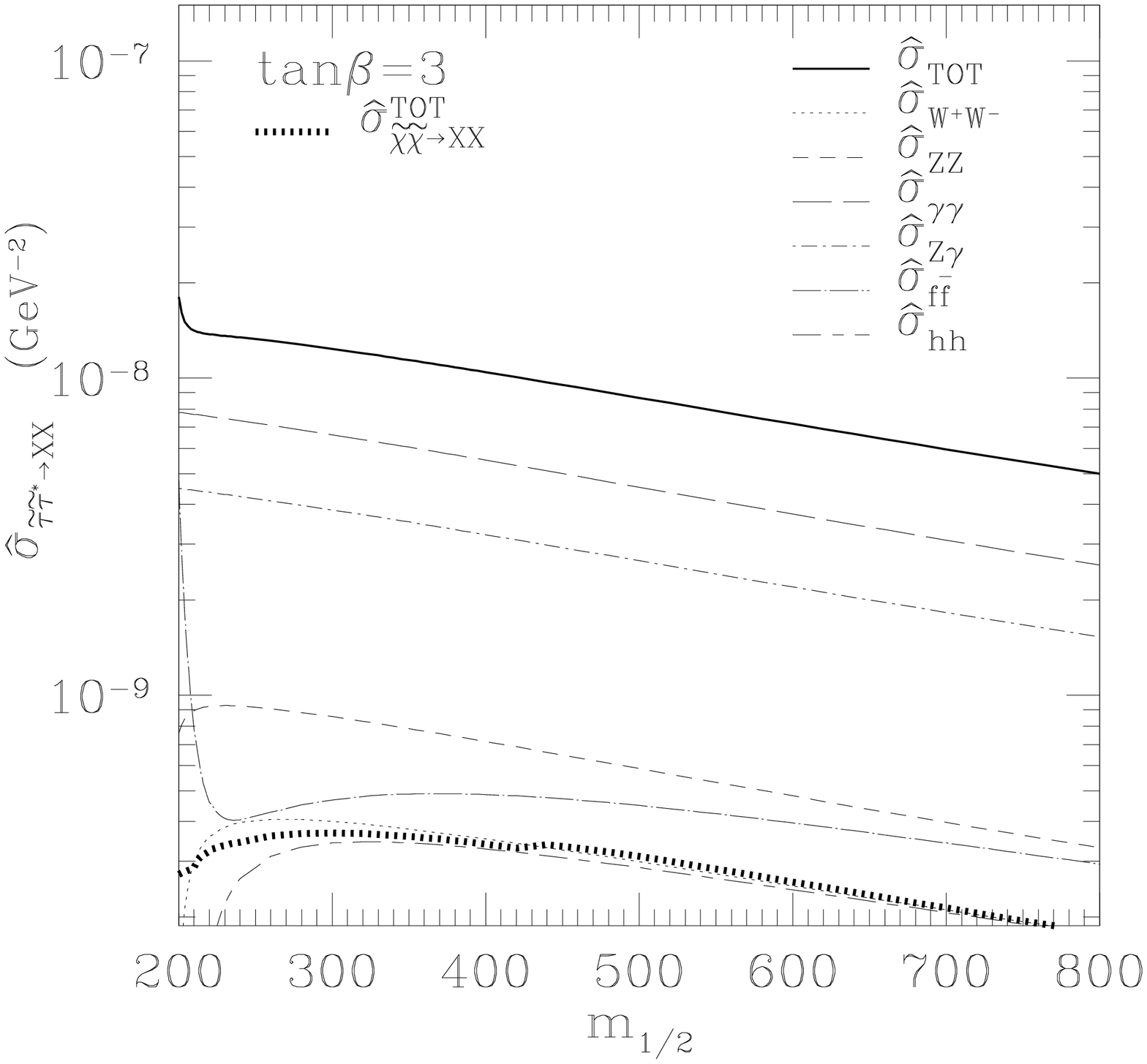,height=3.5in}
\epsfig{file=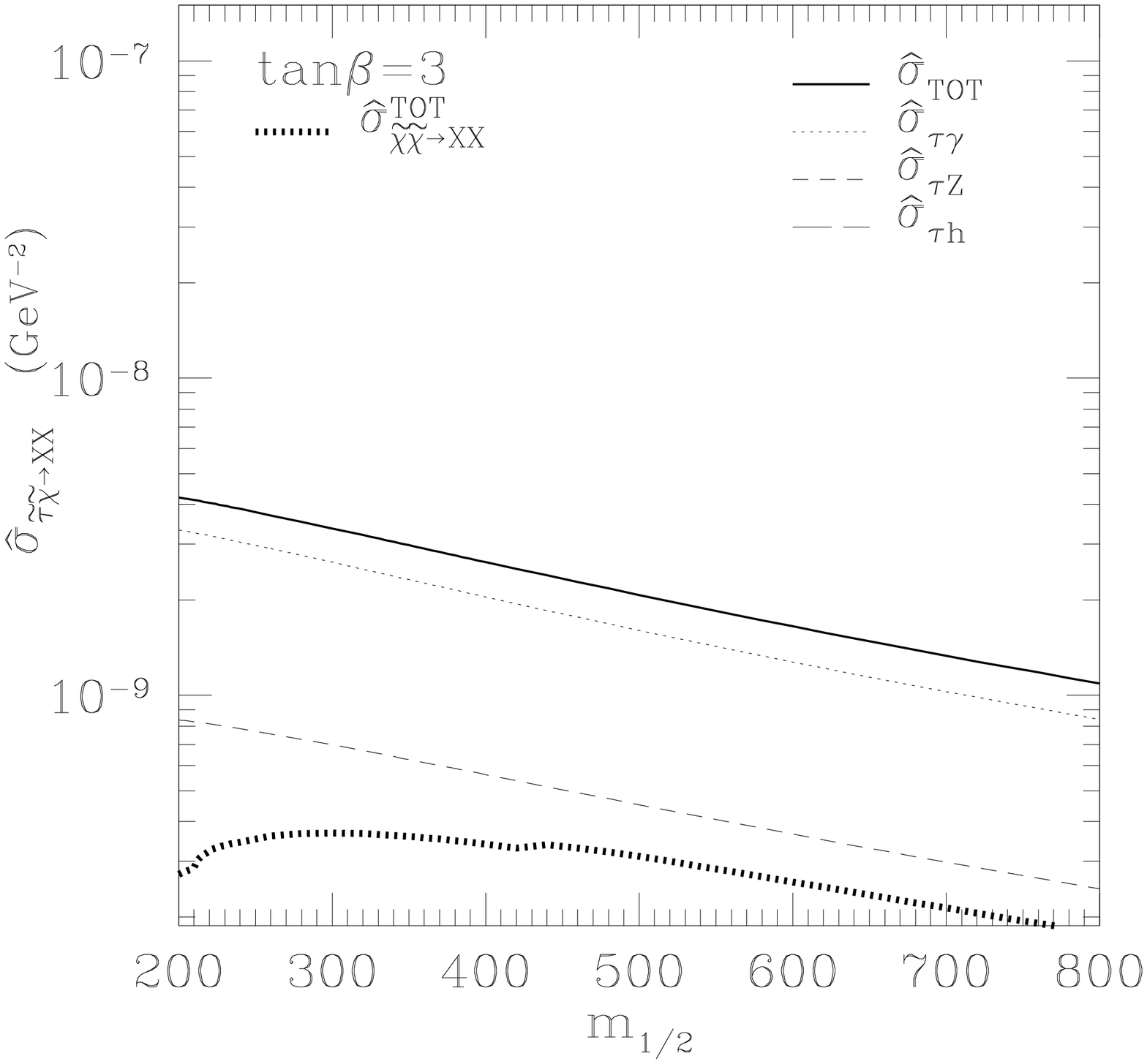,height=3.5in} \hfill
\end{minipage}
\hspace*{1.6in}
\epsfig{file=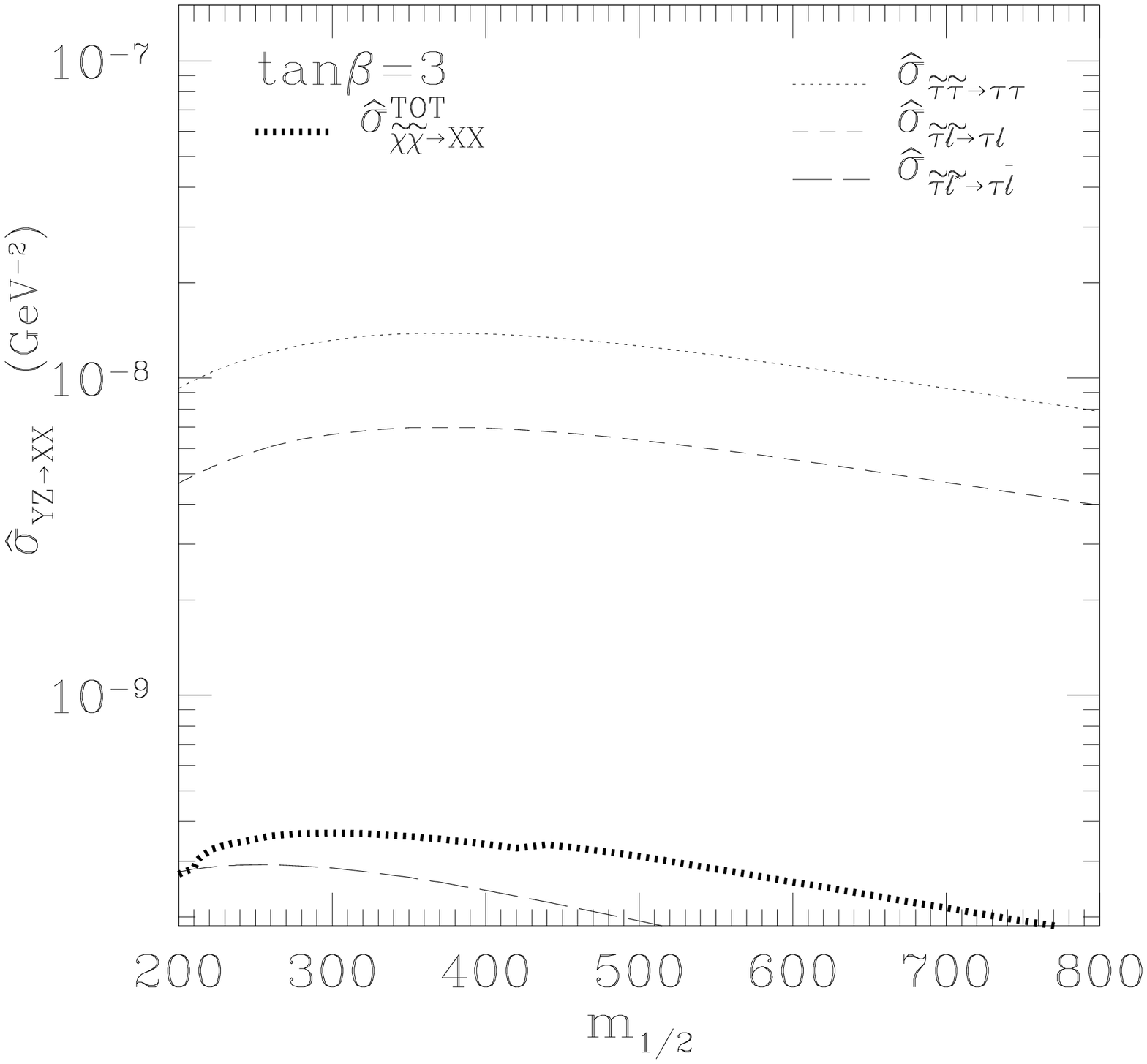,height=3.5in} 
\caption{{\it As in Fig.~\protect\ref{fig:ss}, but for the choice
$m_0=200$~GeV.}\label{fig:ss200}}
\end{figure}

\begin{figure}
\vspace*{-0.5in}
\begin{minipage}[b]{8in}
\epsfig{file=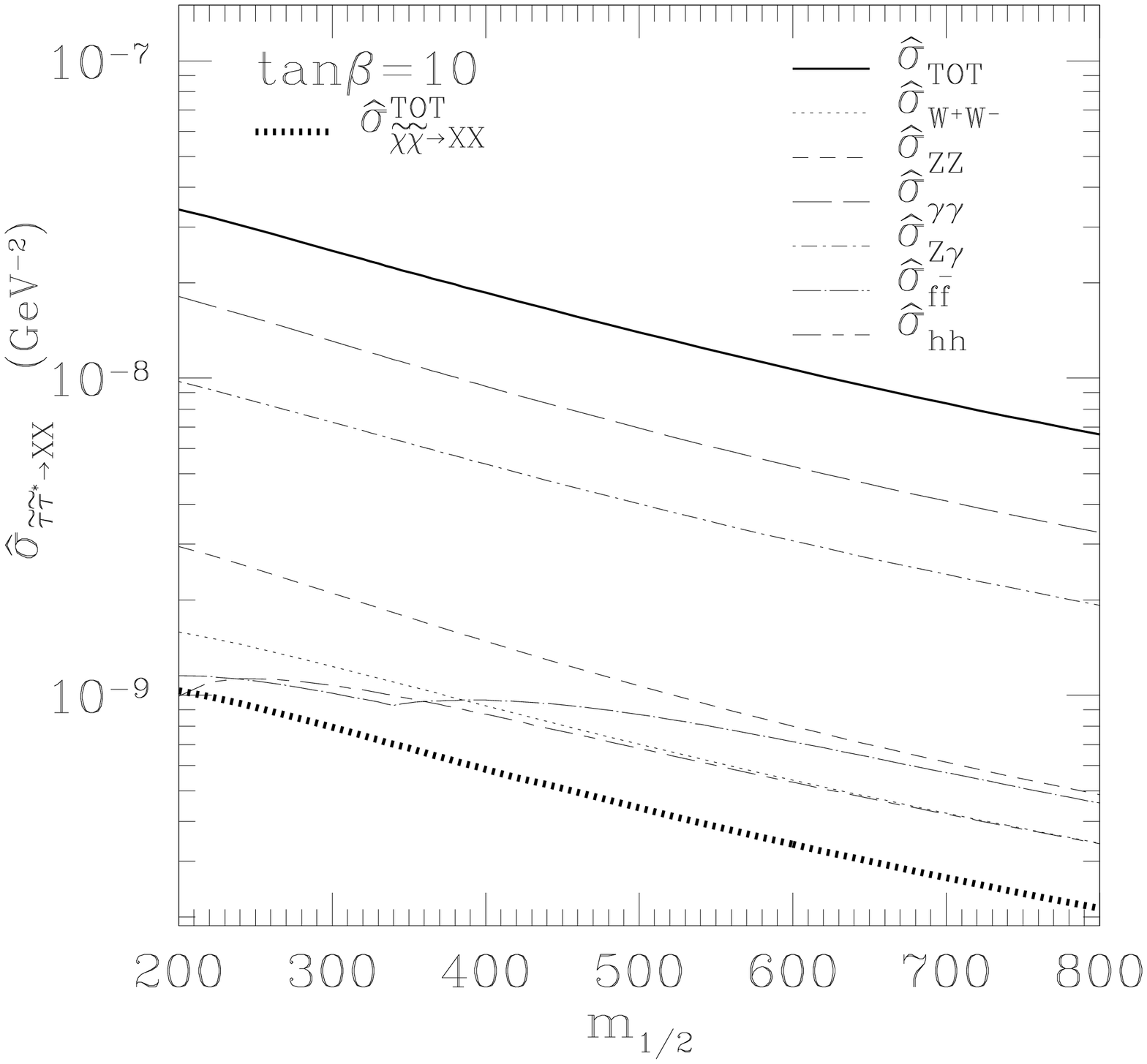,height=3.5in}
\epsfig{file=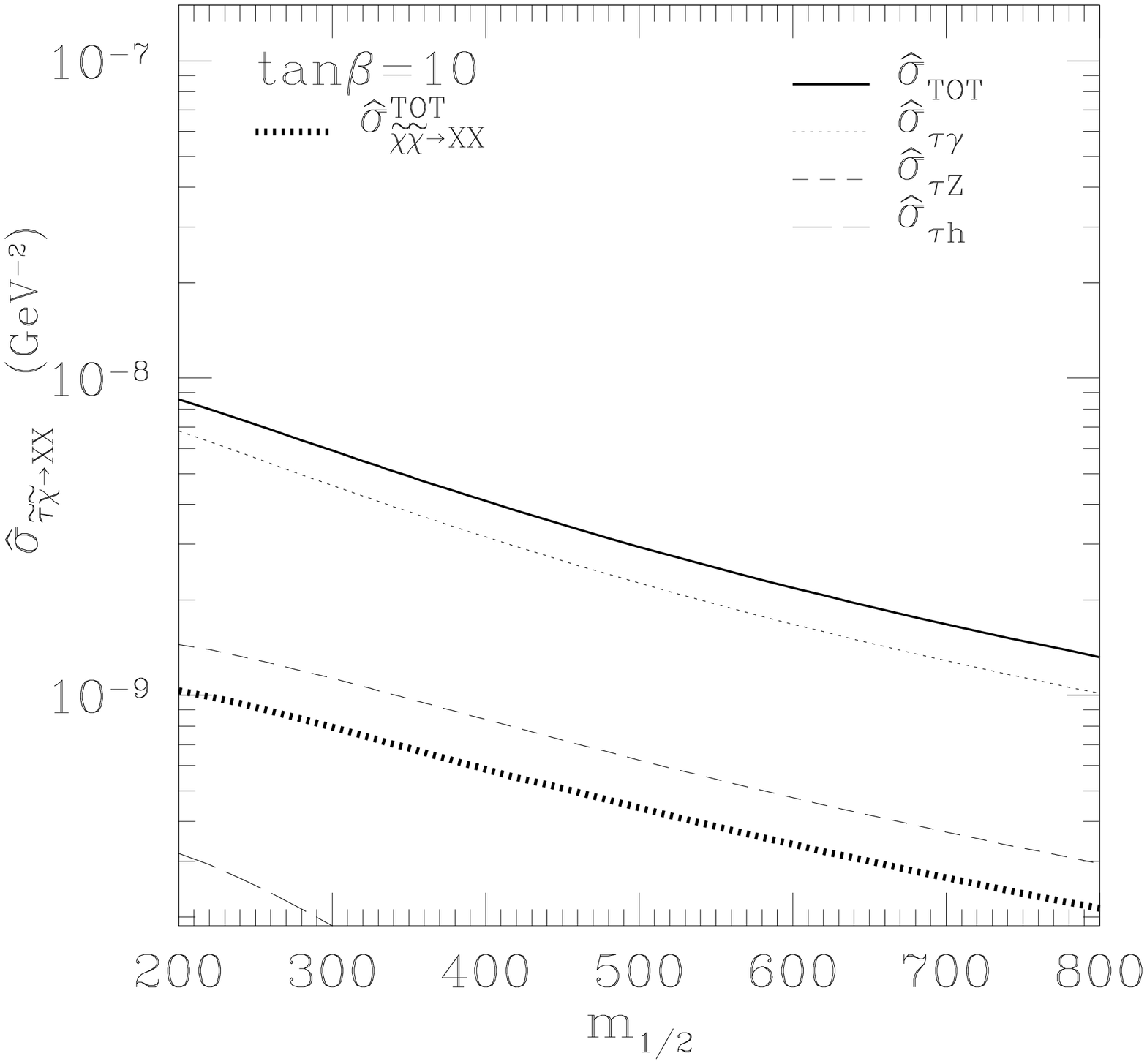,height=3.5in} \hfill
\end{minipage}
\hspace*{1.6in}
\epsfig{file=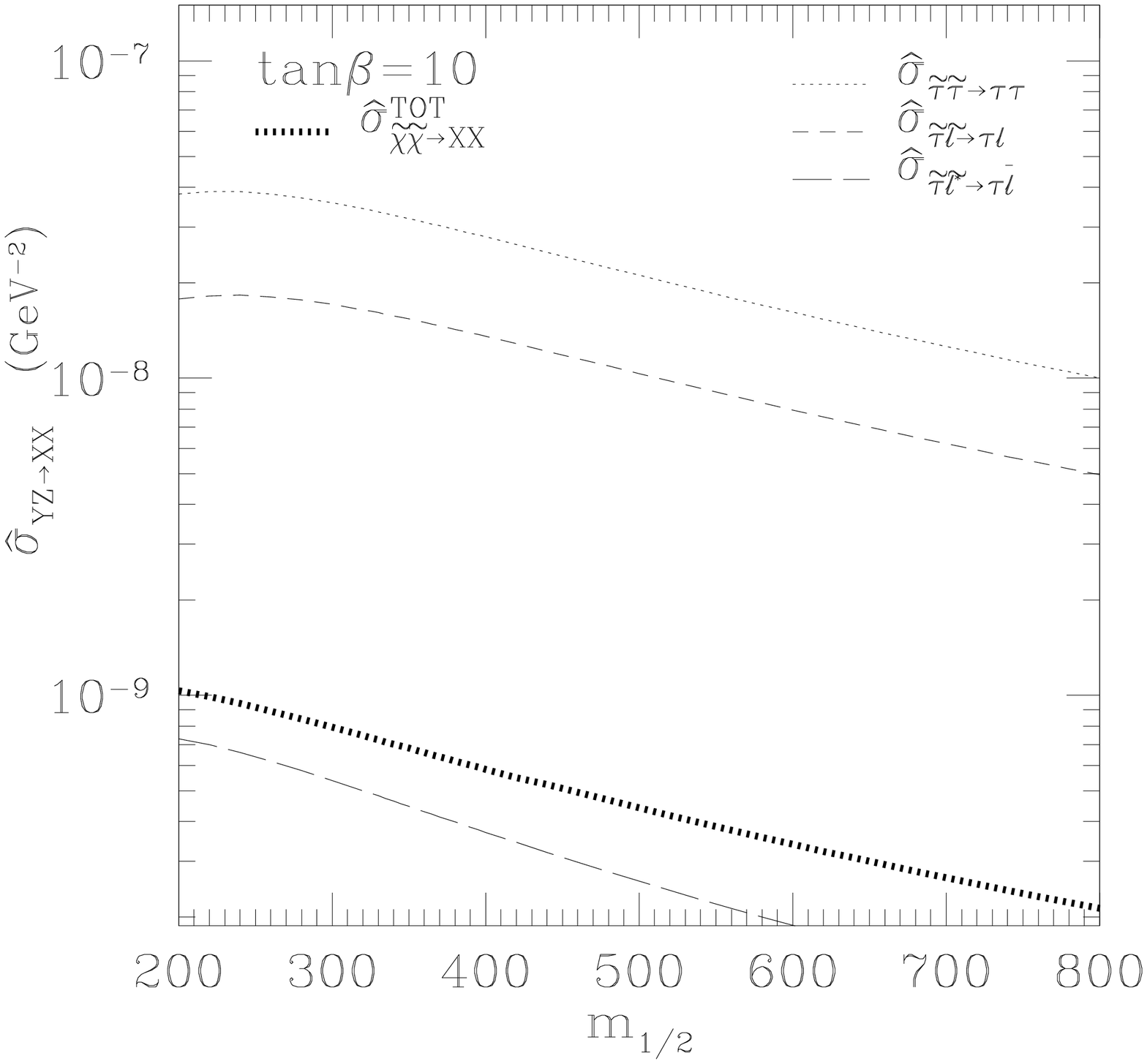,height=3.5in} 
\caption{{\it As in Fig.~\protect\ref{fig:ss}, but for the choice
$\tan\beta=10$.}\label{fig:ss10}}
\end{figure}

\begin{figure}
\vspace*{-0.5in}
\begin{minipage}[b]{8in}
\epsfig{file=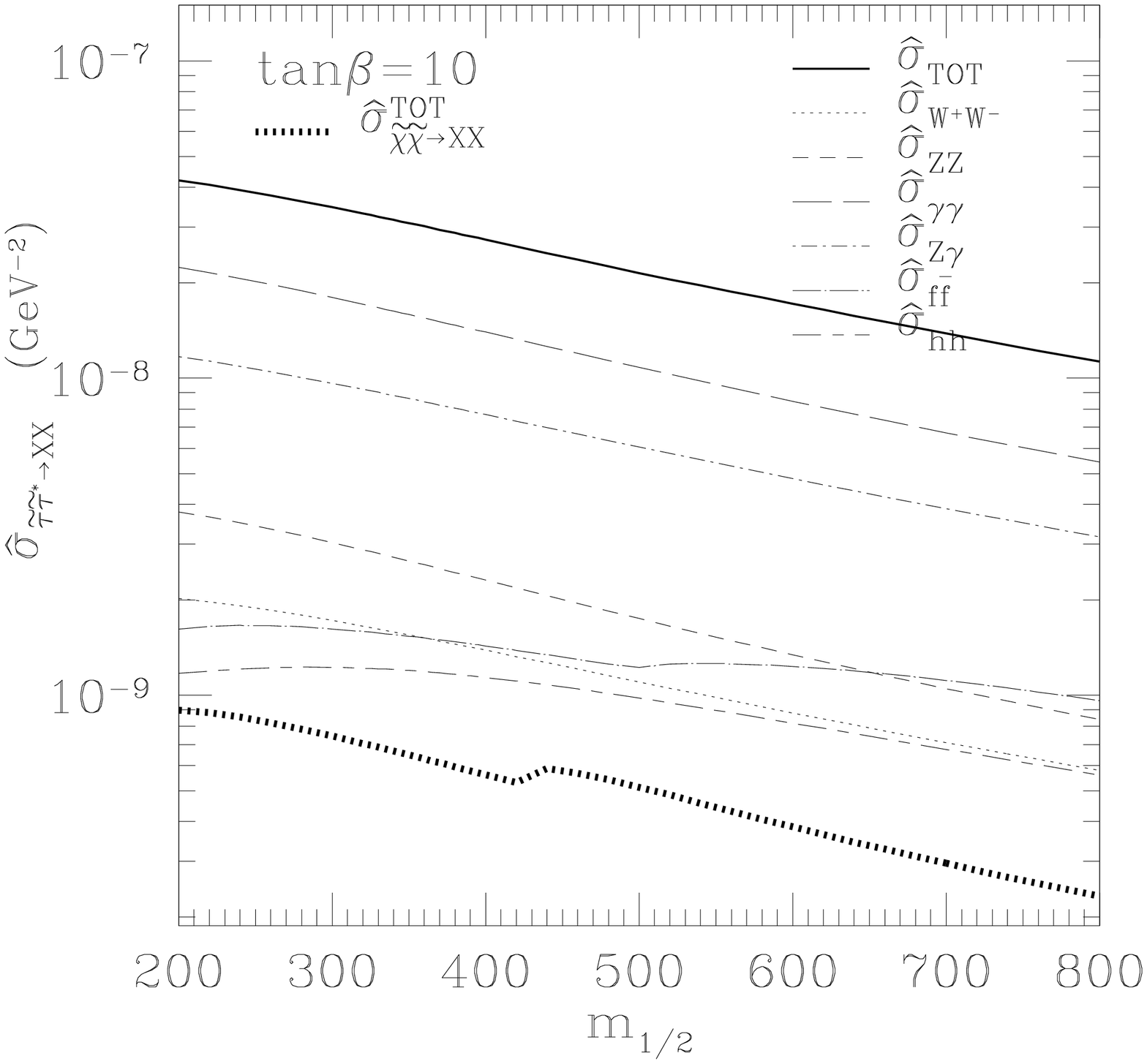,height=3.5in}
\epsfig{file=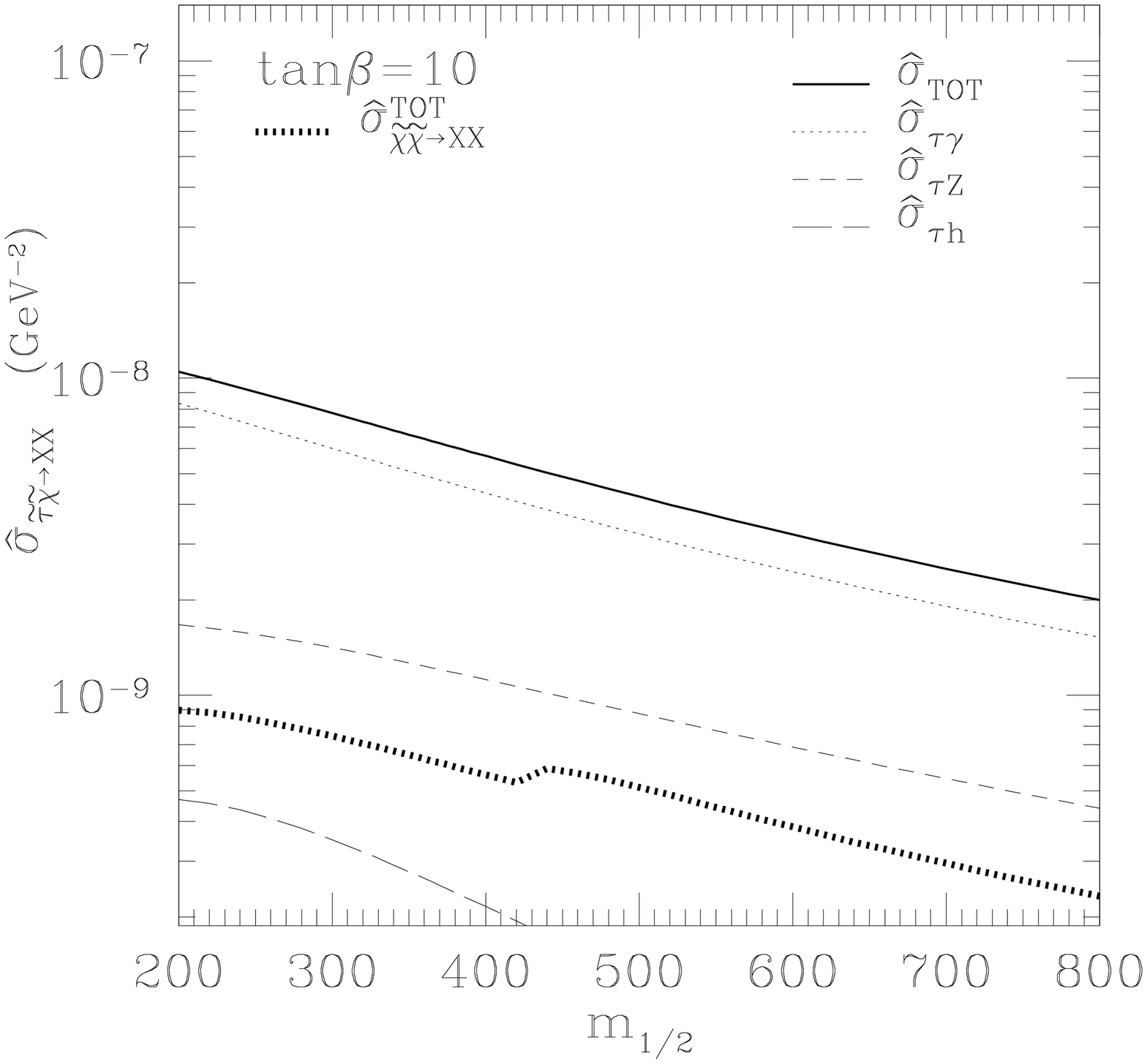,height=3.5in} \hfill
\end{minipage}
\hspace*{1.6in}
\epsfig{file=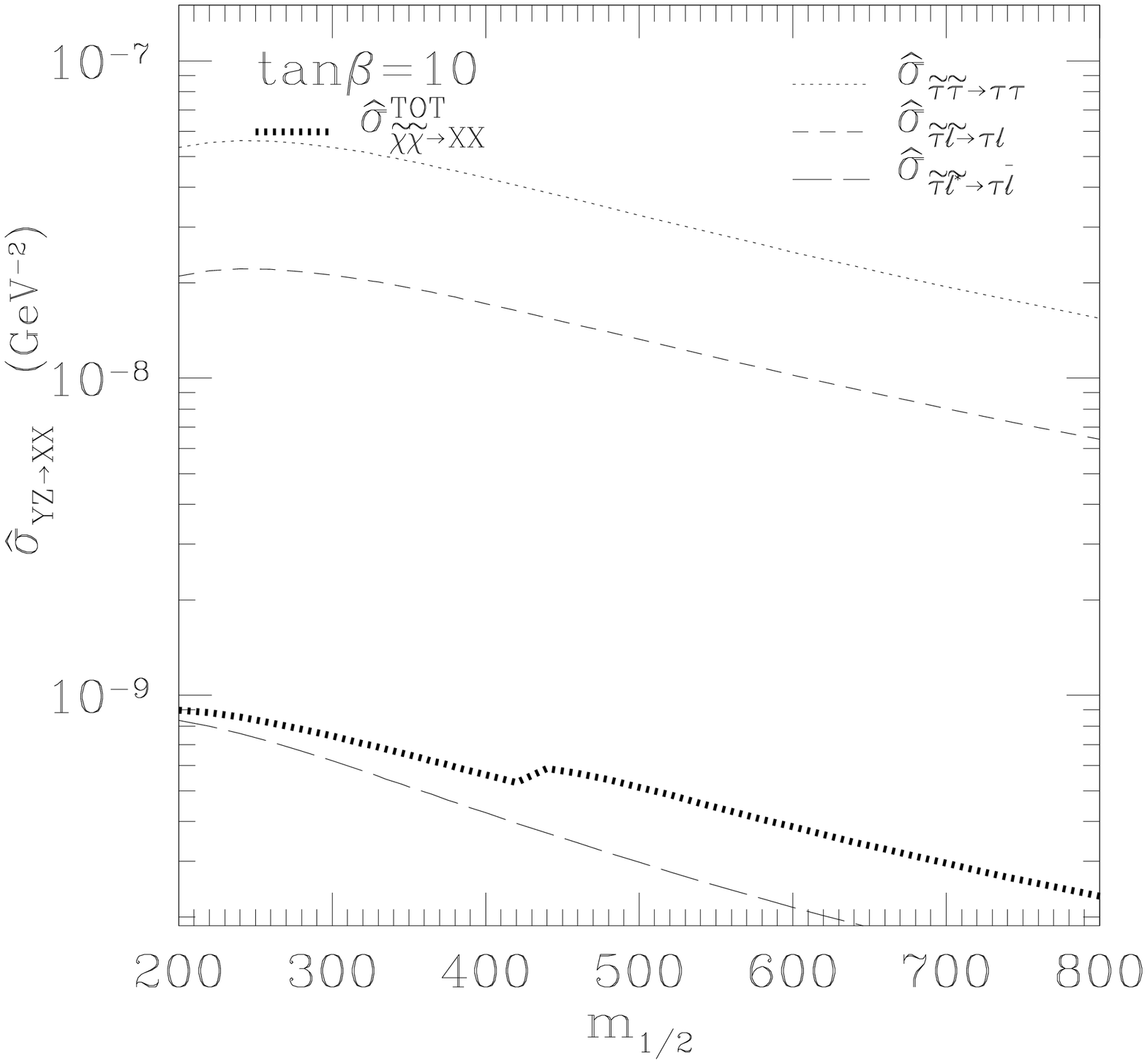,height=3.5in} 
\caption{{\it As in Fig.~\protect\ref{fig:ss}, but for the choice
$\tan\beta=10$, and $A_0=-3\m12$.}\label{fig:ss10.3}}
\end{figure}

We display in Fig.~\ref{fig:ss} numerical values of the contributions to
$\hat\sigma\equiv
a + b x/2$ (see (\ref{eq:ohsq})), for the representative values
$m_0=120\gev, A_0 = 0, x=1/23, \tan\beta=3$, and $\mu>0$, as a function of
$\m12$. For comparison, the
total cross section for $\ch \ch$ annihilation to all final states is
shown as a thick dotted line.  Due to the $P$-wave suppression of the
cross section for $\ch \ch$ annihilation
to fermion pairs, the $\ch \ch$
cross section tends to be an order of magnitude smaller than the
others, which is why coannihilation effects are so
important.  In practice, we find that the dominant contributions to
$\hat\sigma_{\rm eff}$ generally come from annihilations of
$\sl^{\,i}\,\sl^{\,i^{\scs *}}$ to gauge bosons,
$\sl^{\,i}\,\sl^{\,j}$ to lepton pairs, and $\sl^{\,i}\,\ch$ to
$\ell^i \;+\,$ gauge boson.  Due to the momentum dependence of the
$\stau\stau^*Z$ and $\stau\stau^*\gamma$ couplings, the cross sections
for $\stau\stau^*$ annihilation into $Z h, \gamma h$ and $f\!\bar f$
are also $P$-wave suppressed.  The $Z h$ and $\gamma h$ cross sections
are thus off the bottom of the figure, and $f\!\bar f$ is significant
only by dint of the large number of final states.  The top threshold
is visible as a small bump on the $f\!\bar f$ line between $\m12$ of 300
and
400$\gev$.  The curve for $\stau\stau^*\rightarrow W+W^-$ in the first
panel is obscured by the thick dotted line.  
As can be seen by comparing Figs.~\ref{fig:ss} and ~\ref{fig:ss200},
there is a dependence of the cross-sections on $m_0$, and the
variation is largest at low $\m12$, where the slepton masses are most
sensitive to $m_0$.  The sharp rise near $\m12=200$GeV in Fig.~\ref{fig:ss200}a 
is due to heavy Higgs poles.
Figures for $\tan\beta=10$ are very similar to
those for $\tan\beta=3$, as can be seen
in Fig~\ref{fig:ss10}.  There is hardly any dependence on $A_0$ for
$\tan\beta=3$, whilst, for $\tan\beta=10$, a mild dependence can be seen 
in Fig~\ref{fig:ss10.3}, where we have taken $A_0=-3\m12$.

\begin{figure}
\vspace*{-0.5in}
\hspace*{-0.3in}
\begin{minipage}[b]{8in}
\epsfig{file=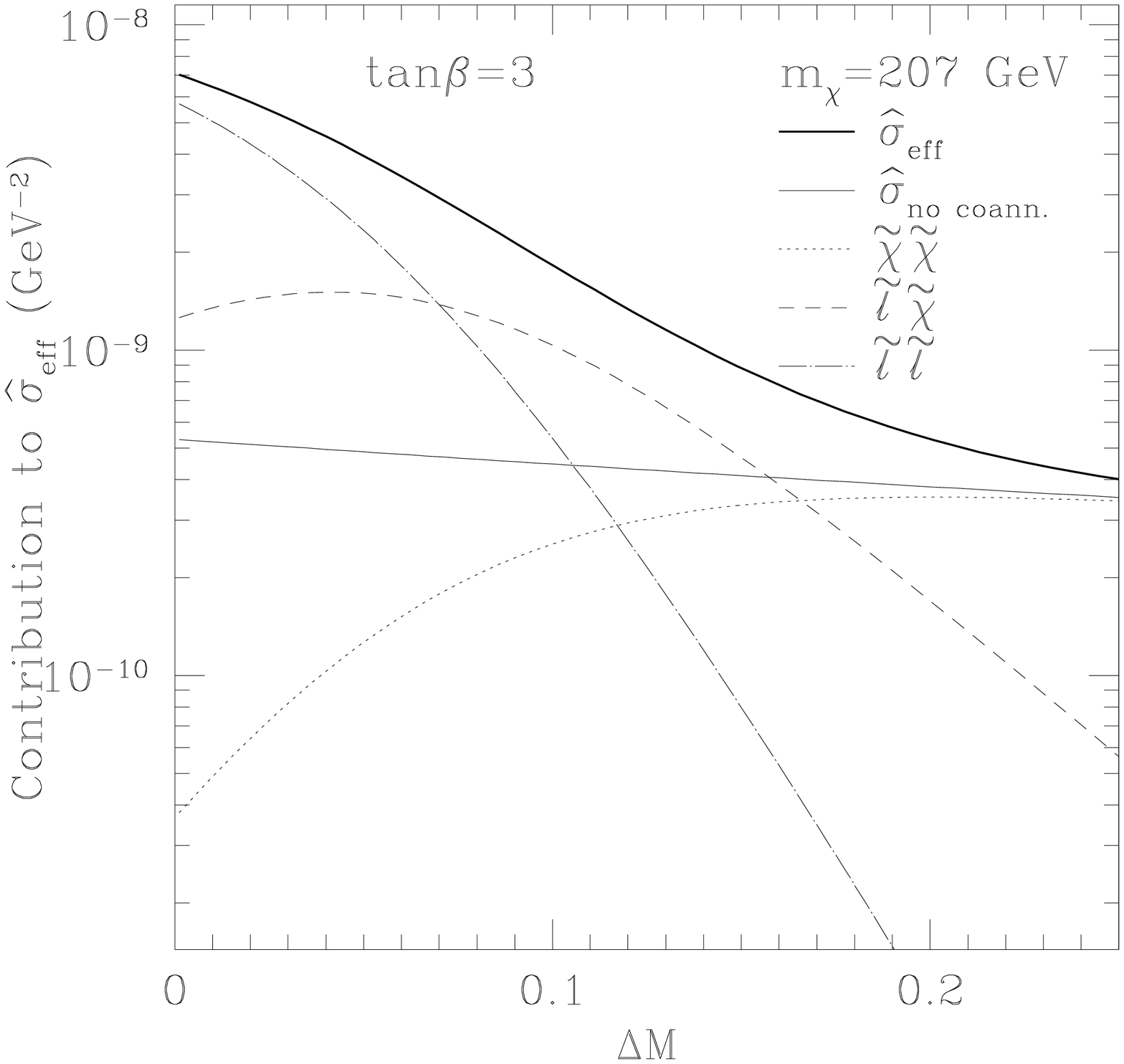,height=3.5in} 
\epsfig{file=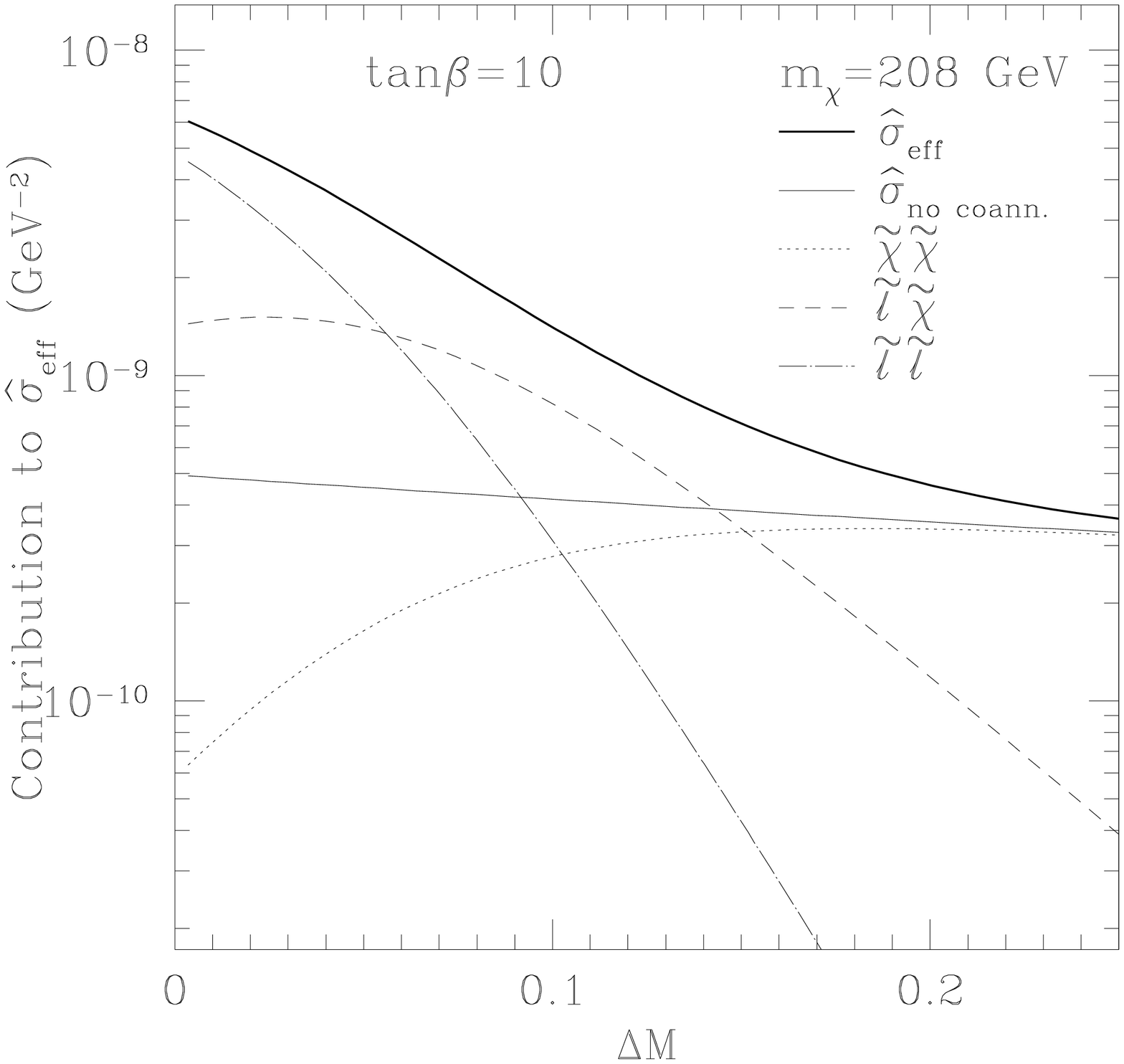,height=3.5in} 
\end{minipage}
\hspace*{-0.3in}
\begin{minipage}[b]{8in}
\epsfig{file=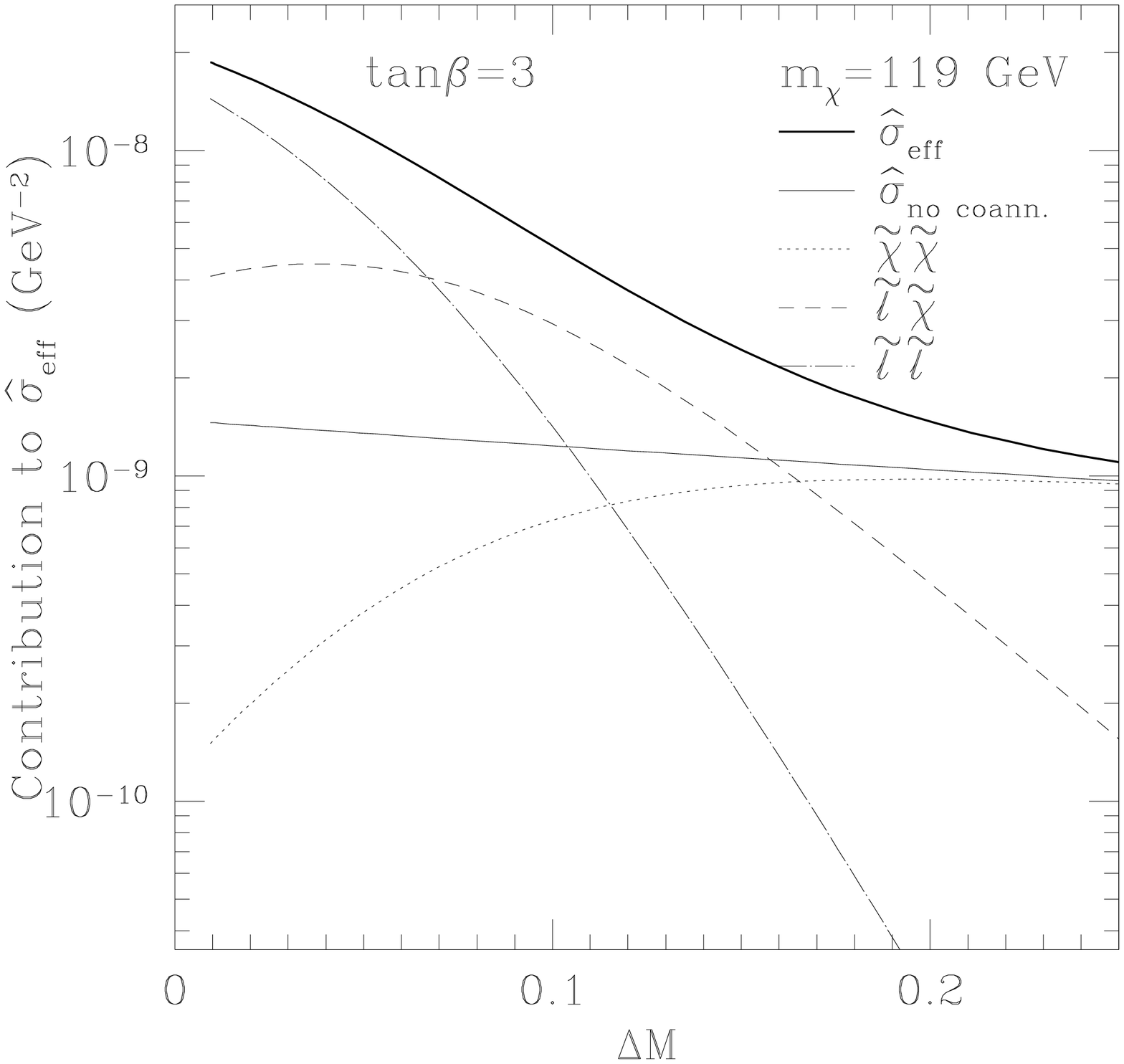,height=3.5in} 
\epsfig{file=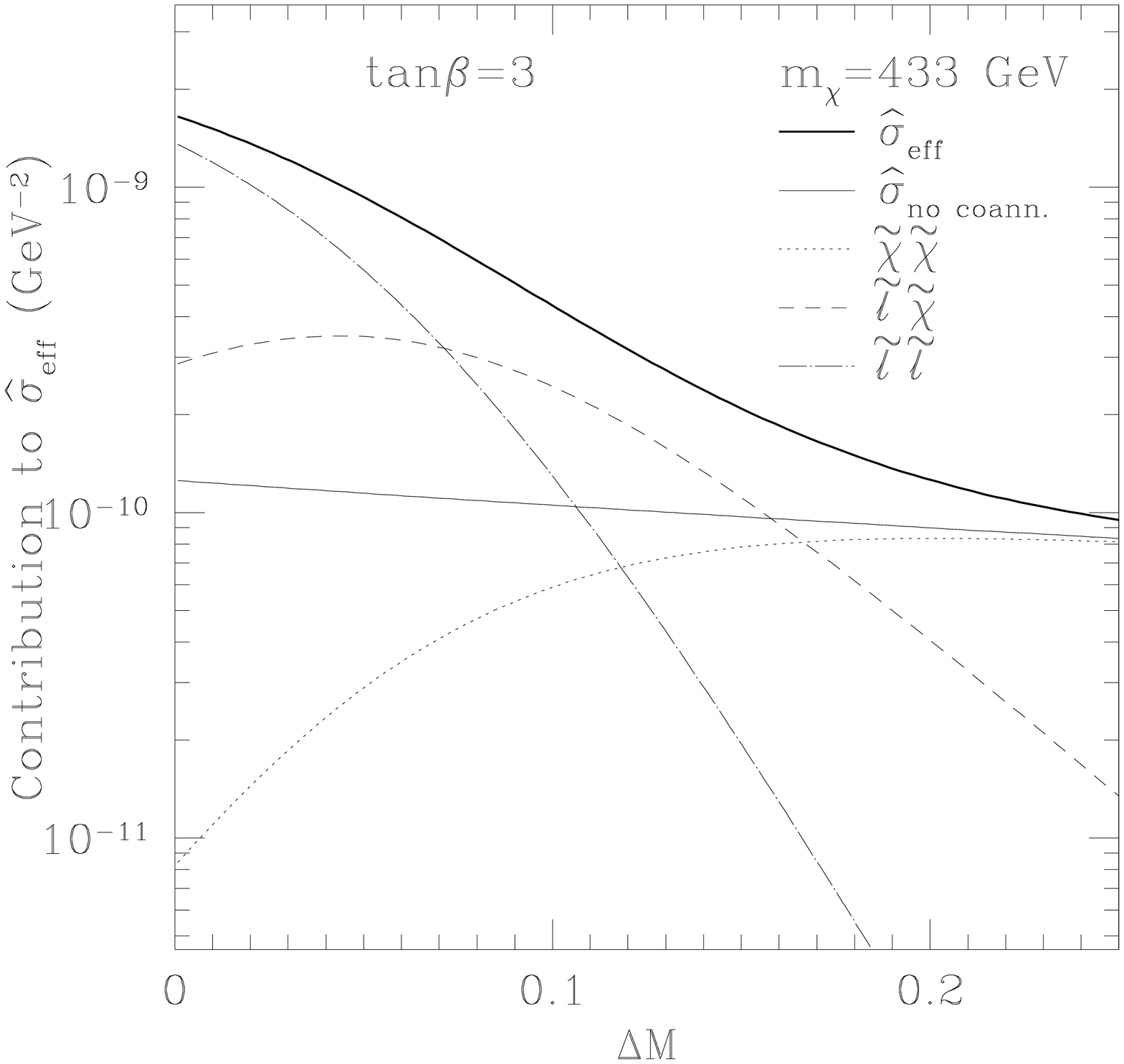,height=3.5in} 
\end{minipage}
\caption{{\it The separate contributions to the cross sections 
  $\hat\sigma_{\rm eff}$ for $x=T/m_\ch=1/23$, as functions
of $\Delta M\equiv(\mst- m_\ch)/m_\ch$, with a) $(\m12, \tan
\beta) = (500$~GeV$, 3)$, b) =$(500$~GeV$,10)$, c) =$(300$~GeV$,3)$, and
d) =$(1000$~GeV$,3)$. } 
\label{fig:svdm}}
\end{figure}
The contributions of the various annihilation channels to $\sigma_{\rm
  eff}$ are weighted by the relative abundances of the $\stau_R,\tilde
e_R,\tilde\mu_R$ and $\ch$.  For sleptons degenerate with the $\ch$,
slepton annihilation and slepton-neutralino coannihilation clearly
dominate the contributions to $\sigma_{\rm eff}$ in (\ref{sv2}), and
the final neutralino relic density is greatly reduced.  As the sleptons
become heavier than the neutralinos, their number densities are
exponentially suppressed, and when the mass differences are no longer
small, the slepton contributions to $\sigma_{\rm eff}$ are negligible.
Fig.~\ref{fig:svdm} shows the sizes of the separate contributions to
$\hat\sigma_{\rm eff}$ from neutralino annihilation,
neutralino-slepton coannihilation and slepton-slepton annihilation and
coannihilation, as functions of the mass difference between the
$\stau$ and $\ch$.  In Fig.~\ref{fig:svdm}a, we
have fixed $\m12=500\gev, \tan\beta=3, \mbox{$A_0=0$}, \mu>0,$ and
computed $\hat\sigma_{\rm eff}$ for varying $m_0$, which amounts to
varying
the mass difference $\Delta M$.  In this case, all the sleptons are closely
degenerate with each other and hence contribute equally to $\hat\sigma_{\rm
eff}$.  The thin solid line is the
$\hat\sigma$ which one would compute if one ignored the slepton states,
i.e., $a_{\chi\chi}+b_{\chi\chi}x/2$.
Note that, in the case of close degeneracy between the $\ch$ and
$\stau_R$, it is
in fact slepton annihilation by itself which dominates $\hat\sigma_{\rm
  eff}$.  Since this contribution is suppressed by two powers of
$n_{\rm eq,\stau_R}$, it drops rapidly with $\Delta M$, and
neutralino-slepton coannihilation takes over at $\Delta M\ga 0.07$.
This contribution in turn falls with one power of $n_{\rm eq,\stau_R}$,
and neutralino annihilation re-emerges as the dominant 
reaction for $\Delta M\ga0.17$. When
$\Delta M\ga 0.25$, the two solid lines and dotted line all merge, and
coannihilation can be neglected.
Figures for other $\m12$ and $\tan\beta$ are shown in
Fig.~\ref{fig:svdm}b-d: the underlying physics is similar, and
they are similar in shape to Fig.~\ref{fig:svdm}a, but with different
normalizations. In Fig.~\ref{fig:svdm}b, we take $\tan \beta = 10$, in
Fig.~\ref{fig:svdm}c,d we take $\tan \beta = 3$, with $\m12 = 300$
and 1000 respectively.  The values of $m_\chi$ shown on the figures is
determined by the choice of $\m12$.

\section{Implications of Coannihilations for the Upper Limit on the LSP 
Mass}

\begin{figure}
\hspace*{-.70in}
\begin{minipage}{8in}
\epsfig{file=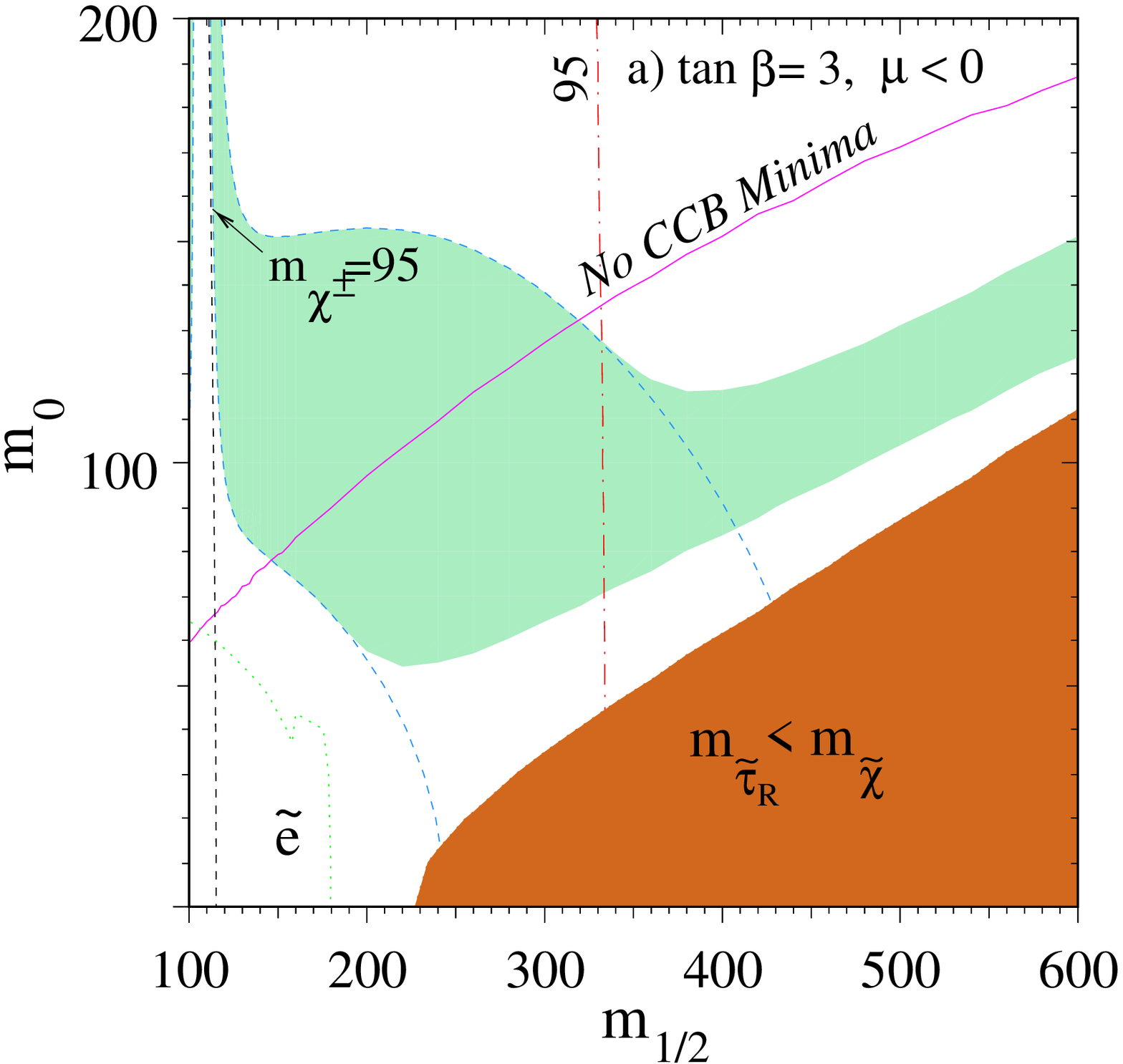,height=3.5in} 
\epsfig{file=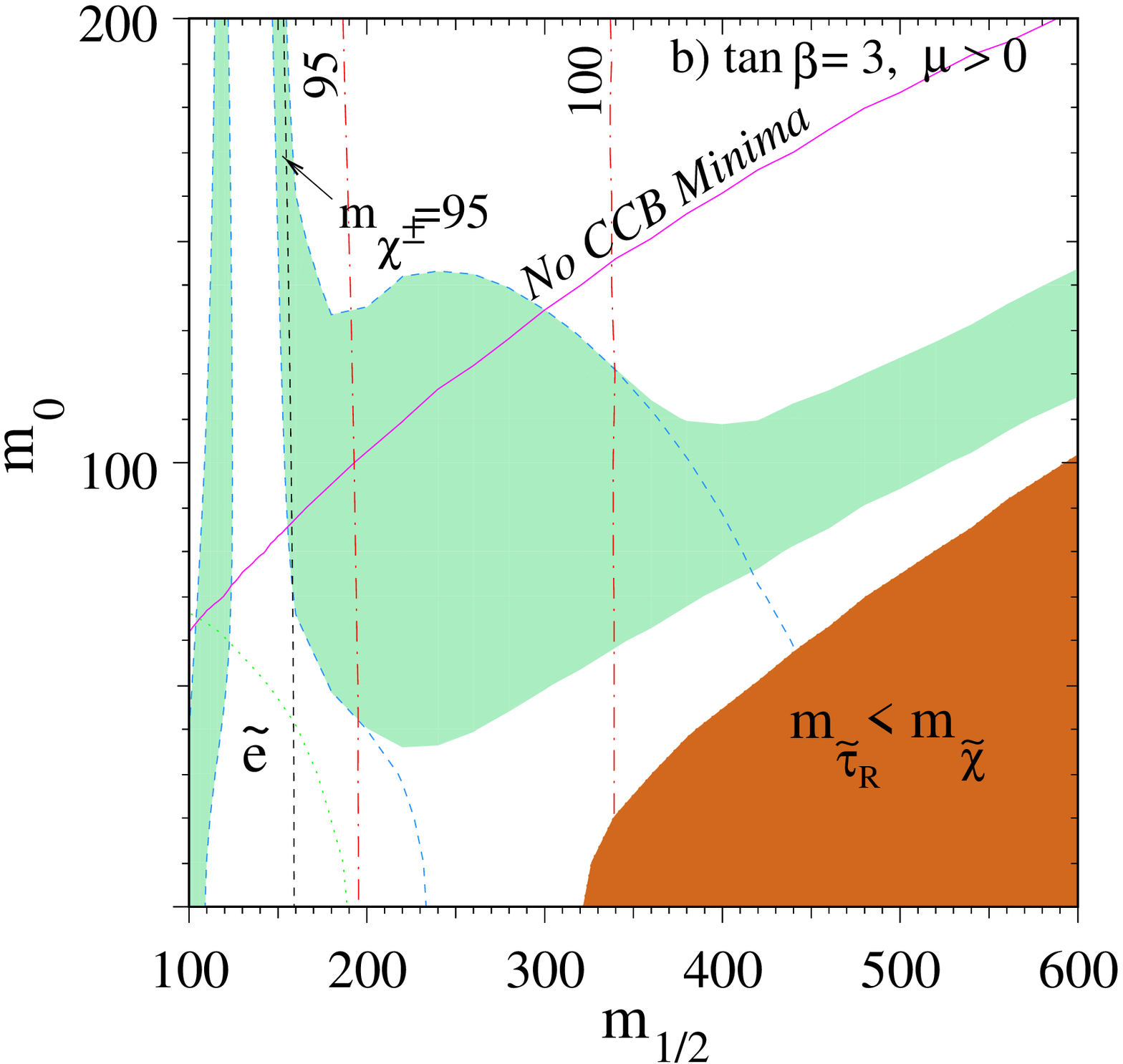,height=3.5in} \hfill
\end{minipage}
\begin{minipage}{8in}
\hspace*{-.70in}
\epsfig{file=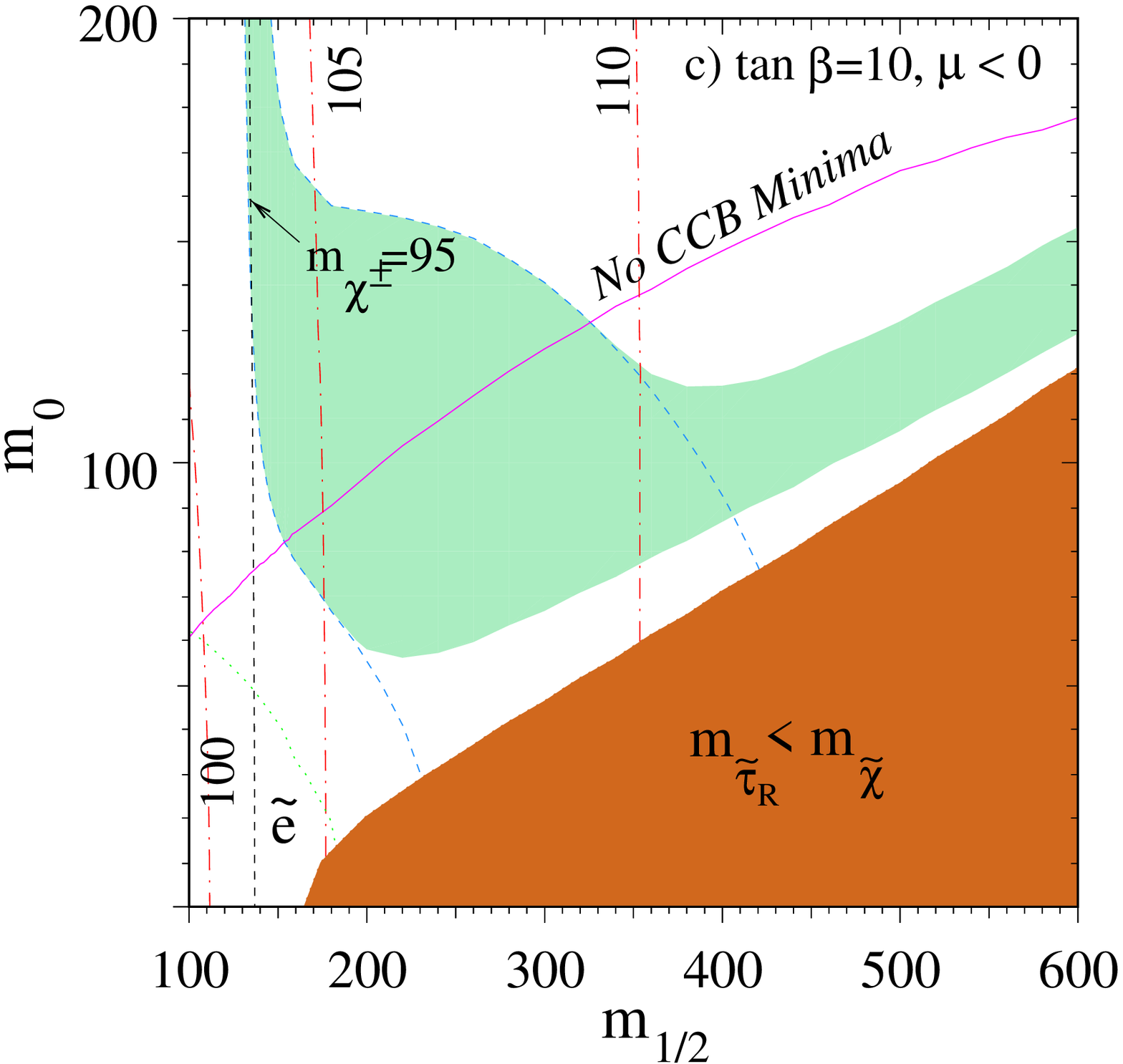,height=3.5in} 
\epsfig{file=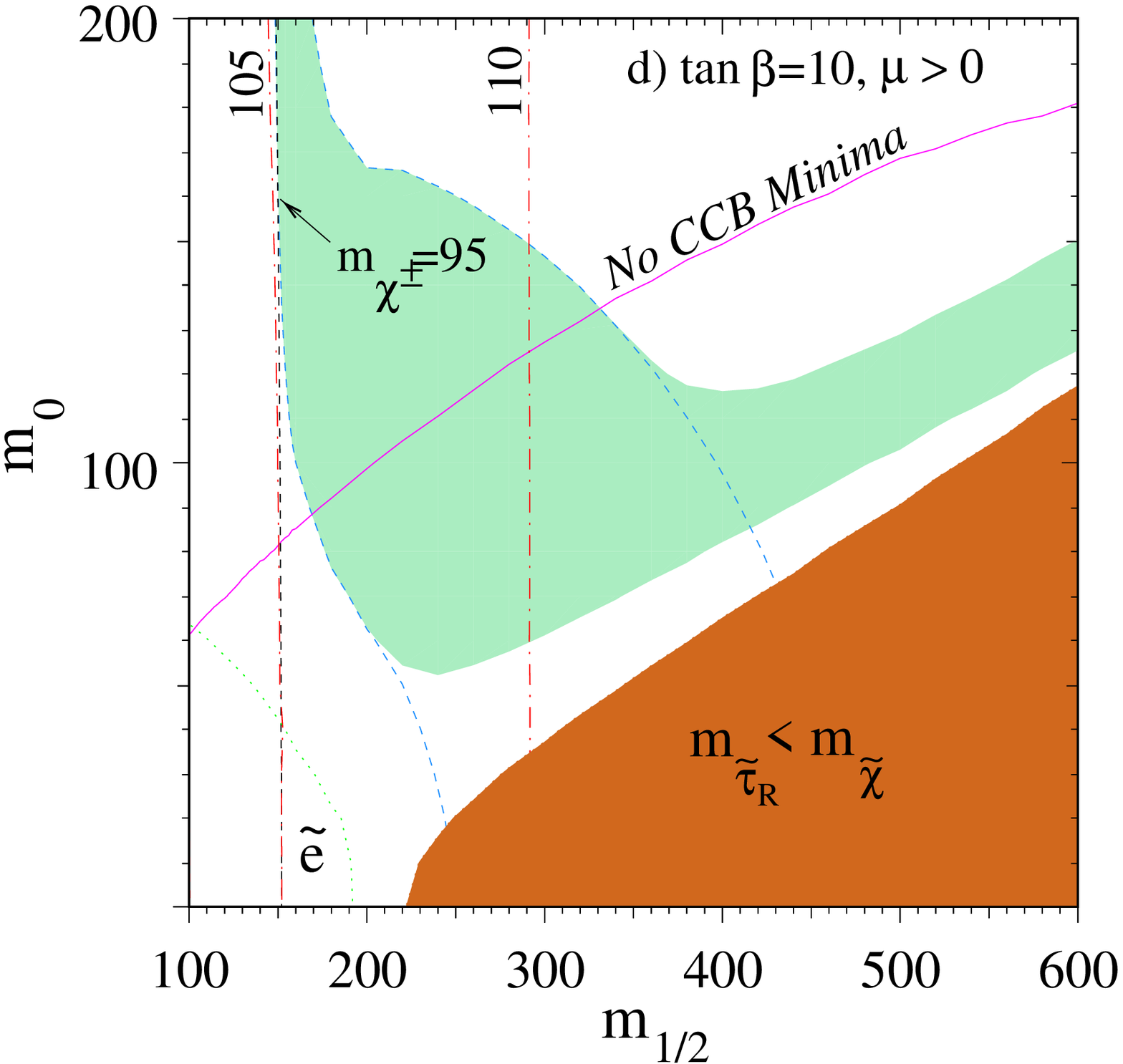,height=3.5in} \hfill
\end{minipage}
\caption{\label{fig:sm}
{\it The light-shaded area is the cosmologically preferred 
region with \protect\mbox{$0.1\leq\ohsq\leq 0.3$}.   The light dashed lines
show the location  of the cosmologically preferred region  if one
ignores  coannihilations with the light sleptons.  
In the dark shaded regions in the bottom right of each panel, the LSP is
the ${\tilde
\tau}_R$, leading to an unacceptable abundance
of charged dark matter.  Also shown are the isomass
contours $m_{\chi^\pm} = 95$~GeV and $m_h = 95,100,105,110$~GeV,
as well as an indication of the slepton bound from
LEP~\protect\cite{LEP}.  In the area below the solid contour, the scalar potential
 contains charge and/or colour breaking minima.}}
\end{figure}

We now explore the consequences of coannihilation for cosmological
bounds in the CMSSM.  In Fig.~\ref{fig:sm}, we display the
cosmologically and experimentally permitted regions of the
$(\m12,m_0)$ plane.  We have chosen the representative points
$\tan\beta=3$ and 10, and present results for both $\mu<0$ and
$\mu>0$. The light shaded
regions correspond to \mbox{$0.1<\ohsq<0.3$}.  The dark shaded regions
have $m_{\st}< m_\ch$ and are excluded by the very stringent bounds on
charged dark matter\cite{ehnos}~\footnote{To be more precise: here and
subsequently, we have included consistently the effects of
${\tilde \tau}_R - {\tilde \tau}_L$ mixing on the mass of the lighter
$\tilde \tau$ (which is mainly $\tilde \tau_R$), both in delineating the
cosmological exclusion domain and in the kinematics of coannihilation.
However, we have not included mixing angle effects in the
(co)annihilation amplitudes, since these are small for the values of
tan$\beta$ studied in this paper.}. The light dashed contours
indicate the corresponding regions in $\ohsq$ if one ignores the effect of
coannihilations.  Neglecting coannihilations, one would find an upper
bound of $\sim450\gev$ on $\m12$, corresponding to an upper bound of
roughly $200\gev$ on $m_{\tilde B}$.  The effect of coannihilations is
to create an allowed band about 25-50 $\gev$ wide in $m_0$ for $\m12 \la
1400\gev$, which tracks above the $\mst=m_\ch$ contour.  Along the
line $\mst= m_\ch$, we find $R\approx10$, as shown numerically in
Fig.~\ref{fig:svdm} and (\ref{eq:R}).  As $m_0$ increases,
$\Delta M$ increases and the slepton contribution to $\hat\sigma_{\rm
  eff}$ falls, as in Fig.~\ref{fig:svdm}, and the relic density rises
abruptly.

We also display bounds from LEP particle searches~\cite{LEP} in
Fig.~\ref{fig:sm}. The chargino
mass bound from LEP essentially
saturates the kinematic limit of $\sim95\gev$, modulo a
small gap which occurs when the chargino is just slightly heavier than
the sneutrino~\cite{LEP}.  The chargino $95\gev$ isomass contour is
displayed as
a dark dashed contour, and it cuts off most of the annihilation pole
zone, which appears as a chimney in the cosmologically preferred
regions at low $\m12$ for the values of $\tan \beta$ considered.  
Representations of slepton
limits from searches~\cite{LEP} for acoplanar lepton pairs at LEP~183
appear as light dotted contours.
The most severe experimental constraint at low
$\tan\beta$ comes from LEP Higgs searches.  At low $\tan\beta$, the tree level
light Higgs mass $m_h\approx \mz|\cos{2\beta}|$ lies well below the
experimental bound $\ga 95\gev$.  Radiative corrections to $m_h$
are known to be large~\cite{MSSMHiggs} and increase logarithmically
with the stop masses, and hence with $\m12$.  Thus, for sufficiently
large $\m12$, the Higgs bound may be satisfied. However, if 
the minimum $\m12$ exceeds its cosmological upper bound
for a given $\tan\beta$,
this value of $\tan\beta$ is excluded.  In the absence of
coannihilations, the lower bound on $\tan\beta$ was computed~\cite{efgos} 
to be 2.0 for $\mu>0$ and 1.65 for $\mu<0$, using mass
limits from LEP~183.  Higgs isomass contours of 95,100,105, and 110
$\gev$ are displayed as dot-dashed lines in Fig.~\ref{fig:sm}.

\begin{figure}
\vspace*{-0.5in}
\hspace*{-.50in}
\begin{minipage}[b]{8in}
\epsfig{file=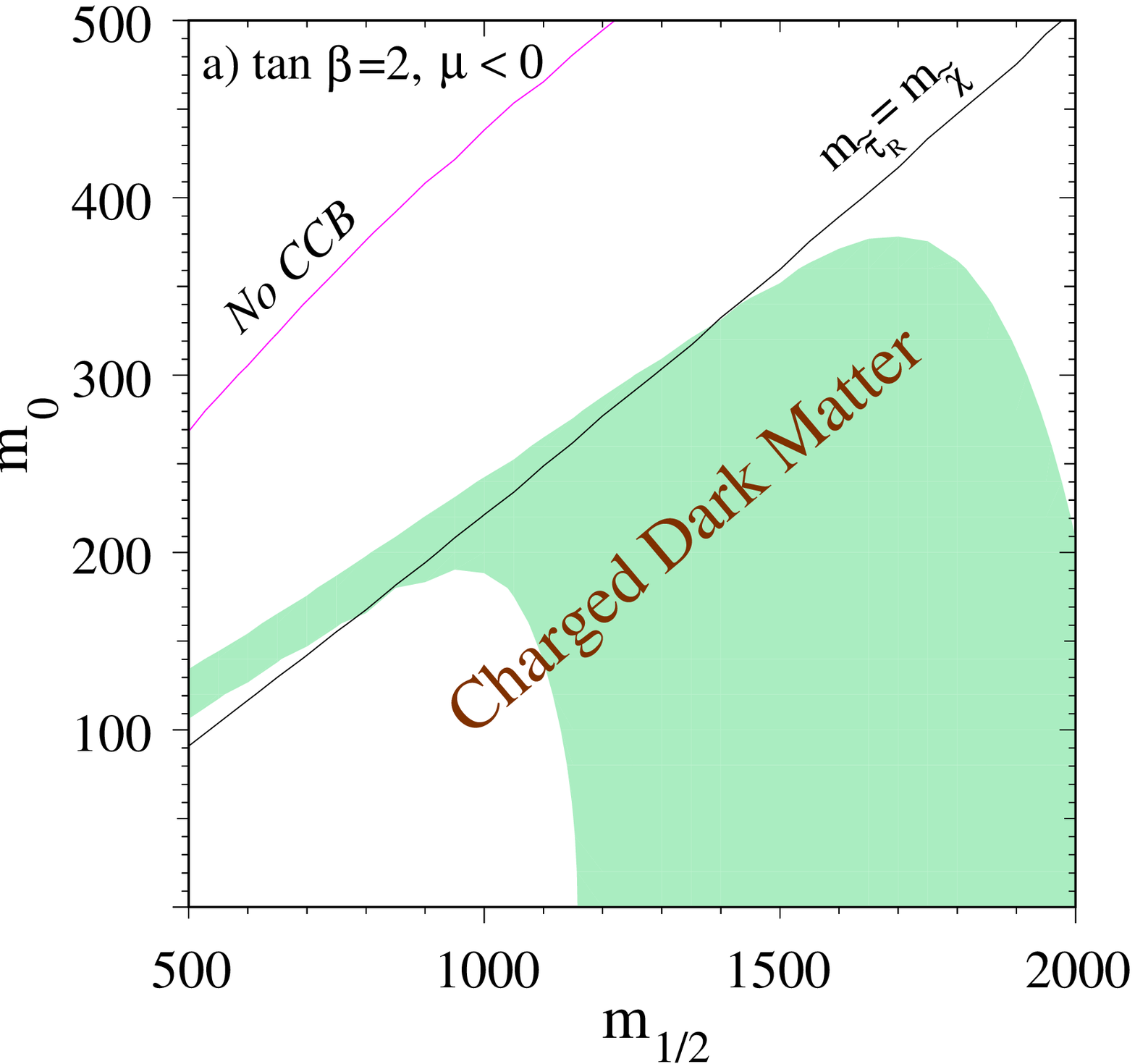,height=3.5in}
\epsfig{file=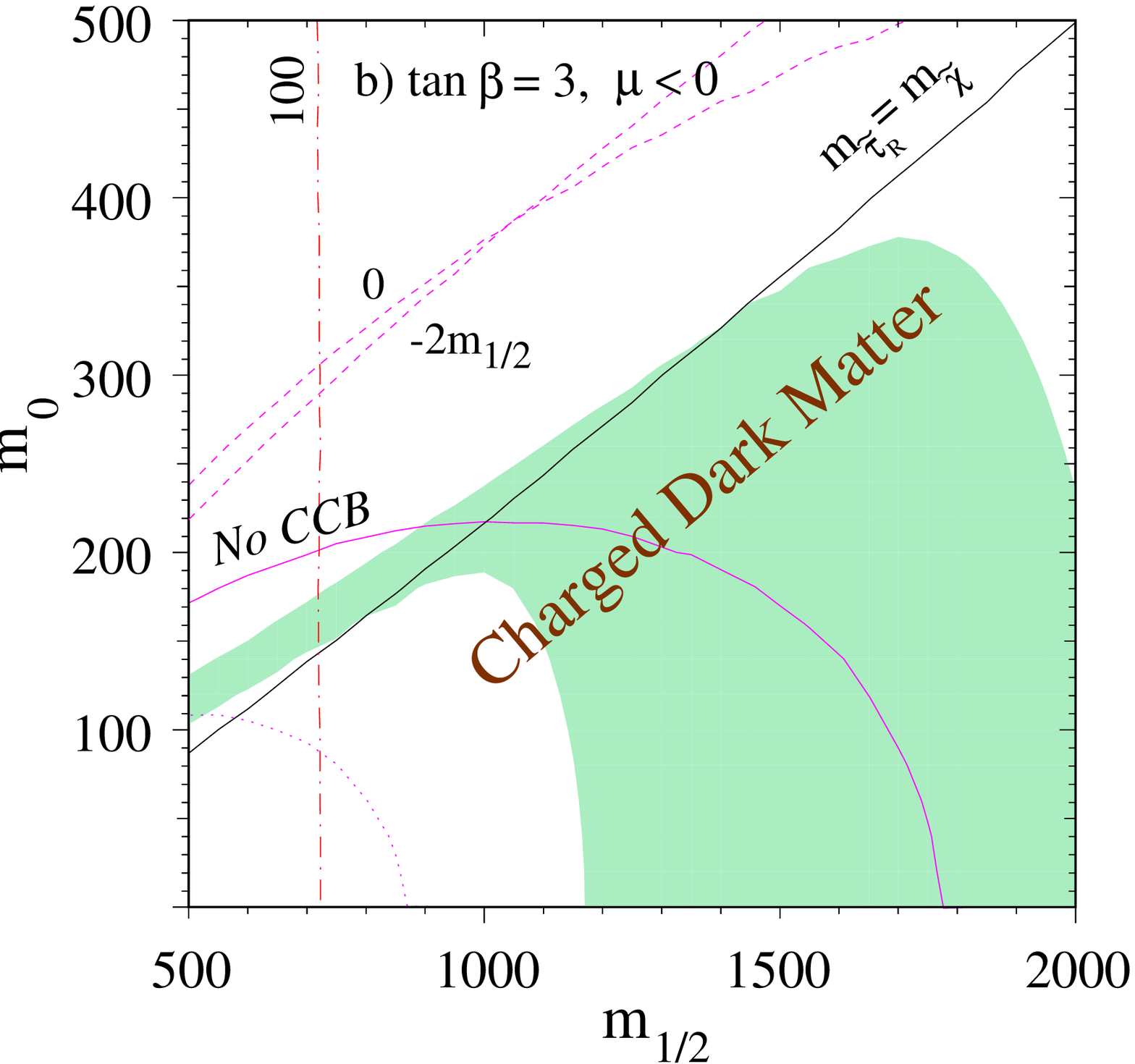,height=3.5in} \hfill
\vspace*{.2in}
\end{minipage}
\hspace*{1.6in}
\epsfig{file=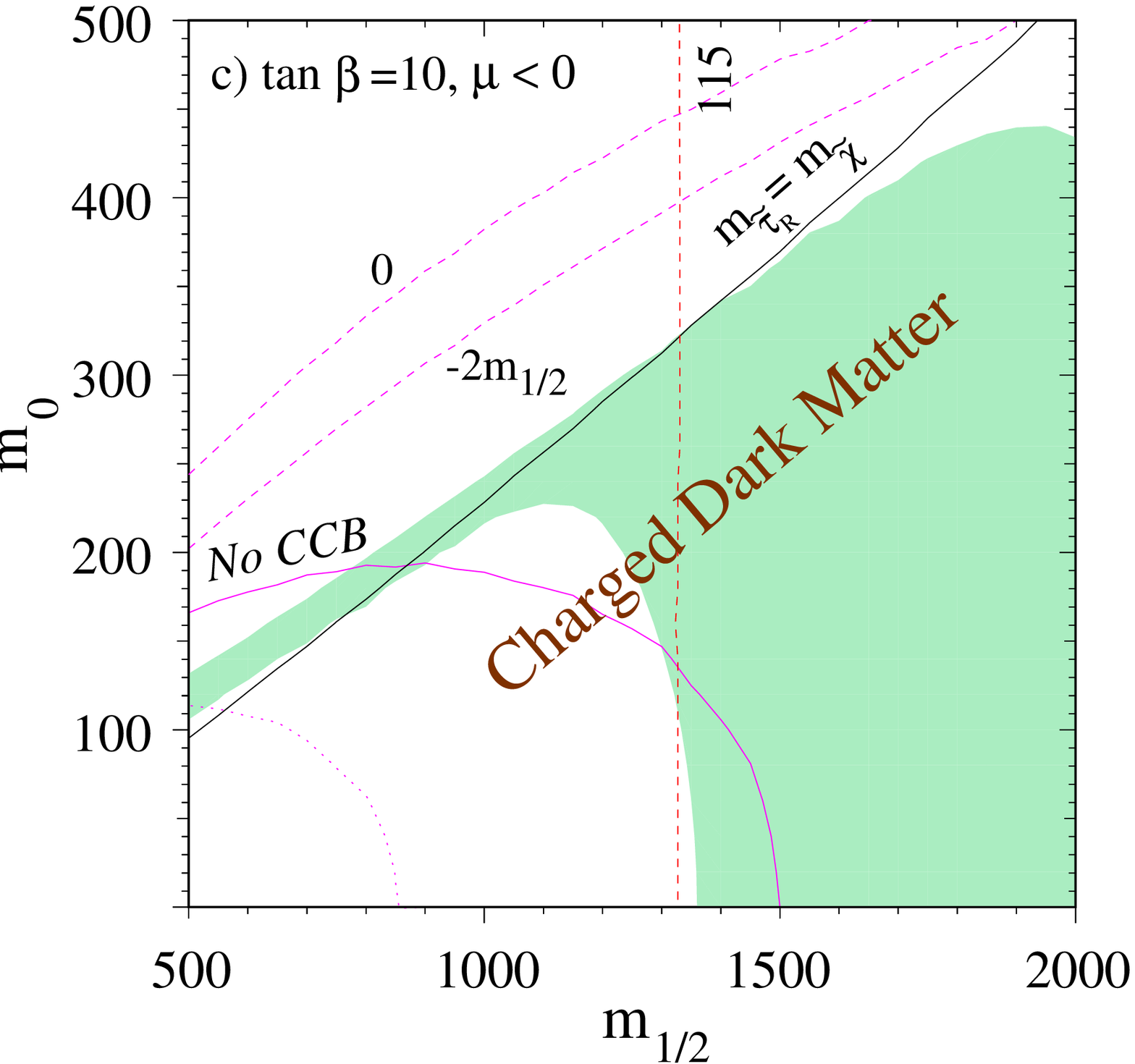,height=3.5in} 
\vspace*{0.1in}
\caption{{\it The
same as Fig.~\protect\ref{fig:sm}, for $\mu<0$, 
but extended to larger $m_{1/2}$.
The dashed contours in panels b) and c) correspond to the solid CCB line for $A_0=0,
-2\m12$, and the dotted contour for $m_t=170$~GeV.}}
\label{fig:bg}
\end{figure}

When we include coannihilations, the cosmological upper bound on $\m12$
weakens considerably, as we have already mentioned.  In Fig.~\ref{fig:bg},
we extend the $\m12$
coverage to larger scales, to show the cross-over point between the
regions with $\ohsq<0.3$ and $m_{\st}< m_\ch$.  The two constraints
together require $\m12\la1450$, corresponding to an upper bound on the
neutralino mass of $m_\ch\la600\gev$.  We take $93\gev$ as our experimental lower limit
on the Higgs mass, to allow for a $2\gev$ uncertainty in the extraction of the radiative
corrections to the Higgs mass~\cite{MSSMHiggs}.  The combined 
lower bound on $\tan\beta$ from the Higgs search and cosmology, 
including the recent LEP~189 limits, indicates that tan$\beta \ga
2.2$ for $\mu<0$ and $ \ga 1.8$ for $\mu>0$~\footnote{The LEP
bounds on the CMSSM parameter space will be considered in more detail
in~\cite{efgoss}, as well as a comparison with the supersymmetric
reach of the Fermilab Tevatron collider.}. 

\begin{figure}
\hspace*{-.70in}
\begin{minipage}{8in}
\epsfig{file=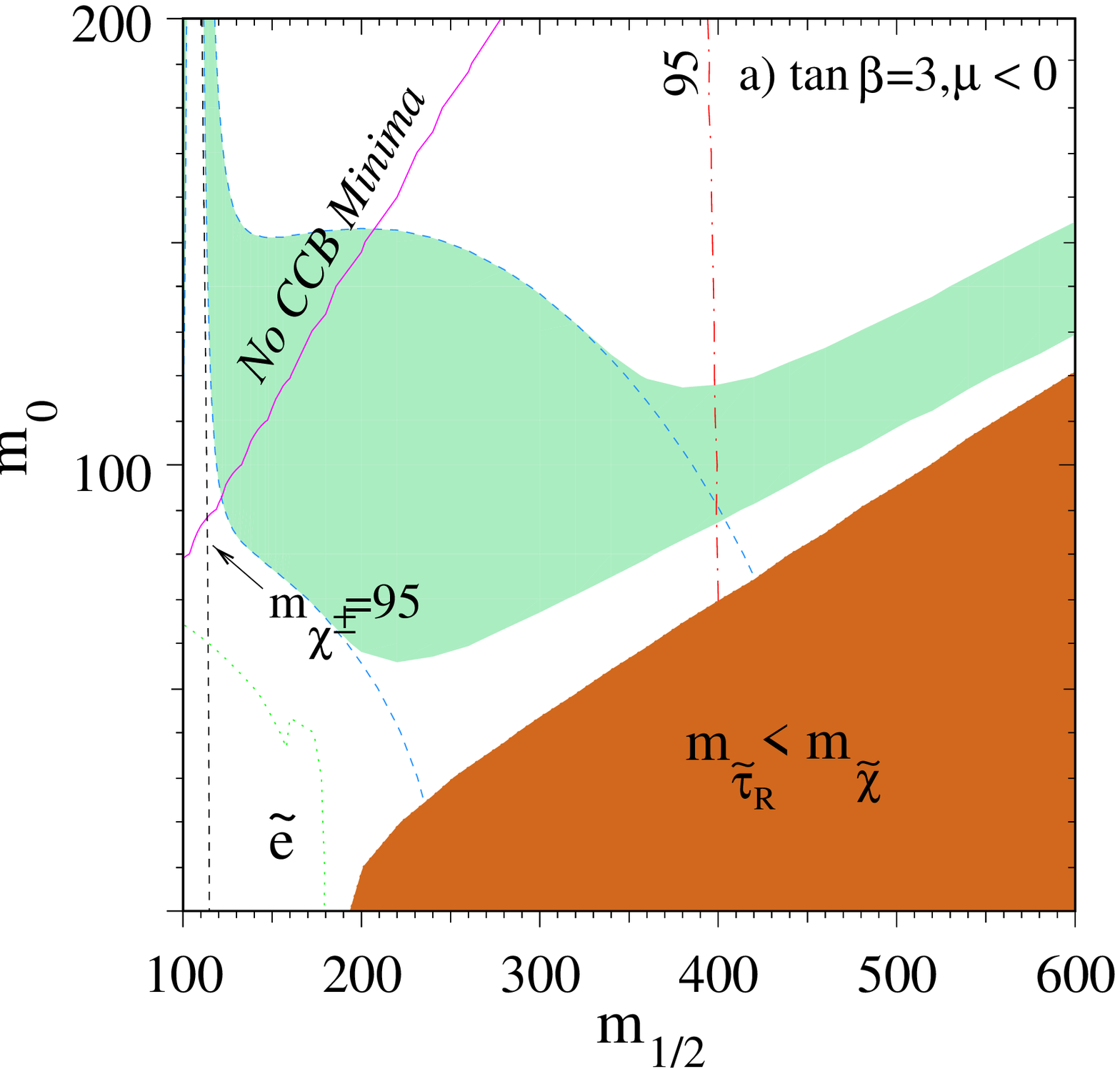,height=3.5in} 
\epsfig{file=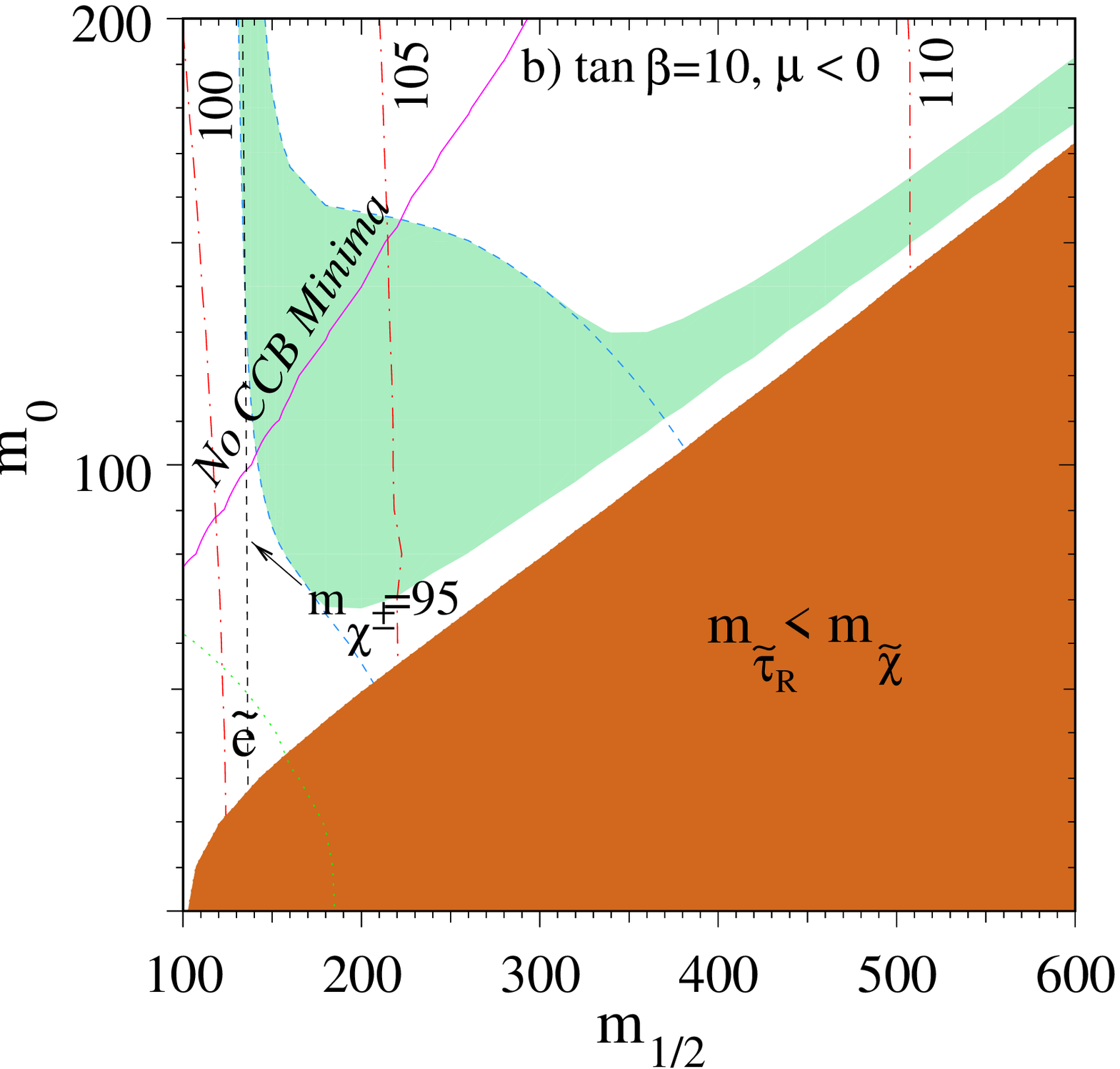,height=3.5in} \hfill
\end{minipage}
\caption{\label{fig:cmpA0}
{\it The same as Fig.~\protect\ref{fig:sm}a and Fig.~\protect\ref{fig:sm}c, for $A_0=-3\m12$.}}
\end{figure}

Figs.~\ref{fig:sm} and \ref{fig:bg} were generated for the particular
choice $A_0=-\m12$, for reasons described below.  Since the stau mass
depends weakly on $A_0$, particularly at larger values of $\tan\beta$,
the position of the $\mst=m_\ch$ contour shifts somewhat with $A_0$.
In Fig.~\ref{fig:cmpA0}, we show figures
corresponding to Figs.~\ref{fig:sm}a and \ref{fig:sm}c for
$A_0=-3\m12$.  For $\tan\beta=3$, the width of the cosmologically
allowed band is the same as in Fig.~\ref{fig:sm}a, but 
the band becomes smaller for $\tan\beta=10$.  The effect for positive
$A_0$ is qualitatively similar.   Likewise, varying $A_0$ has a tiny effect on the 
cosmological upper bound on $\m12$ for $\tan\beta=3$, but can produce a
small
($\sim100\gev$) reduction in the upper bound for $\tan\beta=10$. 

Lastly, we show as a dark solid contour the constraint coming from the
requirement that the global minimum of the scalar potential not break
charge and colour (CCB)~\cite{bbc,as,af}.  
The areas below the light solid line in Figs.~\ref{fig:sm} and
\ref{fig:bg} contain charge and/or colour breaking minima, while the
regions above the lines are free of such minima\footnote{The
  difference between the CCB curves of Fig.~\ref{fig:sm} and \cite{af} 
  is due to our choice here of $m_t=175\gev$, rather than $170\gev$.}, modulo a thin
(\mbox{$\sim10\gev$} wide) strip on top of the solid contour, where
local (but not global) CCB minima 
exist~\cite{af}.
We have chosen $A_0=-\m12$, where the
CCB bounds are weakest~\cite{bbc}.  The bounds are not strongly dependent on the
sign of $\mu$ and are strongest for low $\tan\beta$, where the 
top Yukawa coupling is largest.
In the absence of coannihilations, it is clear from Fig.~\ref{fig:sm}
that a fair fraction of the cosmologically allowed regions contain charge and
colour breaking minima.  In the presence of coannihilations, the
cosmologically allowed region is extended, and as the CCB eventually
falls with $\m12$, CCB free regions with small relic densities may
occur in the proboscis region as well, particularly at the larger values
of
$\tan \beta$ considered.  In Fig.~\ref{fig:bg}a, we see that, for
$\tan\beta=2$, the entire cosmologically allowed region of 
parameter space also contains
CCB minima, even for the conservative choice $A_0=-\m12$.  For
larger $\tan\beta$, the CCB contour bends over and, for $A_0=-\m12$,
exposes a small strip of the shaded region above the $\mst=m_\ch$
line.  For comparison, we also show the corresponding CCB contours for
$A_0=-2\m12$ and 0. For these choices of $A_0$, there is again no
region which satisfies both the CCB and cosmological constraints.  
The CCB bounds are also sensitive to the top mass.
In Figs.~\ref{fig:sm} and \ref{fig:bg} we have taken $m_t=175\gev$;
taking $m_t=170\gev$ produces the dotted CCB curves in
Fig.\ref{fig:bg}a-c, which expose a much larger piece
of the cosmologically allowed band. We recall, however,
that a low value of $m_t$ also reduces the
radiative corrections to the Higgs mass, and 
therefore makes the Higgs constraint
more severe. However, this does not cause a conflict with the
cosmological constraints, even for $\tan\beta$ as low as 3.  

There is one caveat which one must bear in mind when interpreting
the CCB bounds: the tunneling rate
from the charge and colour conserving minimum to the CCB minimum is
very slow, to such an extent that the conserving minimum is essentially
stable over the current lifetime of the universe.  Therefore the
presence of CCB minima may present more of a cosmological problem of
how to populate the physical minimum in preference to the CCB minimum
than a constraint on the particle physics \cite{forss}.

We expect qualitatively similar effects on the corresponding bounds in the
MSSM, that is, when we drop the condition of universality of scalar mass
at the GUT scale. Though there is an upper limit of ${\cal O}$(a few TeV) for
Higgsinos in the MSSM, the limits due to coannihilation that we have
been concerned with apply only to the gaugino limit. In this
case, one must in general take all the squarks and sleptons degenerate
with the neutralino and compute the annihilation and coannihilation
cross sections for all possible combinations of sfermions.  However, if
the rates are the same as for the sleptons, the effect is about a 10-15\%
decrease in $(\ohsq)^0/R$, leading to a similar bound on $m_\ch$ as in
the CMSSM.  In practice, of course the result will depend on which
state is the NLSP.

\section{Conclusions and Open Issues}

We have documented in this paper the importance of
coannihilation effects on the relic density of ${\tilde B}$
dark matter. We have presented a general formalism for including their
effects in relic density calculations, and have provided 
(in the Appendix) analytic approximate formulae for many
coannihilation processes involving $\stau_R, {\tilde e_R}$ and
${\tilde \mu_R}$ sparticles. We have given 
in the Figures numerical examples
that exhibit their relevance for various values of the CMSSM
parameters $m_{1/2}, m_0, A$ and tan$\beta$. We have explained
why coannihilation effects are so important, principally
because of the $P$-wave suppression of 
non-relativistic $\ch \ch$ annihilation.

One immediate physical consequence of these coannihilation
effects is to relax significantly the previous
cosmological upper bound~\cite{up} on the LSP mass, 
from $\ga 200$~GeV to $\ga 600$~GeV. This relaxed upper
limit could have significant implications for strategies
to search for supersymmetric cold dark matter, which have
yet to be explored systematically. Coannihilations also
allow the LSP
mass to approach the boundary of the CMSSM parameter space
within which LHC searches for supersymmetry are
expected to be sensitive~\cite{CMS}. At first glance,
it seems likely that the LHC should still be able to
cover all of the CMSSM parameter space that is consistent
with supersymmetric cold dark matter, but this point
merits further consideration.

Another point to be studied in more detail is the interplay between
coannihilation effects and LEP lower limits on sparticle and higgs
masses. Previously, limits on $m_\ch$ and tan$\beta$ have been derived
from a combination of previous LEP data sets and
cosmology~\cite{efos,efos2,efgos}, but neglecting coannihilation
effects. These limits should now be revisited in the light of more
recent LEP data~\cite{LEP}, as well as coannihilations. We have
commented on these questions in this paper, but a complete study lies
beyond the scope of the present analysis.

We hope that this paper has not only documented the
importance of coannihilation effects, but also provided
the reader with the tools needed to join in the exploration of their
implications for the interesting open issues raised in the previous
paragraph.

\vskip .3in
\vbox{
\noindent{ {\bf Acknowledgments} } \\
\noindent  We are grateful to C. Pallis for pointing out the errors in the Appendix. 
The work of K.O.~was supported in part by DOE grant
DE--FG02--94ER--40823.  The work of T.F.~was supported in part by DOE
grant DE--FG02--95ER--40896, and in part by the University of
Wisconsin Research Committee with funds granted by the Wisconsin
Alumni Research Foundation. The work of M.S.~was supported in part by
NSF grant PHY--97--22022.}

\input APPENDIX

\end{document}

%% file: APPENDIX.tex
\baselineskip=14pt.
\appendix
\setcounter{equation}{0}
\renewcommand{\theequation}{A\arabic{equation}}
\section*{Appendix}
This section contains simplified formulae for the $\stau\stau^*,\stau\chi,
\stau\stau,\stau\tilde\ell$ and $\stau \tilde\ell^*$ annihilation
amplitudes, in the $m_{\stau}\rightarrow 0$ limit.  Expressions for
the $\tilde\ell \tilde\ell^*, \tilde\ell \tilde\ell$ and $\tilde\ell
\chi$ amplitudes can be obtained by taking $\tau\rightarrow\ell$ in
the $\stau\stau^*,\stau\chi, \stau\stau$ formulae below, and $\tilde\mu\tilde e$
and $\tilde e\tilde\mu^*$ by taking $\tau\rightarrow e, \ell\rightarrow\mu$ in the
$\stau\tilde\ell$ and $\stau \tilde\ell^*$ expressions.  The $a$ and $b$ coefficients 
can be simply numerically extracted from the amplitudes, as described in section 2 of the text.

\subsection*{$\stau\stau^*\longrightarrow W^+W^-$}
 I.   s-channel $H$ annihilation\hfill\\
II.  s-channel $h$ annihilation \hfill\\ 
 V.   s-channel $Z$  annihilation \hfill\\
 VI.  s-channel $\gamma$  annihilation \hfill\\
\begin{eqnarray}
\f1 &=& (-g_2 \mw \cos(\beta-\alpha)) (g_2 \mz\sin^{2}\!\theta_{\ss W} 
\cos(\alpha+\beta)/\cos\theta_{\ss W} )   \nonumber \\
      \f2 &=& (-g_2 \mw \sin(\beta-\alpha)) (-g_2 \mz\sin^{2}\!
\theta_{\ss W} \sin(\alpha+\beta)/\cos\theta_{\ss W} )    \nonumber \\
      \f5 &=& (-g_2\sin^{2}\!\theta_{\ss W}/\cos\theta_{\ss W}) (g_2 \cos\theta_{\ss W})    \nonumber \\
      \f6 &=& e^2    \nonumber \\
      {\cal T}_{\rm I}\!\!\times\!\!{\cal T}_{\rm I} &=& (12\mw^4 - 4\mw^2 s + s^2)/
      (4\mw^4 (\mhb^2 - s)^2)    \nonumber \\
      {\cal T}_{\rm II}\!\!\times\!\!{\cal T}_{\rm II} &=& (12 \mw^4 - 4 \mw^2 s + s^2)/
      (4\mw^4 (\mhl^2 - s)^2)    \nonumber \\
      {\cal T}_{\rm V}\!\!\times\!\!{\cal T}_{\rm V} &=&         (128 \mst^2 \mw^4 \mz^4 s - 
     32 \mst^2 \mw^2 \mz^4 s^2 - 32 \mw^4 \mz^4 s^2 + 8 \mw^2 \mz^4 s^3 + 
     12 \mw^4 \mz^4 t^2 - \nl 4 \mw^2 \mz^4 s t^2 + \mz^4 s^2 t^2 - 
     24 \mw^4 \mz^4 t u + 
     8 \mw^2 \mz^4 s t u - 2 \mz^4 s^2 t u + 
     12 \mw^4 \mz^4 u^2 - \nl 4 \mw^2 \mz^4 s u^2 + 
     \mz^4 s^2 u^2)/(4\mw^4 \mz^4 (\mz^2 - s)^2)    \nonumber \\
      {\cal T}_{\rm VI}\!\!\times\!\!{\cal T}_{\rm VI} &=&  (128 \mst^2 \mw^4 s - 32 \mst^2 \mw^2 s^2 - 
     32 \mw^4 s^2 + 8 \mw^2 s^3 + 12 \mw^4 t^2 - 
     4 \mw^2 s t^2 + \nl s^2 t^2 - 24 \mw^4 t u + 
     8 \mw^2 s t u - 2 s^2 t u + 12 \mw^4 u^2 - 
     4 \mw^2 s u^2 + s^2 u^2)/(4\mw^4 s^2)    \nonumber \\
      {\cal T}_{\rm I}\!\!\times\!\!{\cal T}_{\rm II} &=&         (12 \mw^4 - 4 \mw^2 s + s^2)/
   (4\mw^4 (\mhb^2 - s) (\mhl^2 - s))    \nonumber \\
      {\cal T}_{\rm I}\!\!\times\!\!{\cal T}_{\rm V} &=&         (-12 \mw^4 \mz^2 t + \mz^2 s^2 t + 
     12 \mw^4 \mz^2 u - \mz^2 s^2 u)/
   (4\mw^4 \mz^2 (\mhb^2 - s) (\mz^2 - s))    \nonumber \\
      {\cal T}_{\rm II}\!\!\times\!\!{\cal T}_{\rm V} &=&         (-12 \mw^4 \mz^2 t + \mz^2 s^2 t + 
     12 \mw^4 \mz^2 u - \mz^2 s^2 u)/ 
   (4\mw^4 \mz^2 (\mhl^2 - s) (\mz^2 - s))    \nonumber \\
      {\cal T}_{\rm I}\!\!\times\!\!{\cal T}_{\rm VI} &=&         -(-12 \mw^4 t + s^2 t + 12 \mw^4 u - s^2 u)/
   (4\mw^4 (\mhb^2 - s) s)    \nonumber \\
      {\cal T}_{\rm II}\!\!\times\!\!{\cal T}_{\rm VI} &=&         -(-12 \mw^4 t + s^2 t + 12 \mw^4 u - s^2 u)/
   (4\mw^4 (\mhl^2 - s) s)    \nonumber \\
      {\cal T}_{\rm V}\!\!\times\!\!{\cal T}_{\rm VI} &=&         (-128 \mst^2 \mw^4 \mz^2 s + 
     32 \mst^2 \mw^2 \mz^2 s^2 +
     32 \mw^4 \mz^2 s^2 - 8 \mw^2 \mz^2 s^3 -  \nl
     12 \mw^4 \mz^2 t^2 + 4 \mw^2 \mz^2 s t^2 -
     \mz^2 s^2 t^2 + 24 \mw^4 \mz^2 t u - 
     8 \mw^2 \mz^2 s t u +  \nl 2 \mz^2 s^2 t u -
     12 \mw^4 \mz^2 u^2 + 4 \mw^2 \mz^2 s u^2 - 
     \mz^2 s^2 u^2)/(4\mw^4 \mz^2 (\mz^2 - s) s)    \nonumber \\
       \tsq &=&   \f1^2 {\cal T}_{\rm I}\!\!\times\!\!{\cal T}_{\rm I} +  \f2^2 {\cal T}_{\rm II}\!\!\times\!\!{\cal T}_{\rm II} +  \f5^2 {\cal T}_{\rm V}\!\!\times\!\!{\cal T}_{\rm V} + \f6^2 {\cal T}_{\rm VI}\!\!\times\!\!{\cal T}_{\rm VI} +2 \f1 \f2 {\cal T}_{\rm I}\!\!\times\!\!{\cal T}_{\rm II} +
      2 \f1 \f5  {\cal T}_{\rm I}\!\!\times\!\!{\cal T}_{\rm V} + \nl  2 \f1 \f6 {\cal T}_{\rm I}\!\!\times\!\!{\cal T}_{\rm VI} +2 \f2 \f5 {\cal T}_{\rm II}\!\!\times\!\!{\cal T}_{\rm V} +  
      2 \f2 \f6 {\cal T}_{\rm II}\!\!\times\!\!{\cal T}_{\rm VI} +2 \f5 \f6 {\cal T}_{\rm V}\!\!\times\!\!{\cal T}_{\rm VI}
\end{eqnarray}
\subsection*{$\stau\stau^*\longrightarrow ZZ$}
 I.   s-channel $H$ annihilation\hfill\\
 II.  s-channel $h$ annihilation \hfill\\ 
 III. t-channel $\stau$ exchange \hfill\\
 IV. u-channel $\stau$ exchange \hfill\\
 V.  point interaction\hfill\\
\begin{eqnarray}
      \f1 &=& (-g_2 \mz\cos(\beta-\alpha)/\cos\theta_{\ss W} ) (g_2 \mz
      \sin^{2}\!\theta_{\ss W} \cos(\alpha+\beta)/\cos\theta_{\ss W} )    \nonumber \\
      \f2 &=& (-g_2 \mz\sin(\beta-\alpha)/\cos\theta_{\ss W} ) (-g_2 \mz
      \sin^{2}\!\theta_{\ss W} \sin(\alpha+\beta)/\cos\theta_{\ss W} )    \nonumber \\
      \f3 &=& (-g_2\sin^{2}\!\theta_{\ss W}/\cos\theta_{\ss W}  )^2    \nonumber \\
      \f4 &=& (-g_2\sin^{2}\!\theta_{\ss W}/\cos\theta_{\ss W} )^2    \nonumber \\
      \f5 &=& -2 g_2^2\sin^{4}\!\theta_{\ss W} /\cos^{2}\!\theta_{\ss W}    \nonumber \\
     {\cal T}_{\rm I}\!\!\times\!\!{\cal T}_{\rm I} &=& (12 \mz^4 - 4 \mz^2 s + s^2)/(4\mz^4 (\mhb^2 - s)^2)    \nonumber \\
      {\cal T}_{\rm II}\!\!\times\!\!{\cal T}_{\rm II} &=& (12 \mz^4 - 4 \mz^2 s + s^2)/(4\mz^4 (\mhl^2 - s)^2)    \nonumber \\
      {\cal T}_{\rm III}\!\!\times\!\!{\cal T}_{\rm III} &=&         (\mst^8 - 4 \mst^6 \mz^2 + 6 \mst^4 \mz^4 - 
     4 \mst^2 \mz^6 + \mz^8 - 4 \mst^6 t + 
     4 \mst^4 \mz^2 t +  \nl 4 \mst^2 \mz^4 t - 4 \mz^6 t + 
     6 \mst^4 t^2 + 4 \mst^2 \mz^2 t^2 +
     6 \mz^4 t^2 - 4 \mst^2 t^3 - 4 \mz^2 t^3 + t^4
     )/ \nl (\mz^4 (\mst^2 - t)^2)    \nonumber \\
      {\cal T}_{\rm IV}\!\!\times\!\!{\cal T}_{\rm IV} &=&         (\mst^8 - 4 \mst^6 \mz^2 + 6 \mst^4 \mz^4 - 
     4 \mst^2 \mz^6 + \mz^8 - 4 \mst^6 u +
     4 \mst^4 \mz^2 u +  \nl 4 \mst^2 \mz^4 u - 4 \mz^6 u + 
     6 \mst^4 u^2 + 4 \mst^2 \mz^2 u^2 + 
     6 \mz^4 u^2 - 4 \mst^2 u^3 -  4 \mz^2 u^3 + u^4
     )/\nl (\mz^4 (\mst^2 - u)^2)    \nonumber \\
      {\cal T}_{\rm V}\!\!\times\!\!{\cal T}_{\rm V} &=& (12 \mz^4 - 4 \mz^2 s + s^2)/(4\mz^4)    \nonumber \\
      {\cal T}_{\rm I}\!\!\times\!\!{\cal T}_{\rm II} &=&         (12 \mz^4 - 4 \mz^2 s + s^2)/
   (4\mz^4 (\mhb^2 - s) (\mhl^2 - s))    \nonumber \\
      {\cal T}_{\rm I}\!\!\times\!\!{\cal T}_{\rm III} &=&         (-6 \mst^4 \mz^2 - 20 \mst^2 \mz^4 - 6 \mz^6 + 
     \mst^4 s + 2 \mst^2 \mz^2 s + 5 \mz^4 s +
     8 \mst^2 \mz^2 t + \nl 8 \mz^4 t -  2 \mst^2 s t - 
     2 \mz^2 s t - 2 \mz^2 t^2 + s t^2 + 
     4 \mst^2 \mz^2 u + 4 \mz^4 u - 4 \mz^2 t u)/\nl 
   (2\mz^4 (\mhb^2 - s) (\mst^2 - t))    \nonumber \\
      {\cal T}_{\rm I}\!\!\times\!\!{\cal T}_{\rm IV} &=&         (-6 \mst^4 \mz^2 - 20 \mst^2 \mz^4 - 6 \mz^6 + 
     \mst^4 s + 2 \mst^2 \mz^2 s + 5 \mz^4 s + 
     4 \mst^2 \mz^2 t + \nl 4 \mz^4 t + 8 \mst^2 \mz^2 u + 
     8 \mz^4 u - 2 \mst^2 s u - 2 \mz^2 s u -
     4 \mz^2 t u - 2 \mz^2 u^2 + s u^2)/ \nl
   (2\mz^4 (\mhb^2 - s) (\mst^2 - u))    \nonumber \\
      {\cal T}_{\rm I}\!\!\times\!\!{\cal T}_{\rm V} &=& (12 \mz^4 - 4 \mz^2 s + s^2)/(4\mz^4 (\mhb^2 - s))    \nonumber \\
      {\cal T}_{\rm II}\!\!\times\!\!{\cal T}_{\rm III} &=&         (-6 \mst^4 \mz^2 - 20 \mst^2 \mz^4 - 6 \mz^6 + 
     \mst^4 s + 2 \mst^2 \mz^2 s + 5 \mz^4 s + 
     8 \mst^2 \mz^2 t + \nl 8 \mz^4 t - 2 \mst^2 s t - 
     2 \mz^2 s t - 2 \mz^2 t^2 + s t^2 + 
     4 \mst^2 \mz^2 u + 4 \mz^4 u - 4 \mz^2 t u)/\nl
   (2\mz^4 (\mhl^2 - s) (\mst^2 - t))    \nonumber \\
      {\cal T}_{\rm II}\!\!\times\!\!{\cal T}_{\rm IV} &=&         (-6 \mst^4 \mz^2 - 20 \mst^2 \mz^4 - 6 \mz^6 + 
     \mst^4 s + 2 \mst^2 \mz^2 s + 5 \mz^4 s + 
     4 \mst^2 \mz^2 t + \nl 4 \mz^4 t + 8 \mst^2 \mz^2 u + 
     8 \mz^4 u - 2 \mst^2 s u - 2 \mz^2 s u -
     4 \mz^2 t u - 2 \mz^2 u^2 + s u^2)/ \nl
   (2\mz^4 (\mhl^2 - s) (\mst^2 - u))    \nonumber \\
      {\cal T}_{\rm II}\!\!\times\!\!{\cal T}_{\rm V} &=& (12 \mz^4 - 4 \mz^2 s + s^2)/(4\mz^4 (\mhl^2 - s))    \nonumber \\
      {\cal T}_{\rm III}\!\!\times\!\!{\cal T}_{\rm IV} &=&         (\mst^8 + 12 \mst^6 \mz^2 + 38 \mst^4 \mz^4 + 
     12 \mst^2 \mz^6 + \mz^8 - 4 \mst^4 \mz^2 s -
     24 \mst^2 \mz^4 s - \nl 4 \mz^6 s + 4 \mz^4 s^2 - 
     2 \mst^6 t - 14 \mst^4 \mz^2 t - 
     14 \mst^2 \mz^4 t - 2 \mz^6 t + 
     4 \mst^2 \mz^2 s t + \nl 4 \mz^4 s t + \mst^4 t^2 + 
     2 \mst^2 \mz^2 t^2 + \mz^4 t^2 - 2 \mst^6 u - 
     14 \mst^4 \mz^2 u - 14 \mst^2 \mz^4 u - \nl
     2 \mz^6 u + 4 \mst^2 \mz^2 s u + 4 \mz^4 s u + 
     4 \mst^4 t u + 16 \mst^2 \mz^2 t u +
     4 \mz^4 t u - 4 \mz^2 s t u - \nl 2 \mst^2 t^2 u - 
     2 \mz^2 t^2 u + \mst^4 u^2 +
     2 \mst^2 \mz^2 u^2 + \mz^4 u^2 - 
     2 \mst^2 t u^2 - 2 \mz^2 t u^2 + \nl t^2 u^2)/ 
   (\mz^4 (\mst^2 - t) (\mst^2 - u))    \nonumber \\
      {\cal T}_{\rm III}\!\!\times\!\!{\cal T}_{\rm V} &=&         (-6 \mst^4 \mz^2 - 20 \mst^2 \mz^4 - 6 \mz^6 + 
     \mst^4 s + 2 \mst^2 \mz^2 s + 5 \mz^4 s + 
     8 \mst^2 \mz^2 t + \nl 8 \mz^4 t - 2 \mst^2 s t - 
     2 \mz^2 s t - 2 \mz^2 t^2 + s t^2 +
     4 \mst^2 \mz^2 u +  4 \mz^4 u - 4 \mz^2 t u)/\nl 
   (2\mz^4 (\mst^2 - t))    \nonumber \\
      {\cal T}_{\rm IV}\!\!\times\!\!{\cal T}_{\rm V} &=&         (-6 \mst^4 \mz^2 - 20 \mst^2 \mz^4 - 6 \mz^6 + 
     \mst^4 s + 2 \mst^2 \mz^2 s + 5 \mz^4 s + 
     4 \mst^2 \mz^2 t + \nl 4 \mz^4 t + 8 \mst^2 \mz^2 u + 
     8 \mz^4 u - 2 \mst^2 s u - 2 \mz^2 s u -
     4 \mz^2 t u - 2 \mz^2 u^2 + s u^2)/ \nl
   (2\mz^4 (\mst^2 - u))    \nonumber \\
      \tsq &=&   \f1^2 {\cal T}_{\rm I}\!\!\times\!\!{\cal T}_{\rm I} +  \f2^2 {\cal T}_{\rm II}\!\!\times\!\!{\cal T}_{\rm II} + \f3^2 {\cal T}_{\rm III}\!\!\times\!\!{\cal T}_{\rm III} + \f4^2 {\cal T}_{\rm IV}\!\!\times\!\!{\cal T}_{\rm IV} +
      \f5^2 {\cal T}_{\rm V}\!\!\times\!\!{\cal T}_{\rm V} + 2 \f1 \f2 {\cal T}_{\rm I}\!\!\times\!\!{\cal T}_{\rm II} + \nl 2 \f1 \f3 {\cal T}_{\rm I}\!\!\times\!\!{\cal T}_{\rm III} + 
      2 \f1 \f4 {\cal T}_{\rm I}\!\!\times\!\!{\cal T}_{\rm IV} + 2 \f1 \f5  {\cal T}_{\rm I}\!\!\times\!\!{\cal T}_{\rm V} +2 \f2 \f3 {\cal T}_{\rm II}\!\!\times\!\!{\cal T}_{\rm III} +
      2 \f2 \f4 {\cal T}_{\rm II}\!\!\times\!\!{\cal T}_{\rm IV} + \nl 2 \f2 \f5 {\cal T}_{\rm II}\!\!\times\!\!{\cal T}_{\rm V} + 2 \f3 \f4 {\cal T}_{\rm III}\!\!\times\!\!{\cal T}_{\rm IV} +
      2 \f3 \f5 {\cal T}_{\rm III}\!\!\times\!\!{\cal T}_{\rm V} +2 \f4 \f5 {\cal T}_{\rm IV}\!\!\times\!\!{\cal T}_{\rm V}  
\end{eqnarray}
\subsection*{$\stau\stau^*\longrightarrow Z\gamma$}
 I. t-channel $\stau$ exchange \hfill\\
 II. u-channel $\stau$ exchange \hfill\\
 III.  point interaction\hfill\\
\begin{eqnarray}
      \f1 &=& e (-g_2\sin^{2}\!\theta_{\ss W}/\cos\theta_{\ss W})    \nonumber \\
      \f2 &=& e (-g_2\sin^{2}\!\theta_{\ss W}/\cos\theta_{\ss W})    \nonumber \\
      \f3 &=& 2 e g_2\sin^{2}\!\theta_{\ss W}/\cos\theta_{\ss W}     \nonumber \\
      {\cal T}_{\rm I}\!\!\times\!\!{\cal T}_{\rm I} &=&         (-2 \mst^6 + 4 \mst^4 \mz^2 - 2 \mst^2 \mz^4 + 
     2 \mst^4 t + 8 \mst^2 \mz^2 t - 2 \mz^4 t + 
     2 \mst^2 t^2 + \nl 4 \mz^2 t^2 - 2 t^3)/
   (\mz^2 (\mst^2 - t)^2)    \nonumber \\
      {\cal T}_{\rm II}\!\!\times\!\!{\cal T}_{\rm II} &=&         (-2 \mst^6 + 4 \mst^4 \mz^2 - 2 \mst^2 \mz^4 + 
     2 \mst^4 u - 8 \mst^2 \mz^2 u - 2 \mz^4 u + 
     2 \mst^2 u^2 + \nl 4 \mz^2 u^2 - 2 u^3)/
   (\mz^2 (\mst^2 - u)^2)    \nonumber \\
      {\cal T}_{\rm III}\!\!\times\!\!{\cal T}_{\rm III} &=& 3   \nonumber \\
      {\cal T}_{\rm I}\!\!\times\!\!{\cal T}_{\rm II} &=&         (6 \mst^6 + 36 \mst^4 \mz^2 + 6 \mst^2 \mz^4 - 
     2 \mst^4 s - 24 \mst^2 \mz^2 s - 2 \mz^4 s +
     4 \mz^2 s^2 -  \nl 7 \mst^4 t - 12 \mst^2 \mz^2 t - 
     \mz^4 t + 2 \mst^2 s t + 4 \mz^2 s t +
     \mst^2 t^2 + \mz^2 t^2 - 7 \mst^4 u -  \nl
     12 \mst^2 \mz^2 u - \mz^4 u + 2 \mst^2 s u +
     4 \mz^2 s u + 8 \mst^2 t u + 2 \mz^2 t u - 
     2 s t u - t^2 u  \nl + \mst^2 u^2 + \mz^2 u^2 -
     t u^2)/(\mz^2 (\mst^2 - t) (\mst^2 - u))    \nonumber \\
      {\cal T}_{\rm I}\!\!\times\!\!{\cal T}_{\rm III} &=&    (-2 \mst^4 - 15 \mst^2 \mz^2 - 3 \mz^4 + \mst^2 s + 
     5 \mz^2 s + 2 \mst^2 t + 3 \mz^2 t - s t + \nl
     2 \mst^2 u + 4 \mz^2 u - 2 t u)/
   (2\mz^2 (\mst^2 - t))    \nonumber \\
      {\cal T}_{\rm II}\!\!\times\!\!{\cal T}_{\rm III} &=&   (-2 \mst^4 - 15 \mst^2 \mz^2 - 3 \mz^4 + \mst^2 s + 
     5 \mz^2 s + 2 \mst^2 t + 4 \mz^2 t + 2 \mst^2 u + \nl
     3 \mz^2 u - s u - 2 t u)/(2\mz^2 (\mst^2 - u))    \nonumber \\
      \tsq &=&   \f1^2 {\cal T}_{\rm I}\!\!\times\!\!{\cal T}_{\rm I} +  \f2^2 {\cal T}_{\rm II}\!\!\times\!\!{\cal T}_{\rm II} + \f3^2 {\cal T}_{\rm III}\!\!\times\!\!{\cal T}_{\rm III} + 
      2 \f1 \f2 {\cal T}_{\rm I}\!\!\times\!\!{\cal T}_{\rm II} + 2 \f1 \f3 {\cal T}_{\rm I}\!\!\times\!\!{\cal T}_{\rm III} +  \nl 2 \f2 \f3 {\cal T}_{\rm II}\!\!\times\!\!{\cal T}_{\rm III}    
\end{eqnarray}
\subsection*{$\stau\stau^*\longrightarrow \gamma\gamma$}
 I. t-channel $\stau$ exchange \hfill\\
 II. u-channel $\stau$ exchange \hfill\\
 III.  point interaction\hfill\\
\begin{eqnarray}
      \f1 &=& e^2    \nonumber \\
      \f2 &=& e^2    \nonumber \\
      \f3 &=& -2 e^2    \nonumber \\
      {\cal T}_{\rm I}\!\!\times\!\!{\cal T}_{\rm I} &=& (4 \mst^4 + 8 \mst^2 t + 4 t^2)/(\mst^2 - t)^2    \nonumber \\
      {\cal T}_{\rm II}\!\!\times\!\!{\cal T}_{\rm II} &=& (4 \mst^4 + 8 \mst^2 u + 4 u^2)/(\mst^2 - u)^2    \nonumber \\
      {\cal T}_{\rm III}\!\!\times\!\!{\cal T}_{\rm III} &=& 4    \nonumber \\
      {\cal T}_{\rm I}\!\!\times\!\!{\cal T}_{\rm II} &=&   (36 \mst^4 - 24 \mst^2 s + 4 s^2 - 12 \mst^2 t + 
     4 s t + t^2 - 12 \mst^2 u + 4 s u + 2 t u + u^2) /\nl
      ((\mst^2 - t) (\mst^2 - u))    \nonumber \\
      {\cal T}_{\rm I}\!\!\times\!\!{\cal T}_{\rm III} &=& (-12 \mst^2 + 5 s + 4 u)/(2(\mst^2 - t))    \nonumber \\
      {\cal T}_{\rm II}\!\!\times\!\!{\cal T}_{\rm III} &=& (-12 \mst^2 + 5 s + 4 t)/(2(\mst^2 - u))    \nonumber \\
      \tsq &=&   \f1^2 {\cal T}_{\rm I}\!\!\times\!\!{\cal T}_{\rm I} +  \f2^2 {\cal T}_{\rm II}\!\!\times\!\!{\cal T}_{\rm II} + \f3^2 {\cal T}_{\rm III}\!\!\times\!\!{\cal T}_{\rm III} + 
      2 \f1 \f2 {\cal T}_{\rm I}\!\!\times\!\!{\cal T}_{\rm II} + 2 \f1 \f3 {\cal T}_{\rm I}\!\!\times\!\!{\cal T}_{\rm III} +  \nl 2 \f2 \f3 {\cal T}_{\rm II}\!\!\times\!\!{\cal T}_{\rm III}    
\end{eqnarray}
\subsection*{$\stau\stau^*\longrightarrow Z h [H]$}
 I.   t-channel $\stau$ exchange         \hfill\\       
 II.  u-channel $\stau$ exchange      \hfill\\         
 III.  s-channel $Z$ annihilation          \hfill\\     
\begin{eqnarray}
      \f1 &=&  (-g_2\sin^{2}\!\theta_{\ss W}/\cos\theta_{\ss W} ) 
      (-g_2 \mz\sin^{2}\!\theta_{\ss W} \sin[-\cos](\alpha+\beta)/\cos\theta_{\ss W} )    \nonumber \\
      \f2 &=&  -(-g_2\sin^{2}\!\theta_{\ss W}/\cos\theta_{\ss W} ) 
      (-g_2 \mz\sin^{2}\!\theta_{\ss W} \sin[-\cos](\alpha+\beta)/\cos\theta_{\ss W} )    \nonumber \\
      \f3&=& (-g_2\sin^{2}\!\theta_{\ss W}/\cos\theta_{\ss W} ) 
      (-g_2 \mz\sin[\cos](\beta-\alpha)/\cos\theta_{\ss W} )    \nonumber \\
      {\cal T}_{\rm I}\!\!\times\!\!{\cal T}_{\rm I} &=&         (\mst^4 + (\mz^2 - t)^2 - 2 \mst^2 (\mz^2 + t))/
   (\mz^2 (\mst^2 - t)^2)    \nonumber \\
      {\cal T}_{\rm II}\!\!\times\!\!{\cal T}_{\rm II} &=&         (\mst^4 + (\mz^2 - u)^2 - 2 \mst^2 (\mz^2 + u))/
   (\mz^2 (\mst^2 - u)^2)    \nonumber \\
      {\cal T}_{\rm I}\!\!\times\!\!{\cal T}_{\rm II} &=&         (\mst^4 + \mz^4 + \mst^2 (6 \mz^2 - t - u) + t u - 
     \mz^2 (2 s + t + u))/ \nl
   (\mz^2 (\mst^2 - t) (\mst^2 - u))    \nonumber \\
      {\cal T}_{\rm I}\!\!\times\!\!{\cal T}_{\rm III} &=&         (t (t - u) + \mst^2 (-8 \mz^2 - t + u) + 
     \mz^2 (2 s - t + u))/ \nl
   (2\mz^2 (\mz^2 - s) (\mst^2 - t))    \nonumber \\
      {\cal T}_{\rm II}\!\!\times\!\!{\cal T}_{\rm III} &=&         ((t - u) u + \mst^2 (8 \mz^2 - t + u) + 
     \mz^2 (-2 s - t + u))/ \nl
   (2\mz^2 (\mz^2 - s) (\mst^2 - u))    \nonumber \\
      {\cal T}_{\rm III}\!\!\times\!\!{\cal T}_{\rm III} &=&         (-16 \mst^2 \mz^2 + 4 \mz^2 s + (t - u)^2)/
   (4\mz^2(\mz^2 -  s)^2)    \nonumber \\
      \tsq &=&   \f1^2 {\cal T}_{\rm I}\!\!\times\!\!{\cal T}_{\rm I} +  \f2^2 {\cal T}_{\rm II}\!\!\times\!\!{\cal T}_{\rm II} +\f3^2 {\cal T}_{\rm III}\!\!\times\!\!{\cal T}_{\rm III}+2 \f1 \f2 {\cal T}_{\rm I}\!\!\times\!\!{\cal T}_{\rm II} +
      2 \f1 \f3 {\cal T}_{\rm I}\!\!\times\!\!{\cal T}_{\rm III} + \nl 2 \f2 \f3 {\cal T}_{\rm II}\!\!\times\!\!{\cal T}_{\rm III}   
\end{eqnarray}

\subsection*{$\stau\stau^*\longrightarrow \gamma h [H]$}
 I.   t-channel $\stau$ exchange         \hfill\\       
 II.  u-channel $\stau$ exchange      \hfill\\         
\begin{eqnarray}
    \f1 &=&  (e) (-g_2 \mz\sin^{2}\!\theta_{\ss W} \sin[-\cos](\alpha+\beta)/\cos\theta_{\ss W} )  
\nonumber \\
    \f2 &=&  -(e) (-g_2 \mz\sin^{2}\!\theta_{\ss W} \sin[-\cos](\alpha+\beta)/\cos\theta_{\ss W} 
\nonumber )\\
      {\cal T}_{\rm I}\!\!\times\!\!{\cal T}_{\rm I} &=& -2 (\mst^2 + t)/(\mst^2 - t)^2    \nonumber \\
      {\cal T}_{\rm I}\!\!\times\!\!{\cal T}_{\rm II} &=& -(-6 \mst^2 + 2 s + t + u)/((\mst^2 - t) (\mst^2 - u))    \nonumber \\
      {\cal T}_{\rm II}\!\!\times\!\!{\cal T}_{\rm II} &=& -2 (\mst^2 + u)/(\mst^2 - u)^2    \nonumber \\
      \tsq &=&   \f1^2 {\cal T}_{\rm I}\!\!\times\!\!{\cal T}_{\rm I} +  \f2^2 {\cal T}_{\rm II}\!\!\times\!\!{\cal T}_{\rm II} +2 \f1 \f2 {\cal T}_{\rm I}\!\!\times\!\!{\cal T}_{\rm II} 
\end{eqnarray}

\subsection*{$\stau\stau^*\longrightarrow Z A$}
I. s-channel $h$ exchange         \hfill\\       
II. s-channel $H$ exchange         \hfill\\       
\begin{eqnarray}
    \f1 &=& (g_2 \cos(\alpha-\beta)/(2\cos \theta_{\ss W} )(-g_2 \mz
      \sin^{2}\!\theta_{\ss W} \sin(\alpha+\beta)/\cos\theta_{\ss W} )\nonumber\\
    \f2 &=&  (g_2 \sin(\alpha-\beta)/(2\cos \theta_{\ss W} )(g_2 \mz
      \sin^{2}\!\theta_{\ss W} \cos(\alpha+\beta)/\cos\theta_{\ss W} ) \nonumber \\
      {\cal T}_{\rm I}\!\!\times\!\!{\cal T}_{\rm I} &=&         (m_A^4 + (\mz^2 - s)^2 - 2 m_A^2 (\mz^2 + s))/
   (\mz^2 (\mhl^2 - s)^2)    \nonumber \\
      {\cal T}_{\rm II}\!\!\times\!\!{\cal T}_{\rm II} &=&         (m_A^4 + (\mz^2 - s)^2 - 2 m_A^2 (\mz^2 + s))/
   (\mz^2 (\mhb^2 - s)^2)    \nonumber \\
      {\cal T}_{\rm I}\!\!\times\!\!{\cal T}_{\rm II} &=&         (m_A^4 + (\mz^2 - s)^2 - 2 m_A^2 (\mz^2 + s)/
     (\mz^2 (\mhb^2 - s) (\mhl^2 - s)))    \nonumber \\
      \tsq &=&   \f1^2 {\cal T}_{\rm I}\!\!\times\!\!{\cal T}_{\rm I} +  \f2^2 {\cal T}_{\rm II}\!\!\times\!\!{\cal T}_{\rm II} +2 \f1 \f2 {\cal T}_{\rm I}\!\!\times\!\!{\cal T}_{\rm II} 
\end{eqnarray}

\subsection*{$\stau\stau^*\longrightarrow \tau\bar\tau$}
 III.   s-channel $Z$  annihilation \hfill\\
 IV.  s-channel $\gamma$  annihilation \hfill\\
 V. t-channel $\chi$ exchange \hfill\\
\begin{eqnarray}
     f_{3c} &=&   (-g_2\sin^{2}\!\theta_{\ss W}/\cos\theta_{\ss W} ) 
      (g_2(1 - 4 \sin^{2}\!\theta_{\ss W})/(4\cos\!\theta_{\ss W}))    \nonumber \\
      f_{3d} &=& (-g_2 \sin^{2}\!\theta_{\ss W}/\cos\theta_{\ss W} ) 
      (-g_2/(4\cos\!\theta_{\ss W}))   \nonumber \\
      f_{4c} &=& e^2    \nonumber \\
       K&=& g_1 N_{i1} /\sqrt{2}    \nonumber \\
       K'&=& g_1 N_{j1} /\sqrt{2}    \nonumber \\
      {\cal T}_{\rm III}\!\!\times\!\!{\cal T}_{\rm III} &=&      2 (- 4 f_{3c}^2 \mst^2 s - 
       4 f_{3d}^2 \mst^2 s 
       f_{3c}^2 s^2 + f_{3d}^2 s^2 - f_{3c}^2 t^2 - 
       f_{3d}^2 t^2 + 2 f_{3c}^2 t u + 2 f_{3d}^2 t u - \nl
       f_{3c}^2 u^2 - f_{3d}^2 u^2)/(\mz^2 - s)^2    \nonumber \\
      {\cal T}_{\rm III}\!\!\times\!\!{\cal T}_{\rm IV} &=&         2 f_{3c} f_{4c}  (4 \mst^2 s -  s^2 + t^2 - 2 t u + u^2 )/
  ((\mz^2 - s) s) \nonumber \\
      {\cal T}_{\rm IV}\!\!\times\!\!{\cal T}_{\rm IV} &=&         2  f_{4c}^2(- 4 \mst^2 s + s^2  -  t^2 + 2 t u - u^2 )/s^2    \nonumber \\
      {\cal T}_{\rm III}\!\!\times\!\!{\cal T}_{\rm V}&=& -2 (f_{3c} + f_{3d}) K^2 (4 \mst^2 s - s^2 + (t - u)^2)/
   ((\mz^2 - s) ({\mchi}_i^2 - t))    \nonumber \\
      {\cal T}_{\rm IV}\!\!\times\!\!{\cal T}_{\rm V}&=& 2 (f_{4c}) K^2 (4 \mst^2 s - s^2 + (t - u)^2)/
   (s ({\mchi}_i^2 - t))    \nonumber \\
      {\cal T}_{\rm V}\!\!\times\!\!{\cal T}_{\rm V}&=& (16 K^2 K'^2 (\mst^4 - t u ))/(({\mchi}_i^2 - t) (-{\mchi}_j^2 + t))    \nonumber \\
      \tsq&=&  {\cal T}_{\rm III}\!\!\times\!\!{\cal T}_{\rm III} + {\cal T}_{\rm IV}\!\!\times\!\!{\cal T}_{\rm IV} + 2 {\cal T}_{\rm III}\!\!\times\!\!{\cal T}_{\rm IV} + \nl  \sum_{i,j=1}^4\left(({\cal T}_{\rm I}\!\!\times\!\!{\cal T}_{\rm V}+{\cal T}_{\rm II}\!\!\times\!\!{\cal T}_{\rm V}+{\cal T}_{\rm III}\!\!\times\!\!{\cal T}_{\rm V}+{\cal T}_{\rm IV}\!\!\times\!\!{\cal T}_{\rm V})/2 + 
      {\cal T}_{\rm V}\!\!\times\!\!{\cal T}_{\rm V} \right)
\end{eqnarray}
\subsection*{$\stau\stau^*\longrightarrow f\!\bar f$}
 III.   s-channel $Z$  annihilation      \hfill\\     
 IV.   s-channel $\gamma$  annihilation     \hfill\\  
\begin{eqnarray}
         f_{3c} &=&   (-g_2\sin^{2}\!\theta_{\ss W}/\cos\theta_{\ss W}) 
         (g_2(-2T_3^f + 4 Q_f \sin^{2}\!\theta_{\ss W})/(4\cos\!\theta_{\ss W}))    \nonumber \\
         f_{3d} &=& (-g_2\sin^{2}\!\theta_{\ss W}/\cos\theta_{\ss W}) 
         (g_2(2T_3^f)/(4 \cos\!\theta_{\ss W}))    \nonumber \\
         f_{4c} &=& -e_f e^2    \nonumber \\
         {\cal T}_{\rm III}\!\!\times\!\!{\cal T}_{\rm III} &=&      2 (16 f_{3d}^2 \mst^2 m_f^2 - 4 f_{3c}^2 \mst^2 s - 
         4 f_{3d}^2 \mst^2 s - 4 f_{3d}^2 m_f^2 s +
         f_{3c}^2 s^2 + f_{3d}^2 s^2 - \nl f_{3c}^2 t^2 -  
         f_{3d}^2 t^2 + 2 f_{3c}^2 t u + 2 f_{3d}^2 t u -
         f_{3c}^2 u^2 - f_{3d}^2 u^2)/(\mz^2 - s)^2    \nonumber \\
         {\cal T}_{\rm III}\!\!\times\!\!{\cal T}_{\rm IV} &=&         2 f_{3c} f_{4c}(
         4  \mst^2 s - s^2 + t^2 - 2 t u +  u^2 )/((\mz^2 - s) s)    \nonumber \\
         {\cal T}_{\rm IV}\!\!\times\!\!{\cal T}_{\rm IV} &=&         2 f_{4c}^2(- 4 \mst^2 s + s^2 -  t^2 + 2 t u -u^2 )/s^2    \nonumber \\
       \tsq &=&   ({\cal T}_{\rm III}\!\!\times\!\!{\cal T}_{\rm III} + {\cal T}_{\rm IV}\!\!\times\!\!{\cal T}_{\rm IV} + 2 {\cal T}_{\rm III}\!\!\times\!\!{\cal T}_{\rm IV}) (\times3 \;{\rm for \;quarks})
\end{eqnarray}
\subsection*{$\stau\stau^*\longrightarrow t\bar t$}
 I  .   s-channel $H$ annihilation       \hfill\\    
 II.   s-channel $h$ annihilation    \hfill\\       
 III.   s-channel $Z$  annihilation       \hfill\\    
 IV.   s-channel $\gamma$  annihilation    \hfill\\   
\begin{eqnarray}
      f_{1a} &=& (g_2 \mz\sin^{2}\!\theta_{\ss W} 
      \cos(\alpha+\beta)/\cos\theta_{\ss W} )  (-g_2 m_t\sin\alpha/ (2\mw\sin\beta ))  \nonumber \\
      f_{2a} &=& (-g_2 \mz\sin^{2}\!\theta_{\ss W} 
      \sin(\alpha+\beta)/\cos\theta_{\ss W} )(-g_2 m_t\cos\alpha/ (2\mw\sin\beta ))   \nonumber \\
      f_{3c} &=&   (-g_2\sin^{2}\!\theta_{\ss W}/\cos\theta_{\ss W} ) 
      (g_2(-1+4 Q_t \sin^{2}\!\theta_{\ss W})/(4\cos\!\theta_{\ss W}))     \nonumber \\
      f_{3d} &=& (-g_2 \sin^{2}\!\theta_{\ss W}/\cos\theta_{\ss W} ) 
      (g_2/(4\cos\!\theta_{\ss W}))   \nonumber \\
      f_{4c} &=& -e_t e^2    \nonumber \\
      {\cal T}_{\rm I}\!\!\times\!\!{\cal T}_{\rm I} &=& 2 f_{1a}^2 (-4 m_t^2 + s)/(-\mhb^2 + s)^2    \nonumber \\
      {\cal T}_{\rm II}\!\!\times\!\!{\cal T}_{\rm II} &=& 2 f_{2a}^2 (-4 m_t^2 + s)/(-\mhl^2 + s)^2    \nonumber \\
      {\cal T}_{\rm III}\!\!\times\!\!{\cal T}_{\rm III} &=&      2 (16 f_{3d}^2 \mst^2 m_t^2 - 4 f_{3c}^2 \mst^2 s - 
      4 f_{3d}^2 \mst^2 s - 4 f_{3d}^2 m_t^2 s + 
      f_{3c}^2 s^2 + f_{3d}^2 s^2 -  \nl f_{3c}^2 t^2 -
      f_{3d}^2 t^2 + 2 f_{3c}^2 t u + 2 f_{3d}^2 t u -
      f_{3c}^2 u^2 - f_{3d}^2 u^2)/(\mz^2 - s)^2    \nonumber \\
      {\cal T}_{\rm I}\!\!\times\!\!{\cal T}_{\rm II} &=&         2  f_{1a} f_{2a} (-4m_t^2 + s)/((-\mhb^2 + s) (-\mhl^2 + s))    \nonumber \\
      {\cal T}_{\rm I}\!\!\times\!\!{\cal T}_{\rm III} &=&         4 f_{1a} f_{3c} m_t (t - u)/((\mhb^2 - s) (-\mz^2 + s))    \nonumber \\
      {\cal T}_{\rm II}\!\!\times\!\!{\cal T}_{\rm III} &=&         4 f_{2a} f_{3c} m_t( t - u)/((\mhl^2 - s) (-\mz^2 + s))    \nonumber \\
      {\cal T}_{\rm I}\!\!\times\!\!{\cal T}_{\rm IV} &=& 4 f_{1a} f_{4c} m_t (t - u)/((\mhb^2 - s) s)    \nonumber \\
      {\cal T}_{\rm II}\!\!\times\!\!{\cal T}_{\rm IV} &=& 4 f_{2a} f_{4c} m_t (t - u)/((\mhl^2 - s) s)    \nonumber \\
      {\cal T}_{\rm III}\!\!\times\!\!{\cal T}_{\rm IV} &=&         2 f_{3c} f_{4c} (4 \mst^2 s -  s^2 + t^2 -
      2 t u +  u^2 )/((\mz^2 - s) s)    \nonumber \\
      {\cal T}_{\rm IV}\!\!\times\!\!{\cal T}_{\rm IV} &=&         2 f_{4c}^2(- 4  \mst^2 s +s^2 -  t^2 + 2 t u - u^2 )/s^2    \nonumber \\
      \tsq &=&  3\;({\cal T}_{\rm I}\!\!\times\!\!{\cal T}_{\rm I} + {\cal T}_{\rm II}\!\!\times\!\!{\cal T}_{\rm II} + {\cal T}_{\rm III}\!\!\times\!\!{\cal T}_{\rm III} + {\cal T}_{\rm IV}\!\!\times\!\!{\cal T}_{\rm IV} + 2 {\cal T}_{\rm I}\!\!\times\!\!{\cal T}_{\rm II} + 2 {\cal T}_{\rm I}\!\!\times\!\!{\cal T}_{\rm III} 
      +2 {\cal T}_{\rm I}\!\!\times\!\!{\cal T}_{\rm IV} + \nl 2 {\cal T}_{\rm II}\!\!\times\!\!{\cal T}_{\rm III} +  2 {\cal T}_{\rm II}\!\!\times\!\!{\cal T}_{\rm IV} + 2 {\cal T}_{\rm III}\!\!\times\!\!{\cal T}_{\rm IV}) 
\end{eqnarray}
\subsection*{$\stau\stau^*\longrightarrow hh$}
 I.   s-channel $h$ annihilation\hfill\\
 II.  s-channel $H$ annihilation \hfill\\ 
 III.  point interaction\hfill\\
 IV. t-channel $\stau$ exchange \hfill\\
 V. u-channel $\stau$ exchange \hfill\\
\begin{eqnarray}
      \f1 &=& (-g_2 \mz \sin^{2}\!\theta_{\ss W} \sin(\alpha+\beta)/\cos\theta_{\ss W} ) 
      (-3 g_2 \mz \cos2\alpha\sin(\alpha+\beta)/(2\cos\theta_{\ss W}))\nonumber \\
      \f2 &=&  (g_2 \mz\sin^{2}\!\theta_{\ss W} \cos(\alpha+\beta)/\cos\theta_{\ss W}) 
      (g_2 \mz(\cos2\alpha \cos(\alpha+\beta) -\nl
      2 \sin(2 \alpha) \sin(\alpha+\beta))/(2\cos\theta_{\ss W}))    \nonumber \\
      \f3 &=& -g_2^2\cos2\alpha \sin^{2}\!\theta_{\ss W}/(2 \cos^{2}\!\theta_{\ss W})     \nonumber \\
      \f4 &=& (-g_2 \mz\sin^{2}\!\theta_{\ss W} \sin(\alpha+\beta)/\cos\theta_{\ss W}  )^2    \nonumber \\
      \f5 &=& (-g_2 \mz\sin^{2}\!\theta_{\ss W} \sin(\alpha+\beta)/\cos\theta_{\ss W}  )^2    \nonumber \\
      {\cal T}_{\rm I}\!\!\times\!\!{\cal T}_{\rm I} &=& (\mhl^2 - s)^{-2}    \nonumber \\
      {\cal T}_{\rm II}\!\!\times\!\!{\cal T}_{\rm II} &=& (\mhb^2 - s)^{-2}    \nonumber \\
      {\cal T}_{\rm III}\!\!\times\!\!{\cal T}_{\rm III} &=& 1    \nonumber \\
      {\cal T}_{\rm IV}\!\!\times\!\!{\cal T}_{\rm IV} &=& (\mst^2 - t)^{-2}    \nonumber \\
      {\cal T}_{\rm V}\!\!\times\!\!{\cal T}_{\rm V} &=& (\mst^2 - u)^{-2}    \nonumber \\
      {\cal T}_{\rm I}\!\!\times\!\!{\cal T}_{\rm II} &=& 1/((\mhb^2 - s) (\mhl^2 - s))    \nonumber \\
      {\cal T}_{\rm I}\!\!\times\!\!{\cal T}_{\rm III} &=& 1/(\mhl^2 - s)    \nonumber \\
      {\cal T}_{\rm I}\!\!\times\!\!{\cal T}_{\rm IV} &=& 1/((\mhl^2 - s) (\mst^2 - t))    \nonumber \\
      {\cal T}_{\rm I}\!\!\times\!\!{\cal T}_{\rm V} &=& 1/((\mhl^2 - s) (\mst^2 - u))    \nonumber \\
      {\cal T}_{\rm II}\!\!\times\!\!{\cal T}_{\rm III} &=& 1/(\mhb^2 - s)    \nonumber \\
      {\cal T}_{\rm II}\!\!\times\!\!{\cal T}_{\rm IV} &=& 1/((\mhb^2 - s) (\mst^2 - t))    \nonumber \\
      {\cal T}_{\rm II}\!\!\times\!\!{\cal T}_{\rm V} &=& 1/((\mhb^2 - s) (\mst^2 - u))    \nonumber \\
      {\cal T}_{\rm III}\!\!\times\!\!{\cal T}_{\rm IV} &=& 1/(\mst^2 - t)    \nonumber \\
      {\cal T}_{\rm III}\!\!\times\!\!{\cal T}_{\rm V} &=& 1/(\mst^2 - u)    \nonumber \\
      {\cal T}_{\rm IV}\!\!\times\!\!{\cal T}_{\rm V} &=& 1/((\mst^2 - t) (\mst^2 - u))    \nonumber \\
      \tsq &=&   \f1^2 {\cal T}_{\rm I}\!\!\times\!\!{\cal T}_{\rm I} +  \f2^2 {\cal T}_{\rm II}\!\!\times\!\!{\cal T}_{\rm II} + \f3^2 {\cal T}_{\rm III}\!\!\times\!\!{\cal T}_{\rm III} + \f4^2 {\cal T}_{\rm IV}\!\!\times\!\!{\cal T}_{\rm IV} +
      \f5^2 {\cal T}_{\rm V}\!\!\times\!\!{\cal T}_{\rm V} + 2 \f1 \f2 {\cal T}_{\rm I}\!\!\times\!\!{\cal T}_{\rm II} + \nl 2 \f1 \f3 {\cal T}_{\rm I}\!\!\times\!\!{\cal T}_{\rm III} + 
      2 \f1 \f4 {\cal T}_{\rm I}\!\!\times\!\!{\cal T}_{\rm IV} + 2 \f1 \f5  {\cal T}_{\rm I}\!\!\times\!\!{\cal T}_{\rm V} +2 \f2 \f3 {\cal T}_{\rm II}\!\!\times\!\!{\cal T}_{\rm III} +
      2 \f2 \f4 {\cal T}_{\rm II}\!\!\times\!\!{\cal T}_{\rm IV} +\nl 2 \f2 \f5 {\cal T}_{\rm II}\!\!\times\!\!{\cal T}_{\rm V} +  2 \f3 \f4 {\cal T}_{\rm III}\!\!\times\!\!{\cal T}_{\rm IV} + 
      2 \f3 \f5 {\cal T}_{\rm III}\!\!\times\!\!{\cal T}_{\rm V} +2 \f4 \f5 {\cal T}_{\rm IV}\!\!\times\!\!{\cal T}_{\rm V}  
\end{eqnarray}

\subsection*{$\stau\stau^*\longrightarrow hA[HA]$}
 I. s-channel $Z$ annihilation \hfill\\
\begin{eqnarray}
\f1 &=&  (-g_2\sin^{2}\!\theta_{\ss W}/\cos\theta_{\ss W}) 
(g_2\cos[\sin](\alpha-\beta)/(2\cos\theta_{\ss W}))\nonumber \\
{\cal T}_{\rm I}\!\!\times\!\!{\cal T}_{\rm I} &=&    (t - u)^2/(\mz^2 - s)^2    \nonumber \\
\tsq &=&   \f1^2 {\cal T}_{\rm I}\!\!\times\!\!{\cal T}_{\rm I} 
\end{eqnarray}
\subsection*{$\stau\stau^*\longrightarrow W^+H^-$}
 I. s-channel $H$ annihilation \hfill\\
 II. s-channel $h$ annihilation \hfill\\
\begin{eqnarray}
      \f1 &=&   (g_2 \mz
      \sin^{2}\!\theta_{\ss W} \cos(\alpha+\beta)/\cos\theta_{\ss W} )
    (-g_2\sin(\alpha-\beta)/2)\nonumber \\
      \f2 &=&   (-g_2 \mz
      \sin^{2}\!\theta_{\ss W} \sin(\alpha+\beta)/\cos\theta_{\ss W} )
    (-g_2\cos(\alpha-\beta)/2)\nonumber \\
      {\cal T}_{\rm I}\!\!\times\!\!{\cal T}_{\rm I} &=&         (m_{H^+}^4 + (\mw^2 - s)^2 - 2 m_{H^+}^2 (\mw^2 + s))/
   (\mw^2 (\mhb^2 - s)^2)    \nonumber \\
      {\cal T}_{\rm I}\!\!\times\!\!{\cal T}_{\rm II} &=&         (m_{H^+}^4 + (\mw^2 - s)^2 - 2 m_{H^+}^2 (\mw^2 + s))/
     (\mw^2 (\mhb^2 - s) (\mhl^2 - s)))    \nonumber \\
      {\cal T}_{\rm II}\!\!\times\!\!{\cal T}_{\rm II} &=&         (m_{H^+}^4 + (\mw^2 - s)^2 - 2 m_{H^+}^2 (\mw^2 + s))/
   (\mw^2 (\mhl^2 - s)^2)    \nonumber \\
\tsq &=&   \f1^2 {\cal T}_{\rm I}\!\!\times\!\!{\cal T}_{\rm I} +\f2^2 {\cal T}_{\rm I}\!\!\times\!\!{\cal T}_{\rm I} +2 \f1 \f2 {\cal T}_{\rm I}\!\!\times\!\!{\cal T}_{\rm II}
\end{eqnarray}
\subsection*{$\stau\stau^*\longrightarrow A A$}
 I. s-channel $H$ annihilation \hfill\\
 II. s-channel $h$ annihilation \hfill\\
 III. point interaction\hfill\\
\begin{eqnarray}
      \f1  &=&  (g_2 \mz
      \sin^{2}\!\theta_{\ss W} \cos(\alpha+\beta)/\cos\theta_{\ss W} )
    (g_2\mz \cos2\beta \cos(\beta+\alpha)/(2\cos\theta_{\ss W} )\nonumber \\
      \f2 &=&   (-g_2 \mz
      \sin^{2}\!\theta_{\ss W} \sin(\alpha+\beta)/\cos\theta_{\ss W} )
    (-g_2\mz \cos2\beta \sin(\beta+\alpha)/(2\cos\theta_{\ss W} )\nonumber \\
      \f3 &=&   -g_2^2 \cos2\beta \sin^2\theta_{\ss W}/ (2\cos^2\theta_{\ss W})\nonumber \\
      {\cal T}_{\rm I}\!\!\times\!\!{\cal T}_{\rm I} &=& (\mhb^2 - s)^{-2}    \nonumber \\
      {\cal T}_{\rm II}\!\!\times\!\!{\cal T}_{\rm II} &=& (\mhl^2 - s)^{-2}    \nonumber \\
      {\cal T}_{\rm III}\!\!\times\!\!{\cal T}_{\rm III} &=& 1    \nonumber \\
      {\cal T}_{\rm I}\!\!\times\!\!{\cal T}_{\rm II} &=& 1/((\mhb^2 - s) (\mhl^2 - s))    \nonumber \\
      {\cal T}_{\rm I}\!\!\times\!\!{\cal T}_{\rm III} &=& 1/(\mhb^2 - s)    \nonumber \\
      {\cal T}_{\rm II}\!\!\times\!\!{\cal T}_{\rm III} &=& 1/(\mhl^2 - s)    \nonumber \\
      \tsq &=&   \f1^2 {\cal T}_{\rm I}\!\!\times\!\!{\cal T}_{\rm I} +  \f2^2 {\cal T}_{\rm II}\!\!\times\!\!{\cal T}_{\rm II} + \f3^2 {\cal T}_{\rm III}\!\!\times\!\!{\cal T}_{\rm III} + 
      2 \f1 \f2 {\cal T}_{\rm I}\!\!\times\!\!{\cal T}_{\rm II} + 2 \f1 \f3 {\cal T}_{\rm I}\!\!\times\!\!{\cal T}_{\rm III} +  \nl 2 \f2 \f3 {\cal T}_{\rm II}\!\!\times\!\!{\cal T}_{\rm III}    
\end{eqnarray}
\subsection*{$\stau\stau^*\longrightarrow hH$}
 I. s-channel $H$ annihilation \hfill\\
 II. s-channel $h$ annihilation \hfill\\
 III. point interaction\hfill\\
 IV. t-channel $\stau$ exchange \hfill\\
 V. u-channel $\stau$ exchange \hfill\\
\begin{eqnarray}
      \f1  &=&  (g_2 \mz
      \sin^{2}\!\theta_{\ss W} \cos(\alpha+\beta)/\cos\theta_{\ss W} )
   (g_2\mz(2\sin2\alpha \cos(\beta+\alpha)+\nl \sin(\beta+\alpha)\cos2\alpha)/
(2\cos\theta_{\ss W} ))\nonumber \\
      \f2 &=&   (-g_2 \mz
      \sin^{2}\!\theta_{\ss W} \sin(\alpha+\beta)/\cos\theta_{\ss W} )
   (-g_2\mz(2\sin2\alpha \sin(\beta+\alpha)-\nl \cos(\beta+\alpha)\cos2\alpha)/
(2\cos\theta_{\ss W} ))\nonumber \\
       \f3&=&-g_2^2 \sin2\alpha \sin^2\theta_{\ss W}/ (4\cos^2\theta_{\ss W})\nonumber \\
       \f4&=&(g_2 \mz\sin^{2}\!\theta_{\ss W} 
\cos(\alpha+\beta)/\cos\theta_{\ss W} ) (-g_2 \mz\sin^{2}\!
\theta_{\ss W} \sin(\alpha+\beta)/\cos\theta_{\ss W} )    \nonumber \\
       \f5&=&(g_2 \mz\sin^{2}\!\theta_{\ss W} 
\cos(\alpha+\beta)/\cos\theta_{\ss W} ) (-g_2 \mz\sin^{2}\!
\theta_{\ss W} \sin(\alpha+\beta)/\cos\theta_{\ss W} )    \nonumber \\
      {\cal T}_{\rm I}\!\!\times\!\!{\cal T}_{\rm I} &=& (\mhb^2 - s)^{-2}    \nonumber \\
      {\cal T}_{\rm I}\!\!\times\!\!{\cal T}_{\rm II} &=& 1/((\mhb^2 - s) (\mhl^2 - s))    \nonumber \\
      {\cal T}_{\rm I}\!\!\times\!\!{\cal T}_{\rm III} &=& 1/(\mhb^2 - s)    \nonumber \\
      {\cal T}_{\rm I}\!\!\times\!\!{\cal T}_{\rm IV} &=& 1/((\mhb^2 - s) (\mst^2 - t))    \nonumber \\
      {\cal T}_{\rm I}\!\!\times\!\!{\cal T}_{\rm V} &=& 1/((\mhl^2 - s) (\mst^2 - u)) \nonumber\\
      {\cal T}_{\rm II}\!\!\times\!\!{\cal T}_{\rm II} &=& (\mhl^2 - s)^{-2}    \nonumber \\
      {\cal T}_{\rm II}\!\!\times\!\!{\cal T}_{\rm III} &=& 1/(\mhl^2 - s)    \nonumber \\
      {\cal T}_{\rm II}\!\!\times\!\!{\cal T}_{\rm IV} &=& 1/((\mhl^2 - s) (\mst^2 - t))    \nonumber \\
      {\cal T}_{\rm II}\!\!\times\!\!{\cal T}_{\rm V} &=& 1/((\mhb^2 - s) (\mst^2 - u))\nonumber\\
      {\cal T}_{\rm III}\!\!\times\!\!{\cal T}_{\rm III} &=& 1   \nonumber \\
      {\cal T}_{\rm III}\!\!\times\!\!{\cal T}_{\rm IV} &=& 1/(\mst^2 - t)    \nonumber \\
      {\cal T}_{\rm III}\!\!\times\!\!{\cal T}_{\rm V} &=& 1/(\mst^2 - u)    \nonumber \\
      {\cal T}_{\rm IV}\!\!\times\!\!{\cal T}_{\rm IV} &=& (\mst^2 - t)^{-2}    \nonumber \\
      {\cal T}_{\rm IV}\!\!\times\!\!{\cal T}_{\rm V} &=& 1/((\mst^2 - t) (\mst^2 - u)) \nonumber \\
      {\cal T}_{\rm V}\!\!\times\!\!{\cal T}_{\rm V} &=& (\mst^2 - u)^{-2}    \nonumber \\
      \tsq &=&   \f1^2 {\cal T}_{\rm I}\!\!\times\!\!{\cal T}_{\rm I} +  \f2^2 {\cal T}_{\rm II}\!\!\times\!\!{\cal T}_{\rm II} + \f3^2 {\cal T}_{\rm III}\!\!\times\!\!{\cal T}_{\rm 
III} + \f4^2 {\cal T}_{\rm IV}\!\!\times\!\!{\cal T}_{\rm IV} +
      \f5^2 {\cal T}_{\rm V}\!\!\times\!\!{\cal T}_{\rm V} + 2 \f1 \f2 {\cal T}_{\rm I}\!\!
\times\!\!{\cal T}_{\rm II} + \nl 2 \f1 \f3 {\cal T}_{\rm I}\!\!\times\!\!{\cal T}_{\rm III} + 
      2 \f1 \f4 {\cal T}_{\rm I}\!\!\times\!\!{\cal T}_{\rm IV} + 2 \f1 \f5  {\cal T}_{\rm 
I}\!\!\times\!\!{\cal T}_{\rm V} +2 \f2 \f3 {\cal T}_{\rm II}\!\!\times\!\!{\cal T}_{\rm III} +
      2 \f2 \f4 {\cal T}_{\rm II}\!\!\times\!\!{\cal T}_{\rm IV} +\nl 2 \f2 \f5 {\cal T}_{\rm II}\!\!\times\!\!{\cal T}_{\rm V} + 2 \f3 \f4 {\cal T}_{\rm III}\!\!\times\!\!{\cal T}_{
\rm IV} + 2 \f3 \f5 {\cal T}_{\rm III}\!\!\times\!\!{\cal T}_{\rm V} +2 \f4 \f5 {\cal T}_{\rm I
V}\!\!\times\!\!{\cal T}_{\rm V}  
\end{eqnarray}
\subsection*{$\stau\stau^*\longrightarrow HH$}
 I.   s-channel $H$ annihilation\hfill\\
 II.  s-channel $h$ annihilation \hfill\\ 
 III.  point interaction\hfill\\
 IV. t-channel $\stau$ exchange \hfill\\
 V. u-channel $\stau$ exchange \hfill\\
\begin{eqnarray}
      \f1 &=& (g_2 \mz \sin^{2}\!\theta_{\ss W} \cos(\alpha+\beta)/\cos\theta_{\ss W} ) 
      (-3 g_2 \mz \cos2\alpha\cos(\alpha+\beta)/(2\cos\theta_{\ss W}))\nonumber \\
      \f2 &=&  (-g_2 \mz\sin^{2}\!\theta_{\ss W} \sin(\alpha+\beta)/\cos\theta_{\ss W}) 
      (g_2 \mz(\cos2\alpha \sin(\alpha+\beta) +\nl
      2 \sin2\alpha \cos(\alpha+\beta))/(2\cos\theta_{\ss W}))    \nonumber \\
      \f3 &=& g_2^2\cos2\alpha \sin^{2}\!\theta_{\ss W}/(2 \cos^{2}\!\theta_{\ss W})     \nonumber \\
      \f4 &=& (g_2 \mz\sin^{2}\!\theta_{\ss W} \cos(\alpha+\beta)/\cos\theta_{\ss W}  )^2    \nonumber \\
      \f5 &=& (g_2 \mz\sin^{2}\!\theta_{\ss W} \cos(\alpha+\beta)/\cos\theta_{\ss W}  )^2    \nonumber \\
      \tsq &=&   \f1^2 {\cal T}_{\rm I}\!\!\times\!\!{\cal T}_{\rm I} +  \f2^2 {\cal T}_{\rm II}\!\!\times\!\!{\cal T}_{\rm II} + \f3^2 {\cal T}_{\rm III}\!\!\times\!\!{\cal T}_{\rm III} + \f4^2 {\cal T}_{\rm IV}\!\!\times\!\!{\cal T}_{\rm IV} +
      \f5^2 {\cal T}_{\rm V}\!\!\times\!\!{\cal T}_{\rm V} + 2 \f1 \f2 {\cal T}_{\rm I}\!\!\times\!\!{\cal T}_{\rm II} + \nl 2 \f1 \f3 {\cal T}_{\rm I}\!\!\times\!\!{\cal T}_{\rm III} + 
      2 \f1 \f4 {\cal T}_{\rm I}\!\!\times\!\!{\cal T}_{\rm IV} + 2 \f1 \f5  {\cal T}_{\rm I}\!\!\times\!\!{\cal T}_{\rm V} +2 \f2 \f3 {\cal T}_{\rm II}\!\!\times\!\!{\cal T}_{\rm III} +
      2 \f2 \f4 {\cal T}_{\rm II}\!\!\times\!\!{\cal T}_{\rm IV} +\nl 2 \f2 \f5 {\cal T}_{\rm II}\!\!\times\!\!{\cal T}_{\rm V} + 2 \f3 \f4 {\cal T}_{\rm III}\!\!\times\!\!{\cal T}_{\rm IV} + 
      2 \f3 \f5 {\cal T}_{\rm III}\!\!\times\!\!{\cal T}_{\rm V} +2 \f4 \f5 {\cal T}_{\rm IV}\!\!\times\!\!{\cal T}_{\rm V}  
\end{eqnarray}
The ${\cal T}_{\rm I}\!\!\times\!\!{\cal T}_{\rm I}\ldots$ are the same as for $\stau\stau^*\longrightarrow hh$, with 
($\mhl\leftrightarrow\mhb$).
\subsection*{$\stau\stau^*\longrightarrow H^+H^-$}
 I.   s-channel $H$ annihilation\hfill\\
 II.  s-channel $h$ annihilation \hfill\\ 
 III. s-channel $Z$ annihilation \hfill\\
 IV. s-channel $\gamma$ annihilation \hfill\\
 V.  point interaction\hfill\\
\begin{eqnarray}
      \f1 &=& (g_2 \mz \sin^{2}\!\theta_{\ss W} \cos(\alpha+\beta)/\cos\theta_{\ss W} ) 
(-g_2(\mw \cos(\beta-\alpha)-\nl \mz\cos2\beta \cos(\beta+\alpha)/(2\cos\theta_{\ss W})))
\nonumber \\
      \f2 &=&  (-g_2 \mz\sin^{2}\!\theta_{\ss W} \sin(\alpha+\beta)/\cos\theta_{\ss W}) 
(-g_2(\mw \sin(\beta-\alpha)+\nl \mz\cos2\beta \sin(\beta+\alpha)/(2\cos\theta_{\ss W})))
\nonumber \\
      \f3&=& (-g_2\sin^{2}\!\theta_{\ss W}/\cos\theta_{\ss W} ) (-g_2 \cos2\theta_{\ss W}/
(2 \cos\theta_{\ss W}))\nonumber \\
      \f4&=& -e^2\nonumber \\
      \f5&=& -g_2^2\cos2\beta \sin^{2}\!\theta_{\ss W}/(2 \cos^{2}\!\theta_{\ss W} )\nonumber \\
      {\cal T}_{\rm I}\!\!\times\!\!{\cal T}_{\rm I} &=& (\mhb^2 - s)^{-2}    \nonumber \\
      {\cal T}_{\rm I}\!\!\times\!\!{\cal T}_{\rm II} &=& 1/((\mhb^2 - s) (\mhl^2 - s))    \nonumber \\
      {\cal T}_{\rm I}\!\!\times\!\!{\cal T}_{\rm III} &=& (t - u)/((\mhb^2 - s) (\mz^2 - s))    \nonumber \\
      {\cal T}_{\rm I}\!\!\times\!\!{\cal T}_{\rm IV} &=& (t - u)/(-(\mhb^2 s) + s^2)    \nonumber \\
      {\cal T}_{\rm I}\!\!\times\!\!{\cal T}_{\rm V} &=& 1/(\mhb^2 - s)    \nonumber \\
      {\cal T}_{\rm II}\!\!\times\!\!{\cal T}_{\rm II} &=& (\mhl^2 - s)^{-2}    \nonumber \\
      {\cal T}_{\rm II}\!\!\times\!\!{\cal T}_{\rm III} &=& (t - u)/((\mhl^2 - s) (\mz^2 - s))    \nonumber \\
      {\cal T}_{\rm II}\!\!\times\!\!{\cal T}_{\rm IV} &=& (t - u)/(-(\mhl^2 s) + s^2)    \nonumber \\
      {\cal T}_{\rm II}\!\!\times\!\!{\cal T}_{\rm V} &=& 1/(\mhl^2 - s)    \nonumber \\
      {\cal T}_{\rm III}\!\!\times\!\!{\cal T}_{\rm III} &=& (t - u)^2/(\mz^2 - s)^2    \nonumber \\
      {\cal T}_{\rm III}\!\!\times\!\!{\cal T}_{\rm IV} &=& -((t - u)^2/((\mz^2 - s) s))    \nonumber \\
      {\cal T}_{\rm III}\!\!\times\!\!{\cal T}_{\rm V} &=& (t - u)/(\mz^2 - s)    \nonumber \\
      {\cal T}_{\rm IV}\!\!\times\!\!{\cal T}_{\rm IV} &=& (t - u)^2/s^2    \nonumber \\
      {\cal T}_{\rm IV}\!\!\times\!\!{\cal T}_{\rm V} &=& (-t + u)/s    \nonumber \\
      {\cal T}_{\rm V}\!\!\times\!\!{\cal T}_{\rm V} &=& 1\nonumber \\
      \tsq &=&   \f1^2 {\cal T}_{\rm I}\!\!\times\!\!{\cal T}_{\rm I} +  \f2^2 {\cal T}_{\rm II}\!\!\times\!\!{\cal T}_{\rm II} + \f3^2 {\cal T}_{\rm III}\!\!\times\!\!{\cal T}_{\rm III} + \f4^2 {\cal T}_{\rm IV}\!\!\times\!\!{\cal T}_{\rm IV} +
      \f5^2 {\cal T}_{\rm V}\!\!\times\!\!{\cal T}_{\rm V} + 2 \f1 \f2 {\cal T}_{\rm I}\!\!\times\!\!{\cal T}_{\rm II} + \nl 2 \f1 \f3 {\cal T}_{\rm I}\!\!\times\!\!{\cal T}_{\rm III} + 
      2 \f1 \f4 {\cal T}_{\rm I}\!\!\times\!\!{\cal T}_{\rm IV} + 2 \f1 \f5  {\cal T}_{\rm I}\!\!\times\!\!{\cal T}_{\rm V} +2 \f2 \f3 {\cal T}_{\rm II}\!\!\times\!\!{\cal T}_{\rm III} +
      2 \f2 \f4 {\cal T}_{\rm II}\!\!\times\!\!{\cal T}_{\rm IV} + \nl 2 \f2 \f5 {\cal T}_{\rm II}\!\!\times\!\!{\cal T}_{\rm V} + 2 \f3 \f4 {\cal T}_{\rm III}\!\!\times\!\!{\cal T}_{\rm IV} + 
      2 \f3 \f5 {\cal T}_{\rm III}\!\!\times\!\!{\cal T}_{\rm V} +2 \f4 \f5 {\cal T}_{\rm IV}\!\!\times\!\!{\cal T}_{\rm V}  
\end{eqnarray}
\subsection*{$\stau\stau\longrightarrow \tau\tau$}
 I. t-channel $\chi$ exchange \hfill\\
 II. u-channel $\chi$ exchange \hfill\\
\begin{eqnarray}
            K&=& g_1 N_{i1}/\sqrt{2}    \nonumber \\
            K'&=& g_1 N_{j1}/\sqrt{2}    \nonumber \\
            {\cal T}_{\rm I}\!\!\times\!\!{\cal T}_{\rm I}&=& (16 K^2 K'^2 {\mchi}_i {\mchi}_j s)/(({\mchi}_i^2 - t) 
            ({\mchi}_j^2 - t))    \nonumber \\
            {\cal T}_{\rm II}\!\!\times\!\!{\cal T}_{\rm II}&=& (16 K^2 K'^2 {\mchi}_i {\mchi}_j s)/(({\mchi}_i^2 - u) 
            ({\mchi}_j^2 - u))    \nonumber \\
            {\cal T}_{\rm I}\!\!\times\!\!{\cal T}_{\rm II}&=& (16 K^2 K'^2 {\mchi}_i {\mchi}_j s)/(({\mchi}_i^2 - t) 
            ({\mchi}_j^2 - u))    \nonumber \\
            \tsq &=&   \sum_{i,j=1}^4 ({\cal T}_{\rm I}\!\!\times\!\!{\cal T}_{\rm I} +  {\cal T}_{\rm II}\!\!\times\!\!{\cal T}_{\rm II} + 2 {\cal T}_{\rm I}\!\!\times\!\!{\cal T}_{\rm II}    )
\end{eqnarray}
\subsection*{$\stau\tilde\ell^*\longrightarrow \tau\bar\ell$}
 I. t-channel $\chi$ exchange \hfill\\
\begin{eqnarray}
            K&=& g_1 N_{i1}/\sqrt{2}    \nonumber \\
            K'&=& g_1 N_{j1}/\sqrt{2}    \nonumber \\
      {\cal T}_{\rm I}\!\!\times\!\!{\cal T}_{\rm I}&=& -16 K^2 K'^2 (\mst^2m_{\tilde\ell_R}^2 - t u )/(({\mchi}_i^2 - t) ({\mchi}_j^2 - t))    \nonumber \\
      \tsq&=&\sum_{i,j=1}^4 {\cal T}_{\rm I}\!\!\times\!\!{\cal T}_{\rm I}  
\end{eqnarray}
\subsection*{$\stau\tilde\ell\longrightarrow \tau\ell $}
 I. t-channel $\chi$ exchange \hfill\\
\begin{eqnarray}
            K&=& g_1 N_{i1}/\sqrt{2}    \nonumber \\
            K'&=& g_1 N_{j1}/\sqrt{2}    \nonumber \\
            {\cal T}_{\rm I}\!\!\times\!\!{\cal T}_{\rm I}&=& (16 K^2 K'^2 {\mchi}_i {\mchi}_j s)/(({\mchi}_i^2 - t) 
            ({\mchi}_j^2 - t))    \nonumber \\
            \tsq &=&   \sum_{i,j=1}^4 {\cal T}_{\rm I}\!\!\times\!\!{\cal T}_{\rm I}  
\end{eqnarray}
\subsection*{$\stau\chi\longrightarrow Z\tau$}
 I. s-channel $\tau$ annihilation \hfill\\
 II. t-channel $\stau$ exchange \hfill\\
\begin{eqnarray}
       \f1&=& -(g_1 N_{j1} /\sqrt{2}) 
         (-g_2/(2\cos\theta_{\ss W}))    \nonumber \\
       \f2&=& (g_1 N_{j1}/\sqrt{2}) 
          (-g_2\sin^{2}\!\theta_{\ss W}/\cos\theta_{\ss W})    \nonumber \\
     {\cal T}_{\rm I}\!\!\times\!\!{\cal T}_{\rm I}&=& 2 (2 \sin^{2}\!\theta_{\ss W})^2 (\mchi^4 s - \mst^4 s + 
       s (-\mz^4 + \mz^2 s + \mz^2 t - s t - \mz^2 u) + \nl
       \mst^2 (2 \mz^4 - 2 \mz^2 s + s^2 + s t + s u) - 
       \mchi^2 (2 \mz^4 - 2 \mz^2 s + s (t + u)))/ 
      (\mz^2 s^2)    \nonumber \\
      {\cal T}_{\rm II}\!\!\times\!\!{\cal T}_{\rm II}&=&   2 (\mchi^2 - t) (\mst^4 + (\mz^2 - t)^2 - 
       2 \mst^2 (\mz^2 + t))/(\mz^2 (\mst^2 - t)^2)    \nonumber \\
      {\cal T}_{\rm I}\!\!\times\!\!{\cal T}_{\rm II}&=& -(2 \sin^{2}\!\theta_{\ss W}) (\mst^4 (s + t - u) + 
         \mchi^2 (-(\mz^2 s) + \mz^2 t + s t - t^2 + \nl
            \mst^2 (8 \mz^2 - s + t - u) - 5 \mz^2 u + t u) + 
         (\mz^2 - t) (\mz^2 s - s^2 + \mz^2 t - t^2 - \nl
            \mz^2 u + u^2) + 
         \mst^2 (2 \mz^2 s - s^2 - 2 \mz^2 t - s t - 2 t^2 - 
            2 \mz^2 u + t u + u^2))/\nl(\mz^2 s (\mst^2 - t))    \nonumber \\
     \tsq &=&   \f1^2 {\cal T}_{\rm I}\!\!\times\!\!{\cal T}_{\rm I} +  \f2^2 {\cal T}_{\rm II}\!\!\times\!\!{\cal T}_{\rm II} +2 \f1 \f2 {\cal T}_{\rm I}\!\!\times\!\!{\cal T}_{\rm II} 
\end{eqnarray}
\subsection*{$\stau\chi\longrightarrow \gamma\tau$}
 I. s-channel $\tau$ annihilation \hfill\\
 II. t-channel $\stau$ exchange \hfill\\
\begin{eqnarray}
       \f1&=& -(g_1 N_{j1}/\sqrt{2}) (e)    \nonumber \\
       \f2&=& (g_1 N_{j1}/\sqrt{2}) (e)    \nonumber \\
      {\cal T}_{\rm I}\!\!\times\!\!{\cal T}_{\rm I}&=& 4 (\mchi^4 - \mst^4 - s u - \mchi^2 (t + u) + 
       \mst^2 (s + t + u))/s^2    \nonumber \\
      {\cal T}_{\rm II}\!\!\times\!\!{\cal T}_{\rm II}&=& -4 (\mchi^2 - t) (\mst^2 + t)/(\mst^2 - t)^2    \nonumber \\
      {\cal T}_{\rm I}\!\!\times\!\!{\cal T}_{\rm II}&=& (s^2 + t^2 - u^2 + \mst^2 (-s + 3 t + u) + 
     \mchi^2 (-8 \mst^2 + s - t + 5 u))/(s (\mst^2 - t))    \nonumber \\
      \tsq &=&   \f1^2 {\cal T}_{\rm I}\!\!\times\!\!{\cal T}_{\rm I} +  \f2^2 {\cal T}_{\rm II}\!\!\times\!\!{\cal T}_{\rm II} +2 \f1 \f2 {\cal T}_{\rm I}\!\!\times\!\!{\cal T}_{\rm II} 
\end{eqnarray}
\subsection*{$\stau\chi\longrightarrow \tau h [H]$}
 II. t-channel $\stau$ exchange \hfill\\
\begin{eqnarray}
       \f2&=& -(g_1 N_{j1}/\sqrt{2} )
       (-g_2\mz \sin^{2}\!\theta_{\ss W} \sin[-\cos](\alpha+\beta)/\cos\theta_{\ss W} )    \nonumber \\
      {\cal T}_{\rm II}\!\!\times\!\!{\cal T}_{\rm II} &=& 2 (\mchi^2 - t)/(\mst^2 - t)^2   \nonumber \\
      \tsq &=&   \f2^2 {\cal T}_{\rm II}\!\!\times\!\!{\cal T}_{\rm II}  
\end{eqnarray}